\documentclass[12pt,reqno]{amsart}

\usepackage[left=1in, right=1in, top=1.1in,bottom=1.1in]{geometry}
\usepackage[title]{appendix}

\usepackage{graphicx}
\graphicspath{ {./fig/}}

\usepackage[T1]{fontenc}
\usepackage{amsthm,amssymb,amsmath,amsfonts,mathrsfs,amscd}
\usepackage{bm}
\usepackage{enumitem}
\usepackage{stmaryrd}
\usepackage{mathrsfs}

\usepackage[font={scriptsize}]{caption}

\usepackage{hyperref}
\usepackage{apacite}
\usepackage{cancel}
\usepackage{soul}
\usepackage{multirow}
\usepackage{booktabs}
\usepackage{tabularx}
\ifdefined\journalbuild
  \usepackage[doublespacing]{setspace}
\fi
\IfFileExists{placeins.sty}{\usepackage{placeins}}{\providecommand{\FloatBarrier}{\clearpage}}
\hypersetup{
  colorlinks   = true,
  citecolor    = green,
  linkcolor    = red
}

\newtheorem{theorem}{Theorem}[section]

\newtheorem{definition}[theorem]{Definition}

\newtheorem{lemma}[theorem]{Lemma}

\theoremstyle{remark}
\newtheorem{remark}[theorem]{Remark}

\theoremstyle{remark}
\newtheorem{example}[theorem]{Example}

\newtheorem{assumption}[theorem]{Assumption}

\newcommand{\E}{\mathbb E}
\newcommand{\R}{\mathbb R}
\newcommand{\N}{\mathbb N}

\newcommand{\diag}{{\rm diag}}

\newcommand{\cA}{\mathcal{A}}
\newcommand{\cL}{\mathcal{L}}

\newcommand{\bA}{\boldsymbol A}
\newcommand{\be}{\boldsymbol e}
\newcommand{\bh}{\boldsymbol h}
\newcommand{\bM}{\boldsymbol M}
\newcommand{\bU}{\boldsymbol U}
\newcommand{\bu}{\boldsymbol u}

\newcommand{\bV}{\boldsymbol V}
\newcommand{\bW}{\boldsymbol W}

\newcommand{\bGamma}{\boldsymbol\Gamma}
\newcommand{\bLambda}{\boldsymbol\Lambda}
\newcommand{\bSigma}{\boldsymbol\Sigma}
\newcommand{\bsigma}{\boldsymbol\sigma}

\newcommand{\Eof}[1]{\mathbb{E}\left[ #1 \right]}
\newcommand{\p}{\partial}
\DeclareMathOperator{\cov}{Cov}

\DeclareMathOperator*{\argmin}{arg\,min}
\newcommand{\lvr}{{\rm LVR}}
\newcommand{\cV}{\mathcal{V}}

\title[Dynamics of Concentrated Liquidity]{Dynamics of the Liquidity Profile in Concentrated Liquidity Provision: A Stochastic Partial Integro-Differential Equation Approach}

\author[C.-Y. Lee]{Cheuk Yin Lee}
\author[J. Risk]{Jimmy Risk}
\author[S.-N. Tung]{Shen-Ning Tung}
\author[T.-H. Wang]{Tai-Ho Wang}
\date{\today}

\address{Cheuk Yin Lee \newline
School of Science and Engineering, \newline
The Chinese University of Hong Kong, Shenzhen \newline
Shenzhen, China
}
\email{leecheukyin@cuhk.edu.cn}

\address{Jimmy Risk \newline
Department of Mathematics and Statistics, \newline
Cal Poly Pomona\newline
3801 W Temple Ave, Pomona CA 91768
}
\email{jrisk@cpp.edu}

\address{Shen-Ning Tung \newline
Department of Mathematics, \newline
National Tsing Hua University \newline
Hsinchu, Taiwan
}
\email{tung@math.nthu.edu.tw}

\address{Tai-Ho Wang \newline
Department of Mathematics \newline
Baruch College, The City University of New York \newline
1 Bernard Baruch Way, New York, NY 10010
}
\email{tai-ho.wang@baruch.cuny.edu}


\begin{document}

\begin{abstract}
We model the dynamics of the liquidity profile in Uniswap~V3 and other concentrated-liquidity automated market makers with a stochastic partial integro-differential equation (SPIDE). The profile, the aggregate of all provider positions across tick space, evolves under a degenerate Sturm--Liouville operator, which encodes mean-reverting concentration around the market price, coupled with a Hilbert--Schmidt integral operator, which captures nonlocal redistribution driven by provider repositioning. We establish existence and uniqueness of the mild solution in a weighted $L^2$ space built on the Gegenbauer eigensystem, and reduce the SPIDE to a coupled system of linear stochastic differential equations whose low-order modes describe the level, slope, and curvature of the profile. Pairing the SPIDE with a log-price process yields tractable expressions for pool reserves, pool value, and loss-versus-rebalancing under Ornstein--Uhlenbeck-type and Black--Scholes-type noise. A finite-rank calibration in a fixed reference frame, applied to three Uniswap pools, gives interpretable estimates and lowers active-band forecast error relative to a martingale benchmark on the two pools compatible with the stationary specification. On the third pool, out-of-sample diagnostics are consistent with nonstationarity, which motivates price-aware and time-varying extensions.
\end{abstract}

\maketitle
\allowdisplaybreaks

\section{Introduction}\label{sec:introduction}

\textit{Automated Market Makers (AMMs)} have revolutionized decentralized finance by enabling permissionless and continuous trading of digital assets \cite{harvey2021defi, capponi2021adoption, gobet2023decentralized}. A landmark development in this space is Uniswap~V3 \cite{adams2021uniswap}, which introduced \textit{concentrated liquidity}: rather than distributing capital uniformly across all prices as in its predecessor Uniswap~V2 \cite{adams2020uniswap}, V3 allows liquidity providers (LPs) to allocate capital within user-specified price ranges. This mechanism dramatically improves capital efficiency, but it also gives rise to a liquidity profile — the aggregation of all overlapping LP positions — that is highly non-uniform, inherently dynamic, and tightly coupled to market conditions. Understanding the evolution of this profile is the central focus of the present paper; it is equally relevant to LPs seeking to optimize returns, protocol developers refining AMM design, and researchers studying market stability in decentralized venues.

Traditional stochastic processes are typically designed for finite-dimensional systems. The Uniswap~V3 liquidity profile, however, is a continuous function over a shifting, reshaping, and clustering price range in response to provider repositioning and market volatility. Modeling its evolution, therefore, requires an \textit{infinite-dimensional} approach \cite{tung2026mathematical}. Existing models often simplify these dynamics or focus solely on the price process, overlooking the detailed evolution of the liquidity distribution itself.

Recent empirical work by \citeA{risk2025dynamics} has begun to bridge this gap by using \textit{functional principal component analysis} and \textit{dynamic factor methods} to study the evolving Uniswap~V3 liquidity profile. For methodological background on functional principal component analysis, see \citeA{ramsay2005functional} and \citeA{wang2016functional}; for dynamic factor methods, see \citeA{shen2009modeling} and \citeA{hays2012functional}. Using a low-order Legendre polynomial basis, the empirical study illustrates the characteristic shapes of liquidity as it shifts and clusters around the market price. Empirical observation of these surfaces reveals three distinct behaviors that necessitate a sophisticated modeling approach: \textit{clustering} around the market price, \textit{mean reversion} toward a centered shape, and \textit{spatiotemporal correlations} where shifts in one price area affect distant ticks.

In this paper, we introduce a novel mathematical framework for modeling the dynamics of the AMM liquidity profile using a \textit{Stochastic Partial Integro-Differential Equation} (SPIDE). This framework is uniquely suited to capture two critical features of decentralized markets:
\begin{enumerate}
    \item \textit{Nonlocal Interactions}: Capturing how liquidity changes in one price bracket affect distant brackets.
    \item \textit{Continuous Evolution}: Modeling the profile as a fluid, evolving density rather than a static set of bins.
\end{enumerate}
By leveraging \textit{orthonormal eigenfunction expansions}, we show how this infinite-dimensional SPIDE can be systematically transformed into a more tractable system of \textit{linear stochastic differential equations}. This transformation is pivotal; it bridges the gap between complex functional analysis and practical application, opening avenues for efficient parameter estimation and data analysis using techniques adapted for infinite-dimensional systems.
The choice of the SPIDE framework is directly motivated by empirical evidence: persistent level movements, mean-reverting curvature, and cross-mode dependence indicate that liquidity is actively redistributed rather than drifting independently at each tick. Our model incorporates an integral term, the nonlocal component, to represent these cross-mode reallocations while retaining an interpretable differential operator for mean-reverting concentration around the current price.

The use of stochastic partial differential equations (SPDEs) to model the dynamics of order-driven markets has precedent in the \textit{limit order book (LOB)} literature. \citeA{hambly2020limit} derive reflected SPDEs as macroscopic limits of microscopic LOB models via a zero-intelligence scaling procedure, connecting discrete price--volume dynamics to a continuous interface description. \citeA{cont2021stochastic} propose an SPDE with multiplicative noise for the order book density centered at the mid-price and show that finite-dimensional realizations arise under suitable structural assumptions, yielding tractable estimation and reproducing key statistical features of market depth and order flow. Our framework differs from both approaches in several respects. First, while the LOB models describe the supply--demand schedule around a single mid-price in a centralized venue, our SPIDE governs the \textit{aggregated} liquidity profile of a decentralized AMM pool, where providers actively position discrete ranges rather than submit individual limit orders. Second, the degenerate Sturm--Liouville operator in our model---with coefficients tied to the Gegenbauer eigensystem---is specifically designed to encode the mean-reverting, bell-shaped concentration of liquidity observed empirically in Uniswap~V3, a spatial feature absent from LOB density dynamics. Third, the integral (nonlocal) component of our SPIDE captures cross-tick redistribution driven by LP repositioning, whereas the LOB models of \citeA{hambly2020limit} and \citeA{cont2021stochastic} rely on local diffusion or multiplicative noise to generate spatial covariance. Finally, our eigenfunction expansion reduces the SPIDE to a coupled system of stochastic differential equations (SDEs) whose low-order modes carry the same level, slope, and curvature interpretation used in the empirical Legendre diagnostics, while the Gegenbauer order $\alpha$ is allowed to vary in the theoretical and calibration framework.

On the inference side, our setting also falls awkwardly between the existing SPDE estimation molds. Weighted-space spectral estimators often rely on spatio-temporal infill to pre-estimate the parameters defining the basis and then hold it fixed \cite{kaino2021parametric, tonaki2025parametric}; realized-volatility regressions and refining-grid consistency results work directly on physical-grid data \cite{bibinger2023efficient, hildebrandt2021parameter}; generator-eigenfunction martingale methods exploit ensembles of trajectories \cite{pavliotis2022eigenfunction, pavliotis2024method}; and the LOB SPDE work above either estimates on averaged profiles or derives macroscopic limits without a coupled operator-fit objective \cite{cont2021stochastic, hambly2020limit}. None of these routes matches the present calibration problem. We observe one liquidity surface on a fixed tick grid, with a nonlocal relocations operator and a Gegenbauer order $\alpha$ that changes the very space in which spectral residuals are measured. This is why the empirical section calibrates in a fixed Legendre evaluation space rather than in a freely moving parameter-dependent basis.

The rest of the paper is organized as follows. We review in Section \ref{sec:motivation} the notions of concentrated liquidity provision, liquidity profile, and liquidity surface for a liquidity pool in AMM. Section \ref{sec:motivation} also motivates, in more detail, the use of SPIDE for modeling the evolution of the liquidity profile in a concentrated liquidity pool, in line with empirically observed stylized facts. Section \ref{sec:th-bg} provides the necessary theoretical backgrounds and summarizes the proof of the existence and uniqueness of the mild solution to the SPIDE under consideration. Section \ref{sec:app-ls} presents the applications and implications of the SPIDE theory developed in Section \ref{sec:th-bg} for modeling the evolution of the liquidity profile.  An empirical case study on Uniswap V3 pools is conducted in Section \ref{sec:case-study}.
Section \ref{sec:conclusion} concludes the paper with discussions and suggestions on future studies. All technical proofs and derivations are collected in Appendix \ref{sec:math-tech} for the purpose of reading fluency.

\section{Reviews and Motives} \label{sec:motivation}
The evolution from Uniswap v2’s uniform liquidity distribution \cite{adams2020uniswap} to the concentrated liquidity model of v3 \cite{adams2021uniswap} has fundamentally altered the observation and modeling of market depth. Under this new paradigm, liquidity is no longer a static parameter but a dynamic, continuous surface that responds to market volatility and provider sentiment.

\subsection{Fundamentals of Concentrated Liquidity} \label{sec:background-uniswap}
The capital efficiency of Uniswap v3 is rooted in the ability of liquidity providers (LPs) to concentrate capital within a user-defined price range $[p_l, p_r)$. This mechanism is driven by three interconnected components:

\begin{enumerate}
    \item \textit{Liquidity Level $L$:} This represents the "virtual reserves" an LP contributes, analogous to the constant-product formula $L = \sqrt{xy}$ used in Uniswap v2. It determines the LP’s share of trading fees and price exposure.
    \item \textit{Provision Range $[p_l, p_r)$:} Assets are only active for swaps when the current market price $p$ resides within this designated interval. Outside this range, the position is composed entirely of a single asset and earns no fees.
    \item \textit{Aggregated Liquidity $L_{\text{total}}$:} The pool’s total depth is the sum of all overlapping individual positions. At any specific price $q \in (0, \infty)$, the total liquidity available at $q$ is: $$L_{\text{total}}(q) = \sum_{i=1}^N L_i \cdot \mathbf{1}_{[p_{\ell,i}, p_{r,i})}(q)$$ This summation results in a \textit{non-uniform liquidity distribution}, also termed as the {\it liquidity profile}, across the price spectrum, creating a dynamic "landscape" of depth that shifts as LPs adjust their ranges in response to market volatility. At any point in time $t$ and price level $q$, we denote the total liquidity $L_{\text{total}}(q)$ by $L_t(q)$, to emphasize the dependence of liquidity profile on time. 
\end{enumerate}
For more detailed analyses on the relationships among the pool reserve, the pool price, as well as the liquidity profile, we refer the reader to \citeA{tung2026mathematical} and to \citeA{risk2026pricing} for the pricing and hedging of liquidity provision.

\subsection{The Liquidity Surface} \label{sec:liq-sur}
To formalize the dynamics of the pool, we transition from the nominal price level $q$ to the \textit{tick space} $x$ centered around the pool price in log-scale. In fact, ticks are what are reported in the market quotes. Specifically, assume at time $t$, the pool price is $p_t$ and the liquidity profile is $L_t(\cdot)$. Define the spot-centered logarithmic liquidity profile $u_t$ at tick $x \in \mathbb{R}$ by
\begin{equation*}
u_t(x) = \ln L_t(p_t e^x).
\end{equation*}
We refer to the state of the pool as the evolution of the spot-centered log liquidity $u_t(x)$ over time $t$. This log-liquidity coordinate is a modeling convention, not a requirement of the theory developed in Section~\ref{sec:th-bg}; Section~\ref{sec:app-ls} treats both the log-liquidity convention $u_t$ and the raw-liquidity convention $U_t(x)=L_t(p_te^x)$, which equals $e^{u_t(x)}$ for a positive profile; the two conventions lead to alternative models rather than to one solution written in two coordinates. Empirical observations of the surface $u_t(x)$, i.e., $u$ regarded as a function of $t$ and $x$, provided in \citeA{risk2025dynamics} reveal three distinct behaviors that necessitate a sophisticated modeling approach:
\begin{itemize}
    \item \textit{Clustering}: Liquidity tends to concentrate around the current market price, i.e., $u_t(\cdot)$ peaks at tick $x=0$ for the majority of time.
    \item \textit{Mean Reversion}: While the profile shifts, it generally maintains a "bell-like" or "centered" shape relative to the active price.
    \item \textit{Spatiotemporal Correlation}: Changes in liquidity at a specific tick are highly correlated with changes in adjacent or distant ticks, as LPs typically reposition their entire range rather than individual tick spacing.
\end{itemize}

As illustrated in \citeA[Figure 2]{risk2025dynamics}, $u_t(x)$ is not a static curve but a \textit{functional data object} that evolves over time. This empirical reality—where the entire curve evolves as a single entity—provides the primary motivation for an \textit{infinite-dimensional SPIDE approach} rather than a collection of independent scalar SDEs.

\subsection{Dynamic Legendre basis expansion analysis}
In addition to employing functional principal component analysis for the spot-centered log liquidity surface $u_t(x)$, \citeA{risk2025dynamics} also performed a dynamic basis expansion analysis using Legendre polynomials as basis.
The authors projected the liquidity surface $u_t(x)$ onto the linear space spanned by the Legendre polynomials
$$
\hat{v}^{(K)}_t(x) := \sum_{k=1}^K \beta_{t,k} P_{k-1}(x), \; \mbox{ where } \beta_{t,k} = \frac{2k-1}{2}\int_{-1}^1 u_t(\xi) \, P_{k-1}(\xi)  d\xi
$$
and $P_{k-1}(x)$ denotes the Legendre polynomial of degree $k-1$, for $k \geq 1$. Analysis demonstrates that a rank $K=5$ reconstruction captures approximately 90–95\% of the variance. This provides a smooth, low-variance fit that preserves the essential shape of the liquidity profile while effectively filtering high-frequency noise (see \citeNP[Figure 9]{risk2025dynamics}).

The Legendre system is the $\alpha=1/2$ member of the Gegenbauer family used in Section~\ref{sec:th-bg}. Thus the empirical Legendre analysis is not a separate model from the SPIDE below: it supplies the modal interpretation and a stable reference basis, while the theoretical and calibration sections allow the concentration parameter $\alpha$ and the nonlocal relocations operator $V$ to vary.

\subsubsection{Structural Interpretation of Shocks}
Fixing the Legendre polynomials as basis allows us to assign a permanent "market meaning" to each coefficient $\beta_{t,k}$. By applying a one-standard-deviation shock, distinct structural changes in the pool's depth were observed, see \citeA[Figure 10]{risk2025dynamics}. We summarize as follows.
\begin{itemize}
    \item $k=1$ (Level): Corresponds to a uniform vertical shift where liquidity is added or removed across all ticks simultaneously.
    \item $k=2$ (Slope): Induces a tilt effect, increasing liquidity at one end of the price range while decreasing it at the other, signalling a directional shift in market sentiment.
    \item $k=3$ (Curvature): Creates a "flattening" or "peaking" effect. A negative baseline coefficient is standard here to maintain the bell shape; shocks redistribute mass between the center and the tails.
    \item $k=4$ \& $k=5$ (Refinement): These capture complex "mid-level repositioning," where liquidity shifts from intermediate ranges toward both the absolute center and the far edges.
\end{itemize}

\subsubsection{Dynamic Properties and SPIDE Implications}
The time-series behavior of these Legendre coefficients, see \citeA[Figures 11 and 12]{risk2025dynamics}, provides the essential empirical parameters for our stochastic model:

\begin{table}[htbp]
\centering
\caption{Legendre Coefficient Properties}
\footnotesize
\begin{tabularx}{\textwidth}{|l|l|l|>{\raggedright\arraybackslash}X|}
\hline
\textbf{Coefficient} & \textbf{Interpretation} & \textbf{Mean Reversion ($\tau$)} & \textbf{Volatility Characteristic} \\
\hline
$\beta_{1,t}$ (Level)   & Total Pool Depth     & Persistent ($\approx 3$ days) & Highly asymmetric (TARCH/EGARCH) \\
\hline
$\beta_{2,t}$ (Slope)   & Directional Bias     & Slow Reversion                & Split asymmetry \\
\hline
$\beta_{3,t}$ (Curvature) & Center Concentration & Mean-reverting                & Negative baseline, heteroskedastic \\
\hline
$\beta_{4,5,t}$ (Detail) & Local Rebalancing   & Rapid ($\approx 1$ day)       & Symmetric volatility, heavy tails \\
\hline
\end{tabularx}
\end{table}

\subsubsection{The integral operator in SPIDE}
The empirical evidence, specifically the consistent negative mean for $k=3$ (curvature), the distinct heteroskedasticity of $k=1$ (level), and the dependence across low-order coefficients, suggests that liquidity does not merely drift independently across ticks; it is actively redistributed. The integral operator (the nonlocal component) of the SPIDE introduced in Section~\ref{sec:th-bg} is therefore intended as a parameterization of these observed modal interactions. This lets the model represent liquidity mean reversion toward a spot-centered, concentrated profile while allowing provider repositioning to move mass nonlocally across the tick range.

\section{Theoretical Background} \label{sec:th-bg}

We lay out necessary theoretical background in the analysis of stochastic partial integro-differential equations in this section. For reading fluency, certain details of proofs and derivations will be postponed and collected in Appendix \ref{sec:math-tech}.  

Assume that the evolution of the random function $u_t(x)$ is governed by the stochastic partial integro-differential equation (SPIDE) in a fixed interval $[-L, L]$:
\begin{equation}\label{SPIDE}
du_t(x) = \left\{\cL u_t - V_t[u_t] + h(t, x) \right\}dt + \sum_{m=1}^M \left\{ \sigma_m(t,x) + \tilde\sigma_m(t, x)[u_t] \right\} dW^m_t
\end{equation}
with initial condition $u_0(x)$, where $\bW_t = (W^1_t, \dots, W^M_t)$ is an $M$-dimensional Brownian motion, $\cL$ is the linear
second order differential operator 
\begin{align}\label{L}
	\mathcal{L}u = a(L^2-x^2)\partial_x^2 u - b x \partial_x u \quad \text{in $(-L,L)$,}
\end{align}
$h$ is a predictable process, $\sigma_m$ is a function, and 
$V$ and $\tilde\sigma_m$ are linear operators given respectively by kernels
\begin{align}\label{V}
V_t[u] &:= \int V_t(x,y) u(y) w(y) dy, \\
\label{sigma}
\tilde\sigma_m(t,x)[u] &:= \int\tilde\sigma_m(t,x,y) u(y) w(y) dy,
\end{align}
where $w$ is a weight function, specified in \eqref{w}.
The parameters $a$ and $b$ are assumed strictly positive. 
\begin{assumption}\label{a:ab}
$a>0$ and $b>0$.
\end{assumption}
In particular, if $a=1$, $b=2$, and $L=1$, then the operator $\mathcal{L}$ is known as the Legendre operator.
\begin{remark}
We remark that 
\begin{enumerate}
\item The integral operators $V$ and $\tilde\sigma_m$ introduce the integral part to the SPIDE \eqref{SPIDE} and thus make the dynamics nonlocal.
\item Since the differential operator $\mathcal{L}$ is degenerate at the endpoints $x = -L$ and $x = L$, it is thus not uniformly elliptic and the standard theories for elliptic operators and their associated Cauchy problems do not apply directly. 
\item As far as boundary conditions are concerned, 
\begin{itemize}
    \item if $\frac b{2a} \in (0,1)$, we impose the boundary condition $\lim_{x \to \pm L}(L^2-x^2)^{\frac b{2a}} u'(x) = 0$;
    \item if $\frac b{2a} \in [1, \infty)$, no boundary condition is required.
\end{itemize}
The choice of the boundary condition will be justified by Lemma \ref{lem:self-adj} and Remark \ref{rmk:boundary} below.
\end{enumerate}
\end{remark}
In what follows, we study some basic properties of the differential operator $\mathcal{L}$ as well as of the integro-differential operator
\begin{align}\label{A}
	\mathcal{A} = \mathcal{L} - V.
\end{align}
The key property is that $\mathcal A$ generates a $C_0$-semigroup. 
This property will lead to the proof of existence and uniqueness of the mild solution to the SPIDE \eqref{SPIDE} in Section \ref{S:solution}.

\subsection{Eigensystem for \texorpdfstring{$\mathcal{L}$}{L}}

Recall that 
\begin{align}\label{L^ab}
	\mathcal{L} = \mathcal{L}_{L,a,b} = a(L^2-x^2) \partial_x^2 - bx \partial_x \qquad \text{in $(-L,L)$.}
\end{align}
Set
\begin{align}\label{alpha}
	\alpha = \frac{b}{2a}-\frac{1}{2}.
\end{align}
In standard Sturm-Liouville form, the operator $\mathcal{L}$ is written as
\begin{align}\label{L:SL}
	\mathcal{L} u(x) = \frac{1}{w(x)} \left( \frac{d}{dx} \left[ p(x) u'(x) \right] \right),
\end{align}
where
\begin{align}
	\label{w}&w(x) = w_{L,a,b}(x) = (L^2-x^2)^{\frac{b}{2a}-1}= (L^2-x^2)^{\alpha - \frac12},\\
	\label{p}&p(x) = p_{L,a,b}(x) = a(L^2-x^2)^{\frac{b}{2a}}= a(L^2-x^2)^{\alpha + \frac12}.
\end{align}

In order to analyze the operator $\mathcal{L}$, we review the Gegenbauer operator, the Gegenbauer polynomials, and their properties as follows. The Gegenbauer, also known as the {\it ultraspherical}, operator of order $\alpha$  is defined by
\begin{align*}
	\mathcal{D}_{\alpha} = (1-\xi^2) \partial_\xi^2 - (2\alpha+1) \xi \partial_\xi
\end{align*}
in the interval $(-1,1)$. In standard Sturm-Liouville form, it is written as
\begin{align*}
	\mathcal{D}_{\alpha} u(\xi) = \frac{1}{W_{\alpha}(\xi)} \left(\frac{d}{d\xi}\left[ P_{\alpha}(\xi) u'(\xi) \right]\right),
\end{align*}
with
\begin{align*}
	&W_{\alpha}(\xi) = (1-\xi^2)^{\alpha-1/2}, \qquad
	P_{\alpha}(\xi) = (1-\xi^2)^{\alpha+1/2}.
\end{align*}

The $n$th Gegenbauer (or ultraspherical) polynomial of order $\alpha$, denoted by $C^\alpha_n$, is given by the formula
\begin{align*}
    C^\alpha_n(\xi) = \begin{cases}
        \displaystyle \sum_{m=0}^{[n/2]} \frac{(-1)^m}{\Gamma(\alpha)} \frac{\Gamma(\alpha+n-m)}{m!(n-2m)!}(2\xi)^{n-2m} & \mbox{ if } \alpha \ne 0, \\ 
        & \\
       \displaystyle \sum_{m=0}^{[n/2]} (-1)^m \frac{(n-m-1)!}{m!(n-2m)!} (2\xi)^{n-2m} & \mbox{ if } \alpha = 0 \mbox{ and } n \ge 1, \\
        & \\
       1 & \mbox{ if } \alpha = 0 \mbox{ and } n = 0.
    \end{cases}
\end{align*}
For $\alpha = 0$ and $n \ge 1$, $C^0_n = \frac2n T_n$, where $T_n$ is the Chebyshev polynomial of the first kind; that is, $C^0_n(\cos\theta) = \frac2n \cos(n\theta)$.
For example, the first 3 Gegenbauer polynomials of order $\alpha \neq 0$ are
\begin{eqnarray*}
&& C_0^\alpha(\xi) = 1, \quad C_1^\alpha(\xi) = 2\alpha \xi, \quad C_2^\alpha(\xi) = 2\alpha(\alpha + 1) \xi^2  - \alpha.
\end{eqnarray*} 
The family of polynomials $\{C_n^\alpha\}_{n=0}^\infty$ form an orthogonal system of eigenfunctions for the operator $\mathcal{D}_\alpha$ as shown in the following lemma.  

\begin{lemma}\label{lem:ggb}
For any fixed $\alpha > -\frac12$, the Gegenbauer polynomial $C_n^\alpha$ is an eigenfunction for the differential operator $\mathcal{D}_\alpha$ associated with the eigenvalue $\lambda_n^\alpha = n(n+2\alpha)$, i.e., 
\begin{align*}
	\mathcal{D}_\alpha C^\alpha_n = -\lambda^\alpha_n C^\alpha_n.
\end{align*}
The family of polynomials $\{C_n^\alpha\}_{n=0}^\infty$ satisfies the orthogonality relation
\begin{align}\label{orth:ggb}
	\int_{-1}^1 C^\alpha_n(\xi) C^\alpha_m(\xi) W_\alpha(\xi) d\xi
	= K_{\alpha,n}^2 \delta_{mn},
\end{align}
where
\begin{align}\label{orth:K}
	K_{\alpha,n}^2 = \begin{cases}
	    \frac{\pi 2^{1-2\alpha} \Gamma(n+2\alpha)}{n! (n+\alpha) (\Gamma(\alpha))^2} & \mbox{ if }\alpha \ne 0,\bigskip\\
        \frac{2\pi}{n^2} & \mbox{ if } \alpha = 0 \mbox{ and } n \ge 1,\medskip\\
        \pi & \mbox{ if } \alpha = 0 \mbox{ and } n = 0,
	\end{cases}
\end{align}
and $\Gamma(\cdot)$ denotes the gamma function.
\end{lemma}

\begin{proof}
See, e.g., \citeA[\S22]{abramowitz2012handbook}.
\end{proof}

The eigensystem for the operator $\mathcal{L}$ can thus be expressed in terms of the Gegenbauer polynomials and the eigenvalues of the Gegenbauer differential operator as shown below. 
\begin{lemma}\label{lem:eigen:L}
Assume $a>0$ and $b > 0$. For $n \geq 0$, we define the polynomial $e_n(\cdot)$ by
\begin{align}\label{e_n}
	e_n(x) = e_n^{L,a,b}(x) = \frac{1}{L^\alpha K_{\alpha,n}} C^\alpha_n\left( \frac{x}{L} \right), \quad x \in [-L,L],
\end{align}
where $C^\alpha_n(\cdot)$ denotes the $n$th Gegenbauer polynomial of order $\alpha$ and the constant $K_{\alpha,n}$ is defined in \eqref{orth:K}. It follows that the polynomial $e_n$ is an eigenfunction for the operator $\mathcal{L}$ associated with the eigenvalue $\lambda_n = \lambda_n^{a,b} = an^2 + (b-a)n \ge 0$, i.e., 
\begin{align}\label{eigen:L}
	\mathcal{L} e_n(x) = -\lambda_n e_n(x). 
\end{align}
The set of eigenfunctions $\{e_n(\cdot)\}_{n=0}^\infty$ satisfies the orthonormality relation:
\begin{align}\label{orth:L}
	\int_{-L}^L e_n(x) e_m(x) w(x) dx = \delta_{mn}.
\end{align}
\end{lemma}
\begin{proof}
It is clear that $\alpha > -\frac12$ since $\alpha = \frac b{2a}- \frac12$ and $a, b > 0$. 
Then the lemma follows from the change of variable $\xi=x/L$, Lemma \ref{lem:ggb}, and some elementary calculations.
\end{proof}

\subsection{Self-adjointness of the operator \texorpdfstring{$\mathcal L$}{L}} \label{sec:self-adj}
Let the Hilbert space $L^2_w$ be
\begin{align}\label{H}
	L^2_w = L^2([-L,L], w(x)dx)
\end{align}
with the weight function $w$ given by \eqref{w} and the inner product
\begin{align*}
	\langle f, g\rangle_w = \int_{-L}^L f(x) g(x) w(x) dx, \qquad \mbox{ for } f,g \in L^2_w.
\end{align*}
Let $E$ be the linear space spanned by the polynomials $e_n(\cdot)$, for $n \geq 0$, i.e., 
\begin{align*}
	E = \mathrm{span}\{ e_n : n \in \N_0 \}.
\end{align*}
We specify a domain $D(\mathcal{L}) \supset E$ for $\mathcal{L}$ on which $\mathcal{L}$ is self-adjoint.
Let $AC([-L,L])$ and $AC_{\rm loc}((-L,L))$ denote the spaces of absolutely continuous functions on $[-L,L]$ and locally absolutely continuous functions on $(-L,L)$, respectively.
Define
\begin{align*}
	\mathcal{L}: D(\mathcal{L}) \subset L^2_w \to L^2_w \qquad \text{by} \qquad
    \mathcal{L}u = \frac{1}{w}(pu')', \quad u \in D(\mathcal{L}),
\end{align*}
where
\begin{align}\label{D(L)}
	D(\mathcal{L}) = \begin{cases}
    \left\{ u \in L^2_w : u, pu' \in AC([-L,L]), \mathcal{L}u \in L^2_w, \lim\limits_{x \to \pm L} pu' = 0 \right\} & \text{if $\frac b{2a}\in (0,1)$,} \\
    & \\
	\left\{ u \in L^2_w : u, pu' \in AC_{\rm loc}((-L,L)), \mathcal{L}u \in L^2_w \right\} & \text{if $\frac b{2a} \in [1,\infty)$}.
	\end{cases}
\end{align}
Note that for any $u\in D(\mathcal{L})$, the derivative $(pu')'$ exists a.e., so $\mathcal{L}u$ is well-defined.
The proof of Lemma \ref{lem:self-adj} shows that the Sturm-Liouville operator $\mathcal{L}$ is regular when $b/a < 2$ and singular when $b/a \ge 2$. This explains the two cases in \eqref{D(L)}.
We refer, e.g., to \citeA{weidmann1980linear, niessen1992singular, zettl2005sturm, behrndt2020boundary} for the theory of Sturm-Liouville operators, and to \S2 and \S3 of \citeA{gesztesy2020on} for a summary of self-adjoint boundary conditions for Sturm-Liouville operators.

\begin{lemma}\label{lem:self-adj}
Under Assumption \ref{a:ab}, we have the following:
\begin{enumerate}
\item[(i)] $E \subset D(\mathcal{L})$ and $E$ is dense in $L^2_w$, so $\{e_n: n \in \N_0\}$ is an orthonormal basis for $L^2_w$.
\item[(ii)] $\mathcal{L}: D(\mathcal{L}) \to L^2_w$ is self-adjoint. In particular, 
\begin{align}\label{L:symm}
	\langle \mathcal{L} u, v \rangle_w = \langle u, \mathcal{L} v \rangle_w \quad \text{for all $u, v \in D(\mathcal{L})$,}
\end{align}
and $\langle \mathcal{L}u, u \rangle_w= -\sum_{n=0}^\infty \lambda_n |\langle u, e_n \rangle_w|^2 \le 0$ for all $u \in D(\mathcal{L})$.
\item[(iii)] 
$D(\mathcal{L})$ is the closure of $E$ with respect to the inner product
\begin{align}\label{ip:H^2_L}
	\langle u, v \rangle_{H^2_{\mathcal{L}}} := \langle u, v \rangle_w + \langle \mathcal{L} u, \mathcal{L} v \rangle_w \quad \text{for all $u, v \in D(\mathcal{L})$}
\end{align}
and $H^2_{\mathcal{L}} := (D(\mathcal{L}), \langle \cdot, \cdot\rangle_{H^2_{\mathcal{L}}})$ is a Hilbert space for which $\{ f_n := (1+\lambda_n^2)^{-1/2} e_n : n \in \N_0 \}$ is an orthonormal basis and $\mathcal{L} : H^2_{\mathcal{L}} \to L^2_w$ is a bounded operator.
\item[(iv)] 
For any $u \in L^2_w$, 
\begin{align}\label{u:H^2_L}
	u \in H^2_{\mathcal{L}} \quad \text{if and only if}\quad \sum_{n=0}^\infty (1+\lambda_n^2) |\langle u, e_n \rangle_w|^2 < \infty.
\end{align}
When \eqref{u:H^2_L} holds, $\mathcal{L}u = - \sum_{n=0}^\infty \lambda_n \langle u, e_n \rangle_w e_n = \sum_{n=0}^\infty \langle  \mathcal{L}u, e_n \rangle_w e_n$ in $L^2_w$ and $\|u\|_{H^2_{\mathcal{L}}}^2$ equals the value of the series in \eqref{u:H^2_L}.
\end{enumerate}
\end{lemma}
\begin{proof}
See Appendix \ref{pf:self-adj}.
\end{proof}

\begin{remark}\label{rmk:boundary}
Let $X_t$ be the It\^o diffusion whose generator is $\mathcal{L}$, i.e., $dX_t = \mu(X_t)dt+\sigma(X_t)dB_t$, where $\mu(x) = -bx$ and $\sigma(x) = \sqrt{2a(L^2-x^2)}$.
Recall that the scale function $\rho$ is defined by
\[
	\rho(x) := \int_0^x \exp\left( - 2 \int_0^y \frac{\mu(z)}{\sigma^2(z)} dz \right) dy, \quad x \in (-L,L).
\]
Define
\[
	v(x) := \int_0^x dy \, \rho'(y) \int_0^y \frac{2 dz}{\rho'(z) \sigma^2(z)}, \quad x \in (-L,L).
\]
Then
\[
	v(\pm L) \begin{cases}
	< \infty & \text{if $b/a\in (0,2)$,}\\
	= \infty & \text{if $b/a \in [2,\infty)$.}
	\end{cases}
\]
Hence, by Feller's test for explosion \cite[p.348, 5.29]{karatzas1991brownian},
\[
	\mathbb{P}\{ S = \infty \} \begin{cases}
	< 1 & \text{if $b/a \in (0,2)$,}\\
	=1 & \text{if $b/a \in [2,\infty)$,}
	\end{cases}
\]
where $S = \inf\{ t \ge 0 : X_t \not\in (-L,L) \} = \lim_{n\to \infty}\inf\{ t \ge 0: X_t \not\in (-L+\frac1n, L-\frac1n) \}$.
Therefore, the need for a Neumann-type boundary condition in \eqref{D(L)} when $b/a \in (0,2)$ is consistent with the fact that there is a positive probability for $X_t$ to hit the boundaries $\pm L$.
\end{remark}

\subsection{\texorpdfstring{$C_0$}{C0}-semigroups generated by \texorpdfstring{$\mathcal L$}{L} and \texorpdfstring{$\mathcal A$}{A}} \label{S:semigroups}
\subsubsection{\texorpdfstring{$C_0$}{C0}-semigroup for \texorpdfstring{$\mathcal L$}{L}}
Recall the orthonormal basis $\{e_n: n \in \N_0\}$ introduced in Lemma \ref{lem:self-adj}.
The heat kernel, or fundamental solution of $\partial_t - \mathcal{L}$, is given by
\[
	G_t(x,y) = \sum_{n=0}^\infty e^{-\lambda_n t} e_n(x) e_n(y), \quad x,y \in [-L,L],\ t>0.
\]
\begin{lemma}\label{lem:C_0}
Under Assumption \ref{a:ab}, $\mathcal{L}$ generates a $C_0$-semigroup $\mathcal{G}(t) = e^{t\mathcal{L}}: L^2_w \to L^2_w$ given by $\mathcal{G}(0)z = z$ and
\begin{align}\label{G:op}
	(\mathcal{G}(t)z)(x) = \int_{-L}^L G_t(x,y) z(y) w(y) dy = \sum_{n=0}^\infty e^{-\lambda_n t} \langle z, e_n \rangle_w e_n
	\quad \text{for $t>0$,}
\end{align}
which satisfies
\begin{align}\label{G:norm}
	\|\mathcal{G}(t)z\|_w \le \|z\|_w \quad \text{for all $t\ge 0$ and $z \in L^2_w$.}
\end{align}
\end{lemma}
\begin{proof}
See Appendix \ref{pf:C_0}.
\end{proof}
It follows that the $C_0$-semigroup $\mathcal G$ generates the strong solution to the Cauchy problem associated with $\cL$. We summarize the result in the following lemma without proof.
\begin{lemma}\label{lem:cp}
For every $z \in D(\mathcal{L})$, the Cauchy problem 
\begin{align}\label{cp:L}
	\begin{cases}
		\frac{d}{dt}u(t) = \mathcal{L}u(t), & t \ge 0,\\
		u(0)=z
	\end{cases}
\end{align}
has a unique strong solution given by $u(t) = \mathcal{G}(t)z \in C([0,\infty); D(\mathcal{L})) \cap C^1([0,\infty); L^2_w)$.
\end{lemma}

\subsubsection{\texorpdfstring{$C_0$}{C0}-semigroup for \texorpdfstring{$\cA$}{A}}
\begin{assumption}\label{a:V}
$V_t=V$ does not depend on $t$ and 
\begin{align}\label{iint:V}
	\|V\|^2:=\int_{-L}^L \int_{-L}^L |V(x,y)|^2 w(x) w(y) dx dy < \infty.
\end{align}
\end{assumption}

\begin{lemma}\label{lem:V}
Under Assumption \ref{a:V}, 
\[
	V: u \in L^2_w \mapsto V[u] = \int_{-L}^L V(\cdot, y)u(y)w(y)dy \in L^2_w
\]
is a bounded operator on $L^2_w$ with $\|V[u]\|_w \le \|V\| \|u\|_w$ for all $u \in L^2_w$.
\end{lemma}

\begin{remark}
In fact, it is well known that \eqref{iint:V} implies $V$ is a Hilbert-Schmidt operator.
\end{remark}

\begin{proof}[Proof of Lemma \ref{lem:V}]
For any $u \in L^2_w$, Cauchy-Schwarz inequality yields
\begin{align*}
	\|V[u]\|_w^2 &= \int_{-L}^L \left| \int_{-L}^L V(x,y) u(y) w(y)dy\right|^2 w(x) dx\\
	&\le \int_{-L}^L \left( \int_{-L}^L |V(x,y)|^2 w(y)dy \right) \left( \int_{-L}^L |u(y)|^2 w(y) dy \right) w(x) dx= \|V\|^2 \|u\|_w^2.
\end{align*}
This proves the lemma.
\end{proof}

Recall the operator $\cA:D(\cA) \to L^2_w$ in \eqref{A} given by 
\begin{align}\label{A:D(A)}
    \cA u = \mathcal{L}u - V[u]\quad \text{with $D(\cA) := D(\mathcal{L})$,}
\end{align}
where $D(\mathcal{L})$ is defined by \eqref{D(L)}.
\begin{lemma}\label{lem:A}
Under Assumptions \ref{a:ab} and \ref{a:V}, for any $z \in D(\cA)$, the Cauchy problem
\begin{equation}\label{cp:A}
	\begin{cases}
		\frac{d}{dt}u(t) = \cA u(t), \quad t \ge 0,\\
		u(0) = z \in L^2_w.
	\end{cases}
\end{equation}
has a unique strong solution $\mathcal{S}(t)z=u \in C([0,\infty); D(\cA)) \cap C^1([0,\infty); L^2_w)$ and $\mathcal{S}(t)$ defines a $C_0$-semigroup on $L^2_w$.
\end{lemma}
\begin{proof}
See Appendix \ref{pf:lem:A}.
\end{proof}
We conclude the section by showing the relationship between the $C_0$-semigroups $\mathcal G$ and $\mathcal S$. 
\begin{lemma}\label{lem:S}
The relation between $\mathcal{G}$ and $\mathcal{S}$ is given by:
\begin{equation}\label{G:S}
	\mathcal{G}(t) z = \mathcal{S}(t) z + \int_0^t \mathcal{S}(t-s)\, V[\mathcal{G}(s)z]\, ds
    \quad \text{for all $t \ge 0$ and $z \in L^2_w$.}
\end{equation}
\end{lemma}
\begin{proof}
It suffices to show \eqref{G:S} for $z \in D(\cA)=D(\mathcal{L})$.
Note that, for every $z \in D(\cA)=D(\mathcal{L})$, the solution $u(t)=u(t;z)=\mathcal{G}(t)z$ to \eqref{cp:L} is also a solution to the inhomogeneous Cauchy problem (in $v$):
\begin{align}\label{cp:A:inhomo}
	\begin{cases}
		\frac{d}{dt}v(t) = \cA v(t) + V[u(t)], & t\ge 0,\\
		v(0) = z.
	\end{cases}
\end{align}
It is known that, for every $z \in D(\cA)$, the Cauchy problem \eqref{cp:A:inhomo} has a unique solution which is given by Duhamel's principle, or variation of constants:
\begin{align*}
	v(t) = v(t; z) = \mathcal{S}(t)z + \int_0^t \mathcal{S}(t-s) \, V[u(s)] \, ds, \quad t \ge 0;
\end{align*}
see, e.g., \citeA[Proposition A.7]{daprato2014stochastic}.
By uniqueness, $u(t;z) = v(t; z)$, hence
\begin{align*}
	\mathcal{G}(t)z = u(t;z) = v(t;z)
	 &= \mathcal{S}(t)z + \int_0^t \mathcal{S}(t-s) \, V[u(s)] \, ds\\
	 &= \mathcal{S}(t)z + \int_0^t \mathcal{S}(t-s) \, V[\mathcal{G}(s)z] \, ds.
\end{align*}
This completes the proof.
\end{proof}

\subsection{Mild solution to the SPIDE} \label{S:solution}
This subsection is devoted to justifying the meaning of solutions to the SPIDE \eqref{SPIDE} in the sense of mild solutions using the framework of \citeA{daprato2014stochastic} based on semigroups. Let us mention the other alternative approaches for stochastic PDE -- the martingale measure approach \cite{walsh1986introduction} and the variational approach \cite{pardoux1979stochastic, rozovskii1990stochastic, liu2015stochastic}.

Let us first recall the Da Prato-Zabczyk framework and the definition of mild solutions.
Let $\mathsf U, H$ be Hilbert spaces.
Fix $T>0$ and consider the SPDE on a Hilbert space $H$:
\begin{align}\label{spde}
	\begin{cases}
		du(t) = [\cA u(t) + F(t, u(t))] dt + B(t, u(t)) dW_t, \quad t \in [0,T],\\
		u(0)=u_0 \in H,
	\end{cases}
\end{align}
where $W$ is a $Q$-Wiener process with values in a Hilbert space $\mathsf U_1 \supset \mathsf U_0$ into which $\mathsf U_0$ embeds by a Hilbert--Schmidt map ($\mathsf U_1 = \mathsf U$ can be taken when $Q$ is of trace class), defined on a probability space $(\Omega, \mathscr{F}, \mathbb{P})$ with filtration $\mathscr{F}_t$, $\cA:D(\cA) \subset H \to H$ is the generator of a $C_0$-semigroup $\mathcal{S}$, and $F:[0,T] \times \Omega \times H \to H$ and $B: [0,T] \times \Omega \times H \to L^0_2 := L_2(\mathsf U_0, H)$ are measurable mappings with respect to $\mathscr{P}_T \times \mathscr{B}(H)$, where $\mathscr{P}_T$ is the predictable $\sigma$-algebra on $[0,T]\times \Omega$, $\mathscr{B}(H)$ is the Borel $\sigma$-algebra on $H$, and $L_2^0=L_2(\mathsf U_0, H)$ is the space of all Hilbert-Schmidt operators from $\mathsf U_0 := Q^{1/2}\mathsf U$ to $H$.

\begin{definition}\cite[p.187]{daprato2014stochastic}
A mild solution to \eqref{spde} is an $H$-valued predictable process $\{u(t), t \in [0,T]\}$ such that $\int_0^T \|u(s)\|_H^2\, ds <\infty$ a.s., and for arbitrary $t \in [0,T]$,
\begin{equation}\label{mild}
			u(t) = \mathcal{S}(t)u_0 + \int_0^t \mathcal{S}(t-s)F(s,u(s)) ds + \int_0^t \mathcal{S}(t-s)B(s,u(s)) dW_s \quad \text{a.s.}
\end{equation}
where the integrals are well-defined.
\end{definition}

Let $\mathsf U=\mathsf U_0=\mathsf U_1=\R^M$, $Q=I$, and $H = L^2_w$ (see \eqref{H}).
For simplicity, in our equation \eqref{SPIDE} we consider the case that $V_t(x,y) = V(x,y)$, $\tilde\sigma_m(t,x,y) = \tilde\sigma_m(x,y)$, and $\sigma_m(t,x) = \sigma_m(x)$ do not depend on $t$.

\begin{assumption}\label{a:sigma}
$\tilde\sigma_m(t,x,y) = \tilde\sigma_m(x,y)$ and $\sigma_m(t,x) = \sigma_m(x)$ do not depend on $t$,
\begin{align*}
	\|\tilde\sigma_m\|^2 := \int_{-L}^L \int_{-L}^L |\tilde\sigma_m(x,y)|^2 w(x) w(y) dxdy < \infty,
\end{align*}
and $\sigma_m \in L^2_w$.
\end{assumption}

\begin{assumption}\label{a:u0:h}
$u_0 \in L^2_w$ and $h(t) = h(t,\omega)$ is a predictable $L^2_w$-valued process with 
\[
	\int_0^T \left[\E(\|h(s)\|_w^p)\right]^{1/p}\, ds < \infty.
\]
\end{assumption}

Define $\mathcal{A}$ by \eqref{A:D(A)},  
$F(t, \omega, u) = h(t,\omega)$, and $B(t,\omega,u) = B(u) \in L^0_2 = L_2(\R^M, L_w^2)$ by
\begin{align}\label{B}
	B(u) x = \sum_{m=1}^M x_m (\sigma_m+\tilde\sigma_m[u]), \quad x = (x_1,\dots,x_M).
\end{align}
Recall that $W_t = (W_t^1,\dots, W_t^M)$ is an $M$-dimensional Brownian motion.
Fix $T>0$.
Then the SPIDE \eqref{SPIDE} can be recast as
\begin{equation}\label{SPIDE:DZ-form}
	\begin{cases}
		du(t) = [\cA u(t) + h(t)] dt + B(u(t))dW_t, \quad t \in [0,T],\\
		u(0)=u_0.
	\end{cases}
\end{equation}

\begin{theorem}\label{th:mild:sol}
	Under Assumptions \ref{a:ab}, \ref{a:V}, \ref{a:sigma}, and \ref{a:u0:h} for some $p \ge 2$, there exists a mild solution $\{ u(t), t \in [0,T]\}$ to \eqref{SPIDE:DZ-form} that satisfies
	\begin{equation}\label{mild:G}
		u(t) = \mathcal{G}(t)u_0 + \int_0^t \mathcal{G}(t-s)h(s) ds - \int_0^t \mathcal{G}(t-s)V[u(s)]ds  + \int_0^t \mathcal{G}(t-s) B(u(s)) dW_s \text{\ a.s.}
	\end{equation}
	where $\mathcal{G}$ is the semigroup associated to $\mathcal{L}$ (see Lemma \ref{lem:C_0}), and satisfies
	\begin{align}\label{mild:Lp}
		\sup_{t \in [0,T]} \E( \|u(t)\|_w^p ) < \infty.
	\end{align}
	The mild solution is unique among all mild solutions satisfying \eqref{mild:Lp} for $p = 2$.
\end{theorem}
\begin{proof}
See Appendix \ref{pf:mild:sol}.
\end{proof}

\begin{remark}
The mild solution~\eqref{mild:G} is expressed in terms of the $\cL$-semigroup $\mathcal{G}(t)$ with $V$ appearing as an explicit Duhamel correction. By Lemma~\ref{lem:S}, this is equivalent to the more compact form using the $\cA$-semigroup $\mathcal{S}(t)$: $u(t) = \mathcal{S}(t)u_0 + \int_0^t \mathcal{S}(t-s)[h(s)\,ds + B(u(s))\,dW_s]$.
\end{remark}

\subsection{Infinite system of SDEs for the eigenmodes}

Recall that $\{e_n : n \in \N_0\}$ is an orthonormal basis of eigenfunctions of $\mathcal{L}$ for the Hilbert space $L^2_w$ (see Lemma \ref{lem:self-adj}).
For $u_0 \in L^2_w$, consider the eigenfunction expansion of $u_0$
\begin{align}\label{u_0:exp}
u_0(x) = \sum_{n=0}^\infty u^n_0 e_n(x), \quad \text{ where } \quad u_0^n = \int_{-L}^L u_0(x) e_n(x) w(x)dx.
\end{align}
Likewise, the eigenfunction expansions for $h$ and $\sigma_m$ are
\begin{align}
\label{h:exp}
h(t, x) &= \sum_{n=0}^\infty h^n_t e_n(x),\quad \text{where} \quad
h_t^n = \int_{-L}^L h(t,x) e_n(x) w(x)dx,\\
\label{tsigma:exp}
\sigma_m(x) &= \sum_{n=0}^\infty \sigma_m^n e_n(x), \quad \text{where} \quad \sigma_m^n = \int_{-L}^L \sigma_m(x) e_n(x) w(x) dx.
\end{align}
Moreover, the kernels $V$ and $\tilde\sigma_m$ can be expressed in terms of the eigenfunctions as
\begin{align}\label{V:exp}
V(x,y) &= \sum_{i,j=0}^\infty V^{ij} e_i(x) e_j(y), \\
\label{sigma:exp}
\tilde\sigma_m(x,y) &= \sum_{i,j=0}^\infty \tilde\sigma_m^{ij} e_i(x) e_j(y),
\end{align}
where the coefficients are given respectively by
\begin{align*}
V^{ij} &= \int_{-L}^L \int_{-L}^L V(x,y) e_i(x) e_j(y) w(x) w(y) dxdy, \\
\tilde\sigma_m^{ij} &= \int_{-L}^L \int_{-L}^L \tilde\sigma_m(x,y) e_i(x) e_j(y) w(x) w(y) dxdy.
\end{align*}

\begin{remark}\label{rmk:V:sigma}
    Under Assumptions \ref{a:V} and \ref{a:sigma}, respectively, we have
    \[
        \sum_{i,j=0}^\infty |V^{ij}|^2 = \|V\|^2 < \infty,
        \quad
        \sum_{i,j=0}^\infty |\tilde\sigma^{ij}_m|^2 = \|\tilde\sigma_m\|^2 < \infty
        \quad \text{and}\quad
        \sum_{i=0}^\infty |\sigma_m^i|^2 < \infty.
    \]
\end{remark}

Under the assumptions of Theorem \ref{th:mild:sol}, the mild solution $u_t$ to the SPIDE \eqref{SPIDE} takes values in $L^2_w$.
Therefore, for each $t \ge 0$, we may write
\begin{equation}\label{u:exp}
u_t(x) = \sum_{n=0}^\infty u_t^n e_n(x), \quad \text{where} \quad
u_t^n = \int_{-L}^L u_t(x) e_n(x) w(x) dx.
\end{equation}
Moreover, by the orthonormality of $\{e_n : n \in \N_0\}$,
\begin{align}\label{u:l2}
	\sum_{n=0}^\infty |u_t^n|^2 = \|u_t\|_w^2 < \infty.
\end{align}
This implies that, for any $t \geq 0$, the sequence $\{u_t^n\}_{n=0}^\infty \in \ell^2$, where $\ell^2$ denotes the set of all squared summable sequences of real numbers.

\begin{theorem}\label{th:sys:sde}
Under the assumptions of Theorem \ref{th:mild:sol}, the $\ell^2$-valued process $(u_t^n)_{n=0}^\infty$ defined by \eqref{u:exp} satisfies the following coupled infinite system of linear SDEs:
\begin{equation}\label{eq:u^n}
du_t^n 
= \left[-\sum_{k=0}^\infty\left(\lambda_k \delta_{nk} +  V^{nk} \right) u_t^k + h_t^n \right] dt  + \sum_{m=1}^M \left[ \sigma_m^n + \sum_{k=0}^\infty \tilde\sigma_m^{nk} u^k_t \right] dW_t^m
\end{equation}
for $t \ge 0$ and $n \in \N_0$.
\end{theorem}

\begin{proof}
See Appendix \ref{pf:sys:sde}.
\end{proof}

\subsection{Temporal regularities}
In order to prove regularities for the sample functions of $t \mapsto u_t(x)$, we impose Assumptions \ref{a:sigma:2} and \ref{a:u0:h:2} below, which are stronger than Assumptions \ref{a:sigma} and \ref{a:u0:h}, respectively.

\begin{assumption}\label{a:sigma:2}
$\tilde\sigma_m(t,x,y) = \tilde\sigma_m(x,y)$, $\sigma_m(t,x) = \sigma_m(x)$ do not depend on $t$, and satisfy
\begin{align}\label{sigma:l1:l2}
	&\sum_{n=0}^\infty \|\tilde\sigma_m^* [e_n]\|_w =\sum_{n=0}^\infty \left( \sum_{k=0}^\infty |\tilde\sigma_m^{nk}|^2 \right)^{1/2} < \infty,\\
    \label{tsigma:l1}& \sum_{n=0}^\infty |\langle \sigma_m, e_n\rangle_w| = \sum_{n=0}^\infty |\sigma_m^n| < \infty.
\end{align}
Here $\tilde\sigma_m^*$ denotes the adjoint of $\tilde\sigma_m$ on $L^2_w$, the integral operator with kernel $\tilde\sigma_m(y,x)$.
Note that \eqref{sigma:l1:l2} implies that $\tilde\sigma_m$ is a trace-class operator on $L^2_w$.
\end{assumption}

\begin{assumption}\label{a:u0:h:2}
$u_0 \in L^2_w$ and $h(t)$ is a predictable $L^2_w$-valued process with 
\[
	\sup_{t \in [0,T]} \E(\|h(t)\|_w^p) < \infty \quad \text{for every $p \in [2,\infty)$.}
\]
\end{assumption}

The following theorem shows that for fixed $x \in (-L,L)$, the solution $u(t,x)$ has the same temporal regularities as Brownian motion.

\begin{theorem} \label{th:temporal-regularity}
	Under Assumptions \ref{a:ab}, \ref{a:V}, \ref{a:sigma:2}, and \ref{a:u0:h:2},
	the mild solution $u(t,x)$ to the SPIDE \eqref{SPIDE} has the property that for every $x \in (-L,L)$, $t \mapsto u(t,x)$ is almost surely locally $\gamma$-H\"older continuous on $(0,T]$ for every $\gamma\in(0,1/2)$.
\end{theorem}
\begin{proof}
See Appendix \ref{pf:temporal-regularity}.
\end{proof}

\section{Applications to liquidity surface modeling} \label{sec:app-ls}

In this section, we apply the SPIDE framework developed in Section~\ref{sec:th-bg} to construct a market model for the evolution of an AMM liquidity pool with concentrated liquidity provision. As shown in \citeA{tung2026mathematical} and \citeA{risk2026pricing}, the price-liquidity pair provides a convenient mathematical framework for analyzing pool reserves and pool value, as well as the associated pricing and hedging problems.

We proceed as follows. Section~\ref{sec:pool-quantities} reviews the key pool quantities in the price-liquidity coordinate. Section~\ref{sec:log-log} and Section~\ref{sec:log-linear} introduce two alternative coordinate systems---the log-price/spot-centered-log-liquidity pair $(r,u)$ and the log-price/spot-centered-liquidity pair $(r,U)$---and express pool reserves and value in each. Section~\ref{sec:mkt-model} then specifies the market model by coupling a stochastic process for the log price with the SPIDE~\eqref{SPIDE} for the liquidity profile, and derives the resulting dynamics of pool value and loss-versus-rebalancing.

\subsection{Pool quantities in the price-liquidity coordinate} \label{sec:pool-quantities}
Consider a liquidity pool consisting of tokens $X$ and $Y$ (e.g., ETH and USDC). Let the pair $(x,y)$ denote the pool reserve, where $x$ and $y$ are the numbers of $X$ and $Y$ tokens deposited in the pool by LPs, respectively. We adopt token $Y$ as the num\'eraire, so that both the value of token~$X$ and the marked-to-market value of the pool are denominated in $Y$. 

For a given liquidity profile $\ell$ and pool price $p$, the pool reserves are determined by (see \citeNP{tung2026mathematical} or \citeNP{risk2026pricing})
\begin{equation} \label{eq:reserves-price-liq}
 x(p, \ell) = \int_p^\infty \ell(q)\, dq, \quad y(p, \ell) = \int_0^p q \,\ell(q)\, dq.
\end{equation}
To connect with the analysis in Section~\ref{sec:th-bg}, we now re-express these quantities in two alternative coordinate systems.

\subsubsection{Log-price and spot-centered log-liquidity pair \texorpdfstring{$(r,u)$}{(r,u)}} \label{sec:log-log}
Given $(p, \ell)$, define the logarithmic price and spot-centered logarithmic liquidity by
\[
r = \ln p, \qquad u(\xi) = \ln\ell(p e^{\xi}).
\]
Recall that $u$ was introduced as the \emph{spot-centered logarithmic liquidity profile} in Section~\ref{sec:liq-sur}. The inverse relations are $p = e^r$ and $\ell(q) = e^{u(\ln q - r)}$, and the pool reserves become
\begin{equation} \label{eq:reserves-r-u}
 x(r, u) = \int_r^\infty e^{u(\xi - r) + \xi}\, d\xi, \qquad y(r, u) = \int_{-\infty}^r e^{u(\xi - r) + 2\xi}\, d\xi.
\end{equation}

\paragraph{\textit{Pool value}}
The marked-to-market value of the pool is 
\[
\cV := p \, x(p, \ell) + y(p, \ell) = e^r x(r, u) + y(r, u).
\]
Substituting~\eqref{eq:reserves-r-u} and writing $\xi\wedge r = \min\{\xi , r\}$, we obtain
\begin{equation} \label{eq:pool-value-r-u}
\cV(r, u) = \int_{-\infty}^\infty e^{u(\xi - r) + \xi + \xi\wedge r}  \,d\xi. 
\end{equation}

\paragraph{\textit{Finite support}}
When $u$ is supported on the finite interval $[-L, L]$, the change of variable $\xi \mapsto \xi + r$ simplifies the above expressions to
\begin{align}
x(r, u) &= e^r \int_0^L e^{u(\xi) + \xi}\, d\xi, \qquad
y(r, u) = e^{2r}\int_{-L}^0 e^{u(\xi) + 2\xi}\, d\xi, \label{eq:reserves-r-u-finite} \\
\cV(r, u) &= e^{2r}\int_{-L}^{L} e^{u(\xi) + \xi - \xi^-}  \,d\xi. \label{eq:value-r-u-finite}
\end{align}
We note the \emph{exponential-linear} dependence of all these quantities on~$u$: each involves $e^{u(\xi)}$ multiplied by functions of $\xi$ and $r$ alone.

\subsubsection{Log-price and spot-centered liquidity pair \texorpdfstring{$(r,U)$}{(r,U)}} \label{sec:log-linear}
Alternatively, given $(p, \ell)$, define the spot-centered liquidity directly (without the logarithm) by
\[
r = \ln p, \qquad U(\xi) = \ell(p e^{\xi}).
\]
The inverse relations are $p = e^r$ and $\ell(q) = U(\ln q - \ln p)$. The pool quantities in this coordinate system are
\begin{align}
x(r, U) &= e^r \int_0^\infty U(\xi)\, e^{\xi}\, d\xi, \qquad
y(r, U) = e^{2r} \int_{-\infty}^0 U(\xi)\, e^{2\xi}\, d\xi, \label{eq:reserves-r-U} \\
\cV(r, U) &= e^{2r} \int_{-\infty}^\infty U(\xi)\, e^{\xi - \xi^-}\,  d\xi. \label{eq:value-r-U}
\end{align}
In contrast to the $(r,u)$ formulation, the pool reserves and value are \emph{linear} in the spot-centered liquidity~$U$.

When $U$ is supported on $[-L, L]$, these reduce to
\begin{align}
x(r, U) &=  e^r \int_0^L U(\xi)\, e^{\xi}\, d\xi, \qquad
y(r, U) = e^{2r}\int_{-L}^0 U(\xi)\, e^{2\xi}\, d\xi, \label{eq:reserves-r-U-finite} \\
\cV(r, U) &= e^{2r}\int_{-L}^{L} U(\xi)\, e^{\xi - \xi^-} \, d\xi. \label{eq:value-r-U-finite}
\end{align}
 
With these expressions in hand, one can determine the distributions and statistics of pool reserves and value once the dynamics of $(r,u)$ or $(r,U)$ are specified. We turn to this next.

\subsection{The market model} \label{sec:mkt-model}
We propose a market model for the joint evolution of the log price $r_t$ and the liquidity profile. The model couples a standard stochastic process for the log price with the SPIDE~\eqref{SPIDE} for the spot-centered (log-)liquidity profile. Specifically, let $r_t$ and $u_t$ (or~$U_t$) satisfy
\begin{align}
 dr_t &= \left(\mu_t - \frac{\sigma_t^2}2 \right) dt + \sigma_t\, dZ_t, \label{eqn:mkt-model-r} \\
 du_t(\xi) &= \{\cA u_t + h(t, \xi) \}\,dt + \sum_{m=1}^M \left\{\sigma_m(t,\xi) + \tilde\sigma_m(t, \xi)[u_t] \right\}dW^m_t \label{eqn:mkt-model-u}
\end{align}
for $\xi \in (-L, L)$, where the driving Brownian motions $\bW_t$ and $Z_t$ are assumed independent. Note that \eqref{eqn:mkt-model-u} is identical to the SPIDE~\eqref{SPIDE} studied in Section~\ref{sec:th-bg}. Throughout this section we assume that the initial profile ($u_0$ for the $(r,u)$ model, $U_0$ for the $(r,U)$ model) and the source term $h$ are deterministic; Assumption~\ref{a:u0:h} allows a random $h$, but the moment formulas below use deterministic data.

\begin{remark} \label{rmk:model-interpretation}
We collect several remarks on the model~\eqref{eqn:mkt-model-r}--\eqref{eqn:mkt-model-u}.
\begin{enumerate}
\item \textit{Financial interpretation of the operator $\cA$.} Recall the decomposition $\cA = \cL - V$, where the second-order differential operator is
\[
\mathcal{L} = a(L^2 - \xi^2)\partial_\xi^2  - b \xi \partial_\xi.  
\]
The drift term $-b\xi$ (with $b > 0$) captures the tendency for liquidity at tick~$\xi$ to be repositioned toward the center---closer to the current spot price. The diffusion coefficient $a(L^2 - \xi^2)$ models the infinitesimal splitting of a position at $\xi$ into its neighbors. The integral operator $V$ represents the rate (as a fraction of existing liquidity) at which positions at $\xi$ interact with positions at other ticks. Finally, $h(\xi)$ represents the arrival rate of new liquidity at tick~$\xi$. Both models of this section use the relocation operator $V$ of Section~\ref{sec:th-bg}. The case study (Section~\ref{sec:case-study}) estimates $V$, restricted to finite rank, for the additive OU model of Section~\ref{sec:OU-model} in log-liquidity coordinates on the normalized tick-rank grid. The estimates describe $V$ in rank coordinates. Using them in the log-price coordinates of this section, or in the raw-liquidity model of Section~\ref{sec:BS-model}, would require a coordinate map and a separate calibration that the present study does not provide.
\item \textit{Additive and multiplicative noise.} If $\tilde{\sigma}_m \equiv 0$ for all $m$, the noise is purely additive and the model reduces to a Gaussian process of Ornstein-Uhlenbeck (OU) type. Conversely, if $\sigma_m \equiv 0$ for all $m$, the noise is multiplicative, linear in the profile, and gives a model of Black-Scholes (BS) type. Unlike geometric Brownian motion, the nonlocal operators $\tilde\sigma_m$ do not in general make this model lognormal or keep its solution nonnegative (Section~\ref{sec:BS-model}).
\item \textit{Coordinate choice.} Since $u = \ln U$ takes values in $\mathbb{R}$, the OU-type model is naturally paired with the $(r, u)$ coordinate, in which every real-valued $u$ gives a positive profile $e^{u}$. The BS-type model is naturally paired with the $(r, U)$ coordinate, in which the pool quantities are linear in $U$. The two are alternative model specifications: the exponential of a solution of one SPIDE does not in general solve the other.
\item \textit{Independence of driving noises.} The assumption that $Z_t$ and $\bW_t$ are independent is made for convenience. Correlated driving noises can be incorporated at the cost of lengthier expressions.
\end{enumerate}
\end{remark}
\paragraph{\textit{Pointwise values at the spot tick}}
The LVR and its moments evaluate the profile at the spot tick $\xi=0$, whereas the solution of Section~\ref{sec:th-bg} takes values in $L^2_w$, whose elements have no value at a single point. We use the version of Theorem~\ref{th:temporal-regularity}, given by the eigen-series $u_t(0)=\sum_n u_t^n e_n(0)$, and define $m_t(0)$ and $M_t(0)$ below as the means of these values, $\gamma_{t,s}(0,0)$ as their covariances, and $\Gamma_{t,s}(0,0)$ as their cross-moments. The following assumption bounds these series uniformly in time, including as $t\downarrow0$, where Theorem~\ref{th:temporal-regularity} alone gives no control. In the $(r,U)$ model, $u$ and $u_0$ stand for $U$ and $U_0$.
\begin{assumption}\label{a:pointwise}
Assumptions~\ref{a:ab}, \ref{a:V}, and~\ref{a:sigma:2} hold, and:
\begin{enumerate}
\item[(i)] the initial profile and the source term are deterministic, with eigen-coefficients satisfying $\sum_n|u_0^n|<\infty$ and $\sup_{t\le T}\sum_n|h_t^n|<\infty$;
\item[(ii)] $\|V\|_{\ell^1}:=\sup_k\sum_n|V^{nk}|<\infty$;
\item[(iii)] $\mu_t$ and $\sigma_t$ are progressively measurable with respect to the filtration generated by $Z$, and the price weight $Q_t:=e^{2r_t}\sigma_t^2$ satisfies $q_T:=\sup_{t\le T}\Eof{Q_t^2}<\infty$.
\end{enumerate}
\end{assumption}
Condition~(i) implies Assumption~\ref{a:u0:h:2}, since $\|h(t)\|_w\le\sum_n|h^n_t|$. Condition~(iii) makes $(r,\sigma)$ independent of the profile, which is driven by $\bW$ and deterministic data; it holds in the Black--Scholes examples below.
\begin{lemma}\label{lem:pointwise}
Under Assumption~\ref{a:pointwise}, the eigenmodes of the mild solution of~\eqref{SPIDE} satisfy
\[
A_T:=\sup_{t\le T}\sum_{n=0}^\infty\|u_t^n\|_{L^2(\Omega)}<\infty.
\]
Consequently, for every $t\in[0,T]$ and $\delta\in(0,L)$, almost surely the series $\sum_n u_t^n e_n(\xi)$ converges absolutely and uniformly on $[-L+\delta,L-\delta]$ to a continuous function that agrees almost everywhere there with $u_t$. For $t\in(0,T]$, its value $u_t(0)$ agrees almost surely with the version of Theorem~\ref{th:temporal-regularity}, and
\[
\sup_{t\le T}\Eof{|u_t(0)|^2}^{1/2}\le C_0A_T,\qquad C_0:=\sup_n|e_n(0)|<\infty .
\]
\end{lemma}
\begin{proof}
See Appendix~\ref{pf:pointwise}.
\end{proof}
In the LVR integrals below, $s\mapsto u_s(0)$ denotes the modification of Theorem~\ref{th:temporal-regularity}, which is continuous on $(0,T]$.

We now develop each of the two cases in turn: (a)~the OU-type SPIDE for $(r, u)$, and (b)~the BS-type SPIDE for $(r, U)$.

\subsubsection{The Ornstein-Uhlenbeck type SPIDE for \texorpdfstring{$(r, u)$}{(r,u)}} \label{sec:OU-model}
Setting $\tilde{\sigma}_m \equiv 0$ in~\eqref{eqn:mkt-model-u}, the market model becomes
\begin{align}
 dr_t &= \left(\mu_t - \frac{\sigma_t^2}2 \right) dt + \sigma_t\, dZ_t, \label{eqn:mkt-model-r-OU} \\
 du_t(\xi) &= \{\cA u_t + h(t, \xi) \}\,dt + \sum_{m=1}^M \sigma_m(t,\xi)\, dW^m_t. \label{eqn:mkt-model-u-OU}
\end{align}
The mild solution to~\eqref{eqn:mkt-model-u-OU} is
\[
u_t(\xi) = \mathcal{S}(t)[u_0](\xi) + \int_0^t \mathcal{S}(t-s)[h(s,\cdot)](\xi)\, ds + \sum_{m=1}^M \int_0^t \mathcal{S}(t-s)[\sigma_m(s,\cdot)](\xi)\, dW^m_s,
\]
where $\mathcal{S}$ denotes the $C_0$-semigroup generated by $\cA$ (see Lemma~\ref{lem:A}). When both $h$ and the $\sigma_m$ are deterministic, the process $u_t(\cdot)$ is Gaussian with mean
\begin{equation} 
\label{eq:OU-mean}
m_t(\xi) = \mathcal{S}(t)[u_0](\xi) + \int_0^t \mathcal{S}(t-s)[h(s,\cdot)](\xi)\, ds
\end{equation}
and covariance function
\begin{equation} 
\label{eq:OU-cov}
\gamma_{t,s}(\xi,\eta) =  \sum_{m=1}^M \int_0^{t \wedge s} \mathcal{S}(t-\tau)[\sigma_m(\tau,\cdot)](\xi)\, \mathcal{S}(s-\tau)[\sigma_m(\tau,\cdot)](\eta)\, d\tau.
\end{equation}
\paragraph{\textit{Pool quantities under the OU model}}
Under the dynamics~\eqref{eqn:mkt-model-r-OU}--\eqref{eqn:mkt-model-u-OU}, the pool quantities in \eqref{eq:reserves-r-u-finite}--\eqref{eq:value-r-u-finite} evolve as
\begin{align*}
x_t &= e^{r_t}\int_{0}^{L} e^{u_t(\xi) + \xi}\, d\xi, \qquad
y_t = e^{2r_t}\int_{-L}^0 e^{u_t(\xi) + 2\xi}\, d\xi, \label{eq:OU-reserves} \\
\cV_t &= e^{2r_t} \int_{-L}^L e^{u_t(\xi) + \xi - \xi^-} \, d\xi. 
\end{align*}
\paragraph{\textit{Loss-versus-rebalancing}}
Recall that the loss-versus-rebalancing (LVR) of \citeA{milionis2022automated}, for a liquidity profile (see \citeNP[Equation (11)]{tung2026mathematical}), is given by
\[
d\lvr_t = \frac12 \ell(p_t) d\langle p \rangle_t,
\]
where $p_t = e^{r_t}$. Here $\ell$ is the profile of~\eqref{eq:reserves-price-liq}; \citeA{tung2026mathematical} normalize the profile so that $x(p)=\frac12\int_p^\infty \tilde\ell(q)\,q^{-3/2}\,dq$, i.e., $\tilde\ell(q)=2q^{3/2}\ell(q)$, and in that normalization the same quantity reads $\frac14\tilde\ell(p_t)\,p_t^{-3/2}\,d\langle p\rangle_t$. For a time-varying profile, we accumulate this rate along the current profile $\ell_t$: it is the price-arbitrage component of the pool's loss and excludes the cash flows from liquidity additions and removals (see \citeNP[Remark 5]{tung2026mathematical}). Under the OU model, this becomes
\begin{equation*} 
\lvr_t = \frac12 \int_0^t  e^{u_s(0)} e^{2 r_s} \sigma_s^2 ds.
\end{equation*}
\paragraph{\textit{Mean and covariance functions}}
Under Assumption~\ref{a:pointwise}, $u_t$ is independent of $(r_t,\sigma_t)$, and one can compute the mean and covariance functions of $\cV_t$ and $\lvr_t$ in closed form, whenever the expectations involved are finite. For $\lvr_t$ they are, as shown below. For $\cV_t$, Assumption~\ref{a:pointwise} ensures neither the price moments $\Eof{e^{2r_t}}$ and $\Eof{e^{2(r_t+r_s)}}$, since it bounds $Q_t$ rather than $e^{2r_t}$, nor the integrability of the spatial integrands near $\xi=\pm L$. The mean functions are
\begin{align*}
    &\Eof{\cV_t} = \Eof{e^{2r_t}} \int_{-L}^L e^{m_t(\xi)+\frac12 \gamma_{t,t}(\xi,\xi)+\xi-\xi^-}\, d\xi,\\
    &\Eof{\lvr_t} = \frac12 \int_0^t \Eof{e^{2 r_s}\sigma_s^2}\, e^{m_s(0)+\frac12 \gamma_{s,s}(0,0)}\,  ds,
\end{align*}
where $m_t$ and $\gamma_{t,s}$ are given by~\eqref{eq:OU-mean} and~\eqref{eq:OU-cov}, respectively. 
The covariance functions are given by $\cov\left[ \cV_t,\cV_s\right]=\Eof{\cV_t\,\cV_s} - \Eof{\cV_t}\Eof{\cV_s}$ and likewise for $\lvr_t$, where
\begin{align*}
    &\Eof{\cV_t\,\cV_s} = \Eof{e^{2(r_t+r_s)}} \int_{-L}^L \int_{-L}^L e^{m_t(\xi)+m_s(\eta) + \frac12 [\gamma_{t,t}(\xi,\xi)+2\gamma_{t,s}(\xi,\eta) + \gamma_{s,s}(\eta,\eta)]+\xi-\xi^-+\eta-\eta^-} d\xi d\eta,\\
    &\Eof{\lvr_t\,\lvr_s} = \frac{1}{4} \int_0^t \int_0^s \Eof{e^{2(r_{\tau}+r_{\rho})}\sigma_{\tau}^2 \sigma_{\rho}^2} e^{m_{\tau}(0)+m_{\rho}(0)+\frac12[\gamma_{\tau,\tau}(0,0)+2\gamma_{\tau,\rho}(0,0)+\gamma_{\rho,\rho}(0,0)]} d\tau d\rho.
\end{align*}
These LVR moments are finite. Since $u$ is Gaussian, $u_\tau(0)$ and $u_\rho(0)$ are jointly Gaussian as $L^2(\Omega)$-limits of the partial sums, with means $m_\tau(0)$, $m_\rho(0)$ and covariances $\gamma_{\tau,\rho}(0,0)$. By Lemma~\ref{lem:pointwise}, $|m_\tau(0)|\le C_0A_T$ and $\gamma_{\tau,\tau}(0,0)\le C_0^2A_T^2$, so $\Eof{e^{u_\tau(0)+u_\rho(0)}}\le e^{2C_0A_T+2C_0^2A_T^2}$ for all $\tau,\rho\le T$. With $\Eof{Q_\tau Q_\rho}\le q_T$, Tonelli's theorem gives
\[
\Eof{\lvr_t}\le\tfrac12\,t\,q_T^{1/2}e^{C_0A_T+\frac12C_0^2A_T^2},\qquad
\Eof{\lvr_t\,\lvr_s}\le\tfrac14\,ts\,q_T\,e^{2C_0A_T+2C_0^2A_T^2}.
\]
\begin{example}
If we further assume that the price $p_t$ is driven by the Black-Scholes model
\[
\frac{dp_t}{p_t} = \mu dt + \sigma dZ_t,
\]
or equivalently $dr_t = (\mu - \frac{\sigma^2}2) dt + \sigma dZ_t$, where $\sigma > 0$ and $\mu$ are constants, then $u_t$ is indeed independent of $r_t$ and the mean and covariance formulas simplify as follows:
\begin{align*}
    &\Eof{\cV_t} = e^{2r_0+(2\mu+\sigma^2)t} \int_{-L}^L e^{m_t(\xi)+\frac12 \gamma_{t,t}(\xi,\xi)+\xi-\xi^-}\, d\xi,\\
    &\Eof{\lvr_t} = \frac{\sigma^2 e^{2r_0}}{2} \int_0^t e^{(2\mu+\sigma^2)s+m_s(0)+\frac12 \gamma_{s,s}(0,0)}\, ds,
\end{align*}
\begin{align*}
    &\Eof{\cV_t\,\cV_s} = e^{4r_0+(2\mu+\sigma^2)(t+s) + 4 \sigma^2(t\wedge s)} \!\! \int_{-L}^L \int_{-L}^L e^{m_t(\xi)+m_s(\eta) +\xi-\xi^-+\eta-\eta^-} \notag \\
    &\qquad\qquad \times e^{\frac12 [\gamma_{t,t}(\xi,\xi)+2\gamma_{t,s}(\xi,\eta)+ \gamma_{s,s}(\eta,\eta)]}\! d\xi d\eta,\\
    &\Eof{\lvr_t\,\lvr_s} = \frac{\sigma^4 e^{4r_0}}{4} \!\! \int_0^t \int_0^s e^{(2\mu+\sigma^2)(\tau+\rho) + 4 \sigma^2(\tau\wedge \rho)} \notag \\
    &\qquad\qquad \times e^{m_{\tau}(0)+m_{\rho}(0)+\frac12 [\gamma_{\tau,\tau}(0,0)+2\gamma_{\tau,\rho}(0,0)+\gamma_{\rho,\rho}(0,0)]} d\tau d\rho.
\end{align*}
\end{example}

\paragraph{\textit{Eigenfunction expansion of pool quantities}}
The theory of Section~\ref{sec:th-bg} is coordinate-agnostic: the SPIDE state can be instantiated as a log-liquidity field $u_t$ or as a raw-liquidity field $U_t$. In the present OU/log convention, $u_t(\xi)=\log \ell(p_t e^\xi)$, so raw liquidity is $e^{u_t(\xi)}$. Although $u_t$ admits the eigenfunction expansion~\eqref{u:exp}, the pool value in~\eqref{eq:value-r-u-finite} is not linear in the modal coefficients. Substituting the modal expansion gives
\[
\cV_t
=
e^{2r_t}\int_{-L}^{L}
\exp\!\left(\sum_{n=0}^{\infty}u_t^n e_n(\xi)\right)
e^{\xi-\xi^-}\,d\xi,
\]
not a linear sum of the coefficients. Thus an additive eigenmode-level decomposition of pool value is not available in the log coordinate, even though the modal expansion remains useful for representing the SPIDE dynamics. The exact linear modal value decomposition belongs to the raw-liquidity coordinate $U_t$ and is given below.

\subsubsection{The Black-Scholes type SPIDE for \texorpdfstring{$(r, U)$}{(r,U)}} \label{sec:BS-model}
Setting $\sigma_m \equiv 0$ in~\eqref{eqn:mkt-model-u}, the market model for the spot-centered liquidity profile $U_t$ becomes
\begin{align}
 dr_t &= \left(\mu_t - \frac{\sigma_t^2}2 \right) dt + \sigma_t\, dZ_t, \nonumber \\
 dU_t(\xi) &= \{\cA U_t + h(t, \xi) \}\,dt + \sum_{m=1}^M \tilde\sigma_m(t, \xi)[U_t]\, dW^m_t. 
  \label{eqn:mkt-model-u-BS}
\end{align}
Under the nonlocal assumptions of Section~\ref{sec:th-bg} on the operators $\tilde\sigma_m$, the solution $U_t$ need not be nonnegative; precise conditions on $V$ and $\tilde\sigma_m$ under which~\eqref{eqn:mkt-model-u-BS} preserves nonnegativity lie beyond the scope of this paper. We therefore assume throughout this subsection that, almost surely, $U_t(\xi)\ge0$ for almost every $\xi\in[-L,L]$ and every $t\in[0,T]$. This is a condition on the solution, not a consequence of Section~\ref{sec:th-bg}, and the reserve, value, and LVR expressions below have their financial meaning under it. Nonnegativity suffices for these expressions; the relation $u=\ln U$ would require strict positivity, which is not assumed. The LVR uses the value $U_t(0)$, and under Assumption~\ref{a:pointwise} this value is nonnegative as well: by Lemma~\ref{lem:pointwise}, for each $t$ it is the value of a continuous function that agrees almost everywhere near $\xi=0$ with $U_t\ge0$, so $U_t(0)\ge0$ almost surely; by the continuity in $t$ of Theorem~\ref{th:temporal-regularity}, almost surely $U_t(0)\ge0$ for all $t\in(0,T]$.

\paragraph{\textit{Pool quantities under the BS model}}
The pool quantities are
\begin{align*}
x_t &= e^{r_t}\int_{0}^{L} U_t(\xi)\,e^{\xi}\, d\xi, \qquad
y_t = e^{2r_t}\int_{-L}^0 U_t(\xi)\, e^{2\xi}\, d\xi, \\
\cV_t &= e^{2r_t} \int_{-L}^L U_t(\xi)\, e^{\xi - \xi^-}\,  d\xi. 
\end{align*}
The LVR under this model is
\begin{equation*}
\lvr_t = \frac12 \int_0^t  U_s(0)\, e^{2 r_s}\, \sigma_s^2\, ds.
\end{equation*}
Crucially, $\cV_t$ and $\lvr_t$ are both \emph{linear} in the spot-centered liquidity profile~$U_t$. This linearity---absent in the OU formulation where the dependence on~$u$ is exponential---is the principal analytical advantage of the $(r,U)$ coordinate and the BS-type noise specification.

\paragraph{\textit{Eigenfunction expansion of pool value}}
In the $(r,U)$ coordinate the modal decomposition is linear. If $U_t(\xi)=\sum_{n=0}^\infty U_t^n e_n(\xi)$, define
\begin{align*}
 \mathcal V_t^{(n)} &:= e^{2r_t}\int_{-L}^{L} e_n(\xi)\, e^{\xi-\xi^-}\, d\xi.
\end{align*}
Then
\begin{equation} \label{eq:eigen-decomp}
 \cV_t = \sum_{n=0}^\infty U_t^n \mathcal V_t^{(n)}.
\end{equation}
The individual modal terms need not be nonnegative because $e_n$ may change sign for $n\geq 1$; the assumed nonnegativity is a property of the full liquidity profile $U_t$, not of each modal component separately.

In this case, by taking expectations on both sides of \eqref{eqn:mkt-model-u-BS}, the expectation $M_t := \Eof{U_t}$ satisfies the following linear equation  
\[
dM_t(\xi) = \{\cA M_t + h(t, \xi) \}\,dt
\]
whose solution can be obtained by
\begin{equation} 
\label{eq:BS-mean}
M_t(\xi) = \mathcal{S}(t)[M_0](\xi) + \int_0^t \mathcal{S}(t-s)[h(s,\cdot)](\xi)\, ds
\end{equation}
as in \eqref{eq:OU-mean}. Alternatively, $M_t$ can be expressed in terms of eigenfunctions as follows. First, let $\bU_t = (U^n_t)_{n=0}^\infty$ , $\bh_t = (h^n_t)_{n=0}^\infty$ be infinite dimensional vectors and $\bA = (-\lambda_k\delta_{n,k} - V^{n,k}) $, $\bsigma_m = (\tilde\sigma_m^{n,k})$ infinite dimensional square matrices. With $\bM_t := \Eof{\bU_t}$ we rewrite the equation in matrix form as 
\[
\frac d{dt}\bM_t = \bA \bM_t + \bh_t.
\]
We obtain
\[
\bM_t = e^{t\bA}\bM_0 + \int_0^t e^{(t-s)\bA} \bh_s ds,
\]
thus
\begin{align}\label{M_t}
M_t(\xi) = \be^\top(\xi)\bM_t = \be^\top(\xi)e^{t\bA}\bM_0 + \int_0^t \be^\top(\xi) e^{(t-s)\bA} \bh_s ds.
\end{align}

\begin{remark}[Nonnegative mean] \label{rmk:BS-mean-nonneg}
The following sufficient condition for a nonnegative mean profile does not use the assumed nonnegativity of the sample paths. Suppose that $U_0\ge0$ and, for every $t\in[0,T]$, $h(t)\ge0$ almost everywhere on $[-L,L]$, and that the kernel of $V$ satisfies $V(x,y)\le0$ for almost every $(x,y)\in[-L,L]^2$. Then $M_t\ge0$ almost everywhere on $[-L,L]$ for every $t\in[0,T]$.
Indeed, by Theorem~\ref{th:mild:sol}, $\sup_{t\le T}\Eof{\|U_t\|_w^2}<\infty$, so the stochastic integral in~\eqref{mild:G} has mean zero, and taking expectations gives
\[
	M_t = \mathcal{G}(t)U_0 + \int_0^t \mathcal{G}(t-s)h(s)\, ds - \int_0^t \mathcal{G}(t-s)V[M_s]\, ds,
\]
which is equivalent to~\eqref{eq:BS-mean} by Lemma~\ref{lem:S}.
The heat kernel $G_t(x,y)$ is strictly positive. Indeed, with $p:=\alpha-\frac12>-1$ and the change of variables $x=L\eta$, $\mathcal L=-a\,J^{(p,p)}$, where $J^{(p,p)}=-(1-\eta^2)\partial_\eta^2+(2p+2)\eta\,\partial_\eta$ is the Jacobi operator with equal parameters, and $G_t(x,y)=L^{-2\alpha}G^{(p,p)}_{at}(x/L,y/L)$, where $G^{(p,p)}_t$ is the Jacobi heat kernel with respect to $(1-\eta^2)^p\,d\eta$ on $[-1,1]$; the latter has a positive lower bound by \citeA[Theorem~A]{nowak2026new}. Hence $\mathcal{G}(t)$ maps nonnegative functions to nonnegative functions.
So does $-V$, whose kernel is nonnegative. The right-hand side therefore defines a map $\Phi$ that sends nonnegative elements of $C([0,T];L^2_w)$ to nonnegative elements. By~\eqref{G:norm} and $\|V[z]\|_w\le\|V\|\,\|z\|_w$ from~\eqref{iint:V}, $\Phi$ is a contraction for the norm $\sup_{t\le T}e^{-2\|V\|t}\|M_t\|_w$, so its Picard iterates from $\mathcal{G}(t)U_0+\int_0^t\mathcal{G}(t-s)h(s)\,ds\ge0$ remain nonnegative and converge to its unique fixed point $M$. The set of nonnegative functions is closed in $L^2_w$, so $M_t\ge0$.

The condition is restrictive. Since $e_0$ is constant, the level entry $V_{00}=\langle Ve_0,e_0\rangle_w$ of a nonpositive kernel satisfies $V_{00}\le0$: relocations add to the profile level and cannot damp it. The sparse and dense fits of Section~\ref{sec:case-study}, estimated on the rank grid, have $\hat V_{00}>0$ (Table~\ref{tab:fitted-params}), so this remark does not establish a nonnegative mean for those fitted operators.
\end{remark}

For the calculation of the expectation $\Gamma_t(\xi, \eta):=\Eof{U_t(\xi)U_t(\eta)}$, we resort to the eigenfunction expansion of \eqref{eqn:mkt-model-u-BS}. Recall from \eqref{eq:u^n} that, since $\sigma_m = 0$, for all $m$, $U^n_t$ satisfies the infinitely dimensional SDE
\[
dU_t^n 
= \left[-\sum_{k=0}^\infty\left(\lambda_k \delta_{nk} +  V^{nk} \right) U_t^k + h_t^n \right] dt  + \sum_{m,k} \tilde\sigma_m^{nk} U^k_t dW_t^m.
\]
Rewrite the above equation in matrix form as 
\[
d\bU_t = (\bA \bU_t +  \bh_t)dt + \sum_m\bsigma_m  \bU_t dW_t^m.
\]
Writing the last equation in integral form, applying It\^o isometry to the product $\bU_t \bU_s^\top$, and taking expectations shows that the cross-moment $\bGamma_{t,s} := \Eof{\bU_t \bU_s^\top}$ satisfies
\[
    \bGamma_{t,s} = \bM_t \bM_s^\top + \sum_m \int_0^{t \wedge s} e^{(t-\tau)\bA}\bsigma_m \bGamma_{\tau} \bsigma_m^\top e^{(s-\tau)\bA^\top}\, d\tau,
\]
where the symmetric matrix $\bGamma_t = \bGamma_{t,t}:= \Eof{\bU_t\bU_t^\top}$ satisfies the linear differential equation of Lyapunov type
\[
\frac d{dt}\bGamma_t = \bA\bGamma_t + \bGamma_t\bA^\top + \bh_t\bM_t^\top + \bM_t\bh_t^\top + \sum_m\bsigma_m \bGamma_t \bsigma_m^\top.
\]
It follows that
\begin{align}\label{Gamma_t}
\Gamma_{t,s}(\xi, \eta) := \Eof{U_t(\xi)U_s(\eta)} = \Eof{\be^\top(\xi)\bU_t\bU_s^\top\be(\eta)} = \be^\top(\xi)\bGamma_{t,s}\be(\eta).
\end{align}
Under Assumption~\ref{a:pointwise}, $U_t$ is independent of $(r_t,\sigma_t)$, and the mean and covariance functions of $\cV_t$ and $\lvr_t$ can be computed as follows, whenever the expectations involved are finite; as in Section~\ref{sec:OU-model}, this holds for $\lvr_t$ and is not ensured for $\cV_t$:
\begin{align*}
	&\Eof{\cV_t} = \Eof{e^{2r_t}} \int_{-L}^L M_t(\xi) e^{\xi-\xi^-}\, d\xi,\\
	&\Eof{\lvr_t} = \frac12 \int_0^t \Eof{e^{2 r_s}\sigma_s^2}\, M_s(0)\,  ds,\\
	&\Eof{\cV_t\,\cV_s} = \Eof{e^{2(r_t+r_s)}} \int_{-L}^L \int_{-L}^L \Gamma_{t,s}(\xi,\eta)\, e^{\xi-\xi^-+\eta-\eta^-}\, d\xi d\eta,\\
	&\Eof{\lvr_t\,\lvr_s} = \frac{1}{4} \int_0^t \int_0^s \Eof{e^{2(r_{\tau}+r_{\rho})}\sigma_{\tau}^2 \sigma_{\rho}^2} \Gamma_{\tau,\rho}(0,0)\, d\tau d\rho,
\end{align*}
where $M_t$ and $\Gamma_{t,s}$ are given by~\eqref{M_t} and \eqref{Gamma_t}, respectively. The LVR moments are finite: by Lemma~\ref{lem:pointwise} and the Cauchy--Schwarz inequality, $|M_s(0)|\le C_0A_T$ and $|\Gamma_{\tau,\rho}(0,0)|\le C_0^2A_T^2$, so that $\Eof{\lvr_t}\le\frac12\,t\,q_T^{1/2}C_0A_T$ and $\Eof{\lvr_t\,\lvr_s}\le\frac14\,ts\,q_T\,C_0^2A_T^2$.
\begin{example}
If we further assume the Black-Scholes model, i.e., $dr_t = (\mu - \frac{\sigma^2}2) dt + \sigma dZ_t$, where $\sigma > 0$ and $\mu$ are constants, then
\begin{align*}
	&\Eof{\cV_t} = e^{2r_0+(2\mu+\sigma^2)t} \int_{-L}^L M_t(\xi) e^{\xi-\xi^-}\, d\xi,\\
	&\Eof{\lvr_t} = \frac{\sigma^2 e^{2r_0}}{2} \int_0^t e^{(2\mu+\sigma^2)s} M_s(0)\, ds,\\
	&\Eof{\cV_t\,\cV_s} = e^{4r_0+(2\mu+\sigma^2)(t+s)+4\sigma^2(t\wedge s)} \int_{-L}^L \int_{-L}^L \Gamma_{t,s}(\xi,\eta) \, e^{\xi-\xi^-+\eta-\eta^-} \, d\xi d\eta,\\
	&\Eof{\lvr_t\,\lvr_s} = \frac{\sigma^4 e^{4r_0}}{4} \int_0^t \int_0^s e^{(2\mu+\sigma^2)(\tau+\rho)+4\sigma^2(\tau\wedge \rho)}\, \Gamma_{\tau,\rho}(0,0) \, d\tau d\rho.
\end{align*}
\end{example}

\providecommand{\bPhi}{\boldsymbol{\Phi}}
\providecommand{\bs}{\boldsymbol{s}}
\providecommand{\bB}{\boldsymbol{B}}
\providecommand{\bC}{\boldsymbol{C}}
\providecommand{\bT}{\boldsymbol{T}}
\providecommand{\beps}{\boldsymbol{\varepsilon}}

\section{Empirical Case Study}\label{sec:case-study}

This section reads the liquidity dynamics through the theory. The SPIDE framework admits many estimable and restrictable branches; we calibrate one---the stationary additive-OU specification in log-liquidity coordinates---on three Uniswap~V3 pools and read the fitted operator for interpretable features: how fast each profile mean-reverts, which modes the relocation operator damps, and how the pools differ. The aim is to showcase the framework and its restrictions to special cases, in the spirit of \citeA{risk2025dynamics} (Section~\ref{sec:motivation}), not to find the ``best'' dynamics for liquidity. Since the full calibration is involved, we report the procedure in outline and the calibrated results and their implications.

\subsection{Data and Exploratory Profiles}\label{subsec:case-data}

We analyze three Uniswap~V3 pools: WETH--USDC 5\,bps on Arbitrum (ARB5), WETH--USDC 5\,bps on Ethereum Mainnet (ETH5), and WETH--USDC 30\,bps on Mainnet (ETH30). ARB5 and ETH5 differ in chain and trading cadence; ETH5 and ETH30 fix the pair and vary the fee tier; together they give a broad, a concentrated, and a shifting profile. The selection is illustrative rather than a sample of a defined universe. For each pool we retrieve tick-level liquidity snapshots, obtained through the Teahouse Finance API, and price histories and, following \citeA{risk2025dynamics}, represent each snapshot on a uniform $N=201$ grid $x_1<\dots<x_N$ on $[-L,L]$ with $L=1$, whose coordinate is the normalized rank of the sampled liquidity ticks. In the centered preprocessing, sampled liquidity ticks are sorted by tick index; the sampled tick closest to the current pool tick has row index $i$ and rank zero, and row $j$ has $x=(j-i)/100$. Adjacent sampled rows differ by $0.01$ in rank, but rank zero need not coincide exactly with the current tick, and equal rank differences need not be equal price distances. The grid is not calibrated to log-moneyness distances: the fitted rates, the concentration order $\hat\alpha$, the relocation matrices, and the active band below are estimated in these rank coordinates and describe the profile in them, not in price distances. The truncation to $[-L,L]$ trades boundary regularization against the sparse, stale ticks beyond it (Appendix~\ref{app:empirical-calibration}). Snapshots are subsampled at a fixed block stride per chain ($2{,}400$ blocks on Mainnet, $115{,}200$ on Arbitrum), equal to 8 hours at nominal block times; $\Delta$ denotes this step throughout, and all rates, relocation entries and eigenvalues are reported as dimensionless per-step quantities. A common $1200$-observation Full block, starting August 2023, is partitioned into three contiguous $400$-observation windows W1--W3 (half-open in blocks, start inclusive and end exclusive, Table~\ref{tab:pool-summary}); we report W1 and Full.

\begin{figure}[!ht]
\centering
\includegraphics[width=\textwidth]{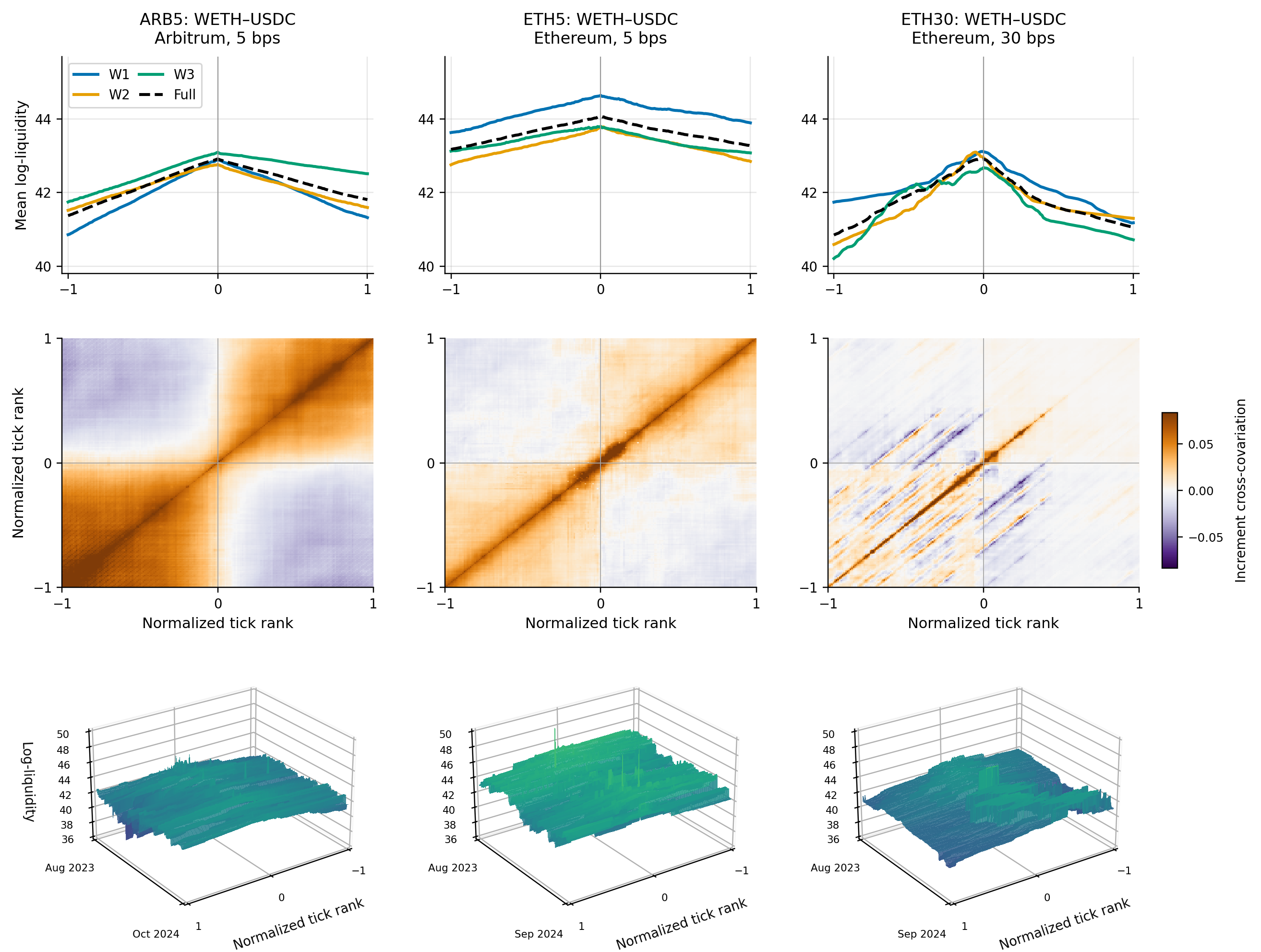}
\caption{Exploratory liquidity diagnostics across pools. Top row: mean log-liquidity profiles for W1--W3 and the Full block. Middle row: empirical cross-covariation of one-step log-profile increments over the Full block, with a common symmetric color scale. Bottom row: connected surface views of subsampled Full-block profiles. All spatial coordinates are normalized tick ranks.}
\label{fig:mean-curves}
\end{figure}

Figure~\ref{fig:mean-curves} shows three qualitatively different surfaces under the same windowing. ARB5 and ETH5 keep a stable mean shape across windows while individual profiles move visibly through time, ETH5 with intermittent spikes near the active region; ETH30 is less regular, with a mean profile that shifts across windows, persistent stale boundary mass, and large intermittent concentration changes. The empirical cross-covariation of one-step increments $\Delta u_j(x_k)=u_{j+1}(x_k)-u_j(x_k)$ (middle row) shows broad same-side dependence on ARB5, covariation concentrated near the active band on ETH5, and sharper diagonal bands with localized bursts on ETH30.

\begin{table}[!ht]
\centering
\caption{Estimation windows by block height. Calendar dates are UTC dates of the displayed block, and each end block is excluded. The Full block is the union of W1--W3.}
\label{tab:pool-summary}
\begin{tabular}{llcc}
\toprule
\textbf{Pool} & \textbf{Window} & \textbf{Start Block (UTC date)} & \textbf{Exclusive End Block (UTC date)} \\
\midrule
\multirow{3}{*}{ARB5} & W1 & \texttt{124163200} (2023-08-23) & \texttt{170243200} (2024-01-14) \\
		                      & W2 & \texttt{170243200} (2024-01-14) & \texttt{216323200} (2024-05-29) \\
		                      & W3 & \texttt{216323200} (2024-05-29) & \texttt{262403200} (2024-10-10) \\
\midrule
\multirow{3}{*}{ETH5} & W1 & \texttt{17973929} (2023-08-23) & \texttt{18933929} (2024-01-04) \\
		                      & W2 & \texttt{18933929} (2024-01-04) & \texttt{19893929} (2024-05-18) \\
		                      & W3 & \texttt{19893929} (2024-05-18) & \texttt{20853929} (2024-09-29) \\
\midrule
\multirow{3}{*}{ETH30} & W1 & \texttt{17973329} (2023-08-22) & \texttt{18933329} (2024-01-04) \\
		                       & W2 & \texttt{18933329} (2024-01-04) & \texttt{19893329} (2024-05-18) \\
		                       & W3 & \texttt{19893329} (2024-05-18) & \texttt{20853329} (2024-09-29) \\
\bottomrule
\end{tabular}
\end{table}

\subsection{Modeling Assumptions}\label{subsec:case-assumptions}

We work in the OU/log-liquidity branch of Section~\ref{sec:OU-model} under two assumptions that select a branch of the family.

\begin{assumption}[Additive OU branch]\label{a:additive}
$\tilde{\sigma}_m \equiv 0$ for all $m$ in \eqref{SPIDE}, and $h(t,x)=h(x)$ is time-invariant.
\end{assumption}

\begin{assumption}[Stationary centered fluctuations]\label{a:stationary}
The profile is well approximated by stationary fluctuations around a stationary mean, for which the temporal mean of the fitting segment, $\bar u(x)=J_{\mathrm{tr}}^{-1}\sum_{j=1}^{J_{\mathrm{tr}}}u_j(x)$, is the finite-sample surrogate.
\end{assumption}

Under Assumption~\ref{a:additive} the noise is additive through the deterministic spatial functions $\sigma_m(x)$, so the one-step conditional expectation is linear in the modal coordinates and the innovations are conditionally Gaussian (Remark~\ref{rmk:model-interpretation}). Under Assumption~\ref{a:stationary}, whose use of $\bar u$ rests on the ergodic theory of infinite-dimensional OU processes \cite{daprato1996ergodicity}, we fit the centered profile $y_j(x)=u_j(x)-\bar u(x)$; centering removes $h$ and leaves a zero-intercept transition, the $O_p(J_{\mathrm{tr}}^{-1/2})$ intercept discrepancy being absorbed into the innovations, and forecasts are mapped back by adding $\bar u$. The failure mode is explicit: under structural drift $\bar u$ ceases to approximate the conditional mean and a systematic out-of-sample bias can appear; the ETH30 held-out diagnostics below are consistent with this mechanism.

\subsection{Calibration Procedure}\label{subsec:case-estimator}\label{subsubsec:case-reduction}

\paragraph{\textit{What the data identify.}}
Under Assumption~\ref{a:additive} the modal system \eqref{eq:u^n} of Theorem~\ref{th:sys:sde} reads $du_t^n=[-\sum_k(\lambda_k\delta_{nk}+V^{nk})u_t^k+h^n]\,dt+\sum_m\sigma^n_m\,dW^m_t$. Fix a rank $K$, let $\boldsymbol c_t\in\R^K$ collect the retained Gegenbauer coefficients, and impose as part of the specification that $V^{nk}=0$ whenever $n\ge K$ or $k\ge K$ (Appendix~\ref{app:empirical-calibration}). The retained block then closes exactly as a finite-dimensional OU process, $d\boldsymbol c_t=(-\bA_K\boldsymbol c_t+\boldsymbol h_K)\,dt+\sum_m\bsigma_m^{(K)}dW^m_t$ with $\bA_K=\bLambda_K+\bV_K$, $\bLambda_K=\diag(\lambda_0,\dots,\lambda_{K-1})$, $\lambda_n=a\,n(n+2\alpha)$, and $\bV_K=(V^{nk})_{n,k<K}$, so its time-$\Delta$ transition is exactly Gaussian and linear: when all eigenvalues of $\bA_K$ have positive real part, the centered coefficients $\tilde{\boldsymbol c}_j:=\boldsymbol c_{j\Delta}-\bA_K^{-1}\boldsymbol h_K$ satisfy
\begin{equation}\label{eq:discrete-centered}
\begin{gathered}
\tilde{\boldsymbol c}_{j+1}=\bPhi_K\tilde{\boldsymbol c}_j+\beps_{j+1},
\qquad
\bPhi_K=\exp(-\bA_K\Delta),
\\
\beps_{j+1}\overset{\text{i.i.d.}}{\sim}N(\boldsymbol 0,\bSigma_\Delta),
\qquad
\bSigma_\Delta=\int_0^\Delta e^{-s\bA_K}\bC e^{-s\bA_K^\top}ds,
\end{gathered}
\end{equation}
with $\bC=\sum_m\bsigma_m^{(K)}\bsigma_m^{(K)\top}$; stationarity is $\rho(\bPhi_K)<1$, where $\rho(\cdot)$ denotes the spectral radius. Under Assumption~\ref{a:stationary} the stationary mean $\bA_K^{-1}\boldsymbol h_K$ is replaced by the projection of $\bar u$. Thus the centered data form a vector autoregression (VAR) of order one, and everything a single trajectory can reveal about $(a,\alpha,\bV_K,\sigma_m)$ passes through the pair $(\bPhi_K,\bSigma_\Delta)$. The calibration problem is to recover $(a,\alpha,\bV_K)$ from $\bPhi_K$; the noise is left unrestricted and profiled out, so the scheme targets the drift only. Since the cadence is fixed we report per-step quantities $a_\Delta:=a\Delta$ and $\bV_\Delta:=\bV_K\Delta$, and write $\bA_\Delta^{(K)}:=\bA_K\Delta$.

The two pieces of the drift have distinct roles. The Sturm--Liouville part damps mode $n$ at the rate $\lambda_n=a\,n(n+2\alpha)$, quadratic in $n$; $\bV_K$ supplies bounded relocations, a diagonal entry adding mode-$n$ damping and an off-diagonal entry $(\bV_K)_{ij}$ letting mode $j$ feed the drift of mode $i$. Since $\lambda_0=0$, the differential operator supplies no damping of the level mode: stationarity of the profile level must come from $\bV_K$, through $V_{00}$ or off-diagonal coupling, and observed level mean reversion is evidence for $V$, not for $\cL$.

\phantomsection\label{subsubsec:case-fixedspace}\label{subsubsec:case-illposed}
The order $\alpha$ sets both the eigenvalue ladder and, through the weight $w_\alpha(x)=(1-x^2)^{\alpha-1/2}$, the Gegenbauer modes and the inner product in which residuals would be measured. Estimating $\alpha$ in each candidate's own coordinates therefore moves the object and the ruler together---$w_\alpha$ up-weights the stale boundary nodes for $\alpha<1/2$ and down-weights them for $\alpha>1/2$---and an optimizer can lower the objective by re-weighting residuals rather than by fitting dynamics. We therefore read the data in one frame shared by all candidates. The polynomials of degree at most $r-1$ form a space $\mathcal P_{r-1}$ that is the same for every $\alpha$ (the Gegenbauer polynomials of every order span it) and is invariant under $\cL$, so the Galerkin truncation of $\cL$ on it is exact. With $\bB\in\R^{N\times r}$ having Euclidean-orthonormal columns spanning the grid samples of the Legendre polynomials $P_0,\dots,P_{r-1}$, the observations enter the fit only through the reference coefficients
\begin{equation}\label{eq:score-map}
\bs_j=\bB^\top(\mathbf u_j-\bar{\mathbf u})\in\R^r,
\end{equation}
formed once and free of $\alpha$. A candidate $\theta_r=(a_\Delta,\bV_\Delta)$ at order $\alpha$ has modal drift $\bA_\Delta^{(r)}=\diag(a_\Delta n(n+2\alpha))_{n<r}+\bV_\Delta$ in its own Gegenbauer basis and reaches the data through the change of basis $\bT(\alpha)$ from the order-$\alpha$ modes to the Legendre frame,
\begin{equation}\label{eq:Dtheta}
\boldsymbol D(\theta_r;\alpha)=\bT(\alpha)\,\bA_\Delta^{(r)}\,\bT(\alpha)^{-1},
\qquad
\bPhi(\theta_r;\alpha)=\exp\big(-\boldsymbol D(\theta_r;\alpha)\big),
\end{equation}
whose differential block is $a_\Delta\boldsymbol M(\alpha)$ with $\boldsymbol M(\alpha)$ affine in $\alpha$ (Lemma~\ref{lem:ladder}); the relocation block is carried over by $\alpha$-weighted analysis up to a quadrature residual (Appendix~\ref{app:score-algebra}). The common Legendre frame keeps the residual coordinates fixed as $\alpha$ varies.

\paragraph{\textit{Objective and estimator.}}\label{subsubsec:case-estdef}
With one-step residuals $\widehat\beps_j(\theta_r;\alpha)=\bs_{j+1}-\bPhi(\theta_r;\alpha)\bs_j$ and their covariance $\widehat\bSigma(\theta_r;\alpha)=(J_{\mathrm{tr}}-1)^{-1}\sum_j\widehat\beps_j\widehat\beps_j^\top$, the objective is the profiled Gaussian quasi-likelihood with a Frobenius penalty on the relocation block,
\begin{equation}\label{eq:objective}
\mathcal J(\theta_r;\alpha)=\log\det\big(\widehat\bSigma(\theta_r;\alpha)+\epsilon\boldsymbol I\big)+\lambda_V\|\bV_\Delta\|_F^2,
\qquad \epsilon>0,\ \lambda_V\ge0 .
\end{equation}
The log-determinant is the Gaussian likelihood with the innovation covariance profiled out. The estimator is
\begin{equation}\label{eq:estimator}
(\hat\alpha,\hat a_\Delta,\widehat\bV_\Delta)=\argmin_{\alpha\in\mathcal G_\alpha}\ \min_{\theta_r\in\Theta_{\mathcal I}}\ \mathcal J(\theta_r;\alpha)
\quad\text{subject to}\quad
\min_k\operatorname{Re}\,\mathrm{eig}_k\big(\bA_\Delta^{(r)}\big)\ge\eta_{\mathrm{stab}},
\end{equation}
where $\mathcal I\subseteq\{0,\dots,r-1\}$ is the relocation support ($\bV_\Delta$ is zero outside $\mathcal I\times\mathcal I$), $\Theta_{\mathcal I}$ the bounded search domain, $\eta_{\mathrm{stab}}>0$ a stability margin, and $\mathcal G_\alpha$ a compact grid in $(-1/2,\infty)$ over which $\alpha$ is profiled; grid, bounds, multistart and optimizer settings are in Appendix~\ref{app:alpha-profile}. The transition is fitted through the exponential in \eqref{eq:Dtheta} rather than by inverting a matrix logarithm of the least-squares estimate of $\bPhi$, which is ill-conditioned for modes with $\lambda_n\Delta\gg1$.

\paragraph{\textit{Identification and forecast map.}}\label{subsubsec:case-alpha}\label{subsubsec:case-forecastmap}
Because $\alpha$ enters only through $\bPhi(\theta_r;\alpha)$ and not through $\bs_j$ or the residual norm, it can help only by explaining the dynamics; but if $\bV_\Delta$ is free on every retained mode, any change in the differential damping can be undone entry by entry and $\alpha$ is not separately identified. What separates $\cL$ from $V$ is Assumption~\ref{a:V} made quantitative: $\|V\|^2=\sum_{n,k}|V^{nk}|^2<\infty$ forces the relocation diagonal to vanish along $n$ while $\lambda_n$ grows quadratically. In finite rank this is imposed rather than observed, through the penalty $\lambda_V\|\bV_\Delta\|_F^2$ and by restricting the support $\mathcal I$ to the low-order profile modes, so that the retained modes outside $\mathcal I$ identify the differential part; $\hat\alpha$ and $\widehat\bV_\Delta$ must therefore be read jointly. Estimation and forecasting use two coordinate systems: the estimator lives in the fixed frame, and forecasts are produced in the Gegenbauer coordinates at the fitted order $\hat\alpha$, for a forecast rank $K\ge r$ with $\mathcal I\subseteq\{0,\dots,K-1\}$, by projecting the profiles and $\bar u$ onto the first $K$ modes at $\hat\alpha$, propagating one step by \eqref{eq:discrete-centered} with $\widehat\bV_\Delta$ extended by zero (modes $r\le n<K$ move by their eigenvalues alone), and reconstructing on the grid with $\bar u$ added back.

\subsection{Evaluation and Discussion}\label{subsec:case-results}\label{subsubsec:case-models}\label{subsubsec:case-evaluation}

\paragraph{\textit{Configurations.}}
Unless stated otherwise we use $r=6$ reference coefficients, forecast with $K=7$ Gegenbauer modes, and take the dense relocation support $\mathcal I=\{0,1,2\}$ (level, tilt, concentration). Each configuration isolates one question about what the data pin down.
\begin{itemize}
\item \textbf{Dense-3}: $\alpha$ profiled on $\mathcal G_\alpha$, full nonsymmetric $\bV_\Delta$ on modes $\{0,1,2\}$, $\lambda_V=0$.
\item \textbf{Dense-3 + penalty}: the same with $\lambda_V=3.0$ (Appendix~\ref{app:empirical-calibration}).
\item \textbf{Sparse SPIDE}: $\alpha=1/2$ (Legendre case), $\bV_\Delta$ restricted to the diagonal entries $(0,0)$ and $(1,1)$: how much of the dense block do the data require?
\item \textbf{Legendre OU, $V{=}0$}: $\alpha=1/2$, no relocations: can the differential operator alone carry the dynamics? Since $\lambda_0=0$, the level mode is a random walk by construction and $\rho(\bPhi_K)=1$; the scalar $a_\Delta$ is fitted by the log-determinant criterion on the $K=7$ Legendre coefficients with the level mode held at the unit root.
\end{itemize}
Two non-SPIDE references: the \textbf{martingale null} $\hat{\bu}_{j+1}=\bu_j$, which defines the denominators of all ratio diagnostics; and a \textbf{Factor VAR}, $\boldsymbol z_{j+1}=\boldsymbol\Psi\boldsymbol z_j+\boldsymbol\eta_{j+1}$ on the first $K=7$ coefficients of a mass-weighted principal component analysis (PCA) fitted on the training half, $\boldsymbol\Psi$ by ridge least squares, a generic low-rank model carrying no operator structure.

\paragraph{\textit{Protocols and metrics.}}
The \emph{forecasting} protocol fits once on the first half of a window and rolls one-step-ahead raw log-profile forecasts $\hat\bu_{j|j-1}$ over the held-out second half; the \emph{in-sample} protocol refits on the whole window and examines residual whiteness. All diagnostics use the same evaluation setup across configurations, independently of fitted parameters: the common rank grid of Section~\ref{subsec:case-data}, a shared mass-weighted PCA evaluation space fitted once per window, and ratios against the common martingale null. We report the active-band pointwise error ratio $R_A$ (mass-weighted root-mean-square error, RMSE, on the central band $|x|\le0.5$ of the rank grid), the value-weighted profile ratio $R_V$ (an RMSE ratio for a standardized value-weighted profile score, lognormal-corrected, spot price normalized out, so it diagnoses the profile forecast rather than the joint price--liquidity model; because the value weights are applied to the rank coordinate, it is not a pool-value forecast error), and the Hosking whiteness severity $S_H=Q_H/(d^2H)$ of the one-step residuals in the $d=9$ evaluation space with $H=10$ lags, scaled to be near $1$ under whiteness; definitions are in Appendix~\ref{app:empirical-metrics}. Values of $R_A,R_V$ below one mean the fitted structure lowered that error out of sample. There is no single natural loss for a spatial forecast, so the panel is read as diagnostics of the fitted structure, not as a model competition.

\paragraph{\textit{Results.}}\label{subsubsec:case-main-results}
Table~\ref{tab:forecasting-results} reports the diagnostics and Table~\ref{tab:fitted-params} the fitted parameters for W1 and Full; full Dense-3 relocation matrices are in Appendix~\ref{app:V-matrices}.

\begin{table}[!ht]
\centering
\scriptsize
\renewcommand{\arraystretch}{0.93}
\setlength{\tabcolsep}{2.4pt}
\caption{In-sample whiteness and out-of-sample forecasting diagnostics ($K=7$). Ratio columns are relative to the martingale null (equal to $1.000$ by definition and omitted); values below $1$ indicate that the fitted structure lowered that error out of sample. Hosking columns report $S_H=Q_H/(d^2H)$ and are not ratios. Dashes mark diagnostics outside the reported protocol for that model.}
\label{tab:forecasting-results}
\begin{tabular}{ll @{\hspace{2pt}} c @{\vrule width 1pt\hspace{3pt}} ccc}
\toprule
 & & \multicolumn{1}{c}{In-sample} & \multicolumn{3}{c}{Out-of-sample} \\
\cmidrule(lr){3-3}\cmidrule(lr){4-6}
\textbf{Pool} & \textbf{Config} & $S_H^{\mathrm{IS}}$ & $R_A$ & $R_V$ & $S_H^{\mathrm{OOS}}$ \\
\midrule
\multirow{4}{*}{\shortstack[l]{ARB5\\W1}}
 & Legendre OU, $V{=}0$ & 0.854 & 0.955 & 0.984 & 1.350 \\
 & Sparse SPIDE & 0.862 & 0.958 & 0.965 & 1.332 \\
 & Dense-3 & 0.849 & 0.987 & 1.075 & 1.314 \\
 & Dense-3 + pen & 0.846 & 0.957 & 0.985 & 1.326 \\
\midrule
 & Factor VAR & --- & 1.006 & 1.056 & 1.641 \\
\midrule
\multirow{4}{*}{\shortstack[l]{ARB5\\Full}}
 & Legendre OU, $V{=}0$ & 1.171 & 0.931 & 0.864 & 2.063 \\
 & Sparse SPIDE & 1.166 & 0.926 & 0.780 & 2.033 \\
 & Dense-3 & 1.190 & 0.948 & 1.150 & 2.276 \\
 & Dense-3 + pen & 1.208 & 0.929 & 0.807 & 2.018 \\
\midrule
 & Factor VAR & --- & 0.966 & 1.214 & 2.879 \\
\midrule
\multirow{4}{*}{\shortstack[l]{ETH5\\W1}}
 & Legendre OU, $V{=}0$ & 0.744 & 0.863 & 1.062 & 1.403 \\
 & Sparse SPIDE & 0.746 & 0.869 & 1.080 & 1.434 \\
 & Dense-3 & 0.751 & 0.861 & 1.224 & 1.944 \\
 & Dense-3 + pen & 0.749 & 0.852 & 1.079 & 1.581 \\
\midrule
 & Factor VAR & --- & 1.174 & 2.123 & 3.024 \\
\midrule
\multirow{4}{*}{\shortstack[l]{ETH5\\Full}}
 & Legendre OU, $V{=}0$ & 1.134 & 0.843 & 0.954 & 1.894 \\
 & Sparse SPIDE & 1.135 & 0.842 & 0.954 & 1.909 \\
 & Dense-3 & 1.133 & 0.854 & 0.964 & 1.953 \\
 & Dense-3 + pen & 1.133 & 0.847 & 0.957 & 1.928 \\
\midrule
 & Factor VAR & --- & 0.881 & 0.956 & 2.496 \\
\midrule
\multirow{4}{*}{\shortstack[l]{ETH30\\W1}}
 & Legendre OU, $V{=}0$ & 0.790 & 1.588 & 1.043 & 4.385 \\
 & Sparse SPIDE & 0.787 & 1.603 & 1.102 & 4.393 \\
 & Dense-3 & 0.897 & 1.866 & 1.310 & 4.650 \\
 & Dense-3 + pen & 0.916 & 1.850 & 1.226 & 4.640 \\
\midrule
 & Factor VAR & --- & 1.698 & 1.722 & 4.128 \\
\midrule
\multirow{4}{*}{\shortstack[l]{ETH30\\Full}}
 & Legendre OU, $V{=}0$ & 1.200 & 1.322 & 2.527 & 15.065 \\
 & Sparse SPIDE & 1.203 & 1.325 & 2.573 & 15.235 \\
 & Dense-3 & 1.666 & 1.399 & 3.079 & 17.255 \\
 & Dense-3 + pen & 1.569 & 1.394 & 3.051 & 17.147 \\
\midrule
 & Factor VAR & --- & 1.369 & 2.663 & 12.415 \\
\bottomrule
\end{tabular}
\end{table}
\renewcommand{\arraystretch}{1.0}

\begin{table}[!ht]
\centering
\scriptsize
\setlength{\tabcolsep}{3.2pt}
\caption{Fitted parameters (forecasting-protocol fits, first half of each block). $\hat\alpha$ is the concentration order; $\hat a_\Delta$ the rate per sampling step (nominally 8 hours); $\hat V_{00},\hat V_{11}$ the level and tilt self-feedback entries of $\widehat\bV_\Delta$ per step; $\rho(\hat\bPhi_K)$ the spectral radius of the fitted transition. Legendre OU and Sparse SPIDE rows fix $\alpha=0.5$; only the Dense-3 rows profile it. Legendre OU rows have no relocation entries and hold the level mode at a unit root.}
\label{tab:fitted-params}
\begin{tabular}{llccccc}
\toprule
\textbf{Pool} & \textbf{Config} & $\hat\alpha$ & $\hat a_\Delta$ & $\hat{V}_{00}$ & $\hat{V}_{11}$ & $\rho(\hat{\boldsymbol{\Phi}}_K)$ \\
\midrule
\multirow{4}{*}{ARB5 W1}
 & Legendre OU, $V{=}0$ & 0.50 & 0.0096 & --- & --- & 1.000 \\
 & Sparse SPIDE & 0.50 & 0.0129 & 0.088 & 0.060 & 0.925 \\
 & Dense-3 & 0.55 & 0.0132 & 0.118 & 0.067 & 0.946 \\
 & Dense-3 + pen & 0.80 & 0.0118 & 0.063 & 0.036 & 0.950 \\
\midrule
\multirow{4}{*}{ARB5 Full}
 & Legendre OU, $V{=}0$ & 0.50 & 0.0152 & --- & --- & 1.000 \\
 & Sparse SPIDE & 0.50 & 0.0160 & 0.090 & 0.020 & 0.950 \\
 & Dense-3 & 0.35 & 0.0177 & 0.112 & 0.045 & 0.961 \\
 & Dense-3 + pen & 0.35 & 0.0173 & 0.068 & 0.025 & 0.964 \\
\midrule
\multirow{4}{*}{ETH5 W1}
 & Legendre OU, $V{=}0$ & 0.50 & 0.0089 & --- & --- & 1.000 \\
 & Sparse SPIDE & 0.50 & 0.0105 & 0.090 & 0.020 & 0.960 \\
 & Dense-3 & 1.15 & 0.0092 & 0.158 & 0.042 & 0.964 \\
 & Dense-3 + pen & 1.25 & 0.0085 & 0.081 & 0.030 & 0.963 \\
\midrule
\multirow{4}{*}{ETH5 Full}
 & Legendre OU, $V{=}0$ & 0.50 & 0.0130 & --- & --- & 1.000 \\
 & Sparse SPIDE & 0.50 & 0.0144 & 0.007 & 0.021 & 0.993 \\
 & Dense-3 & 0.40 & 0.0155 & 0.031 & 0.016 & 0.983 \\
 & Dense-3 + pen & 0.40 & 0.0155 & 0.021 & 0.015 & 0.984 \\
\midrule
\multirow{4}{*}{ETH30 W1}
 & Legendre OU, $V{=}0$ & 0.50 & 0.0001 & --- & --- & 1.000 \\
 & Sparse SPIDE & 0.50 & 0.0012 & 0.008 & 0.035 & 0.993 \\
 & Dense-3 & $-0.45$ & 0.0050 & 0.092 & 0.101 & 0.987 \\
 & Dense-3 + pen & $-0.45$ & 0.0039 & 0.051 & 0.076 & 0.982 \\
\midrule
\multirow{4}{*}{ETH30 Full}
 & Legendre OU, $V{=}0$ & 0.50 & 0.0013 & --- & --- & 1.000 \\
 & Sparse SPIDE & 0.50 & 0.0019 & 0.002 & 0.002 & 0.998 \\
 & Dense-3 & $-0.30$ & 0.0031 & 0.015 & 0.016 & 0.991 \\
 & Dense-3 + pen & $-0.25$ & 0.0030 & 0.013 & 0.015 & 0.994 \\
\bottomrule
\end{tabular}
\end{table}

Read as diagnostics, Table~\ref{tab:forecasting-results} says four things. First, on ARB5 and ETH5---the pools compatible with the stationary branch, hereafter the \emph{branch-compatible pools}---the active-band ratios sit below one for every SPIDE configuration ($R_A\approx0.93$--$0.99$ on ARB5, $0.84$--$0.87$ on ETH5) and are nearly common across them, consistent with out-of-sample mean reversion carried by the low modes that all configurations share. Second,\label{subsubsec:case-vpenalty} the value-weighted profile ratio sits near or above one on the shorter W1 slices. The instructive exception is the unpenalized dense fit on the broad ARB5 pool, which over-parameterizes ($R_V=1.150$ on Full); activating the finite-rank Frobenius penalty motivated by the Hilbert--Schmidt restriction restores it ($R_V=0.807$) while shrinking the off-diagonal relocation entries (Appendix~\ref{app:V-matrices}) and moving $\hat\alpha$ only within the flat band of its profile (Appendix~\ref{app:alpha-profile}). On ETH5 Full the penalty changes the dense fit little, whereas on ETH5 W1 it lowers $R_V$ from $1.224$ to $1.079$ and out-of-sample severity from $1.944$ to $1.581$.

Third, the in-sample severity sits near unity and is common across SPIDE configurations, so none is misspecified in sample in the whiteness sense the objective does not enforce; out of sample it inflates moderately on the branch-compatible pools ($1.3$--$1.9$ on W1, $1.9$--$2.3$ on Full), consistent with mild parameter drift, and the Factor VAR shows larger out-of-sample severity than every SPIDE configuration on ARB5 and ETH5. Fourth,\label{subsubsec:case-eth30} ETH30 departs from all of this: $R_A$ sits well above one on both blocks ($1.32$--$1.87$) and $R_V$ reaches $2.5$--$3.1$ on Full, though only $1.04$--$1.31$ on W1, and the severity inflates out of sample to $\approx4$ on W1 and $15$--$17$ on Full against in-sample values of $0.8$--$1.7$---consistent with an operator that fits in sample but deteriorates on the held-out data, possibly under drift. The penalty leaves the ETH30 diagnostics essentially unchanged, which is consistent with strain on Assumption~\ref{a:stationary} rather than with an unbounded relocation block, although it is not a formal test; as documented by \citeA{risk2025dynamics}, wider fee tiers are dominated by slow, massive position adjustments and quote staleness.

\paragraph{\textit{Fitted parameters.}}\label{subsec:case-params}
For the Legendre and sparse configurations, the per-step damping of mode $n$ is $\lambda_{n,\Delta}+\hat V_{nn}$ with $\lambda_{n,\Delta}=\hat a_\Delta n(n+1)$, and its half-life is $\ln 2/(\lambda_{n,\Delta}+\hat V_{nn})$ steps. The $V{=}0$ fits leave the level a random walk ($\rho(\hat\bPhi_K)=1$); the sparse and dense fits assign positive $\hat V_{00}$, supplying the level damping that $\cL$ cannot, and yield $\rho(\hat\bPhi_K)<1$. On ARB5 the level relocation is sizeable ($\hat V_{00}\approx0.06$--$0.12$, exceeding the tilt damping), giving a level half-life of a few days in the sparse fits (on W1, $\hat V_{00}=0.088$ gives about $7.9$ steps: about $2.8$ days at the window's realized average of $8.6$ hours per step, or $2.6$ days at the nominal $8$ hours), the same order as the $\approx3$-day level persistence documented independently by \citeA{risk2025dynamics}: a broad pool whose dominant fitted feature is fluctuation of the rank-grid profile level. On ETH5 the level relocation is weaker on the Full block ($\hat V_{00}=0.007$--$0.031$ against $0.068$--$0.112$ on ARB5 Full) but reaches ARB5 magnitudes on W1.

On ETH30 the fitted differential reversion rate is small ($\hat a_\Delta\le0.005$ in every fit, $\approx10^{-4}$ for $V{=}0$ on W1, against $0.013$--$0.018$ on the branch-compatible Full blocks). The differential rate alone implies mode-1 half-lives of months for ETH30; the full drift, including relocation feedback, can decay at a different rate. The transition stays close to the unit-root boundary ($\rho(\hat\bPhi_K)\approx0.99$), as on ETH5 Full, and the dense fits push $\hat\alpha$ into the lower region of the grid ($-0.45$, at the edge, on W1; $-0.30$ and $-0.25$ on Full): the small differential rate is consistent with the shifting ETH30 profiles, and a training-mean mismatch, as anticipated after Assumption~\ref{a:stationary}, is a possible explanation of the out-of-sample bias.

The dense fits also reveal off-diagonal structure that only they can deliver; on ARB5 W1 the largest entry is the curvature-to-level coupling $\hat V_{02}\approx-0.18$ per step (concentration-to-level feedback in the rank-grid profile), shrunk to $\approx-0.04$ by the penalty. The profiled $\hat\alpha$ varies across windows and configurations ($0.35$--$0.80$ on ARB5, $0.40$--$1.25$ on ETH5), with nearly flat objective profiles (Appendix~\ref{app:alpha-profile}).

\paragraph{\textit{Discussion.}}\label{subsec:case-discussion}\label{subsubsec:case-future}
The case study supports the following reading: the operator family of Section~\ref{sec:th-bg} admits a disciplined calibration, and the stationary additive-OU branch yields interpretable estimates whose scales match what is known about these pools independently. Within one fixed frame each ingredient---the rate $a$, the order $\alpha$, the support and magnitude of $\bV_K$---may be freed or restricted, and the restrictions answer structural questions rather than compete for a ranking: the level mode needs the relocation operator; the data require only a sparse, bounded part of the dense block; the order is meaningful only jointly with that block. The fit leans toward the Legendre case with sparse bounded relocations, which we read as a regularized outcome rather than as evidence that $\alpha=1/2$. We are equally explicit about what is not established: a unique concentration order for the rank-grid model, since the objective profiles are nearly flat (Remark~\ref{rmk:boundary}); uniform forecasting gains over the martingale null (the active-band gains are consistent, but the value-weighted gains depend on the window and configuration); and the noise operator, which the scheme does not calibrate---reproducing full-resolution quadratic covariation on the grid is dominated by high-frequency and boundary structure that a smooth rank-$K$ truncation does not represent. ETH30 illustrates the framework's diagnostic value: $\hat a_\Delta$ measures the differential reversion rate directly; its small value, together with the out-of-sample deterioration in Table~\ref{tab:forecasting-results}, signals that stationarity is strained on that window.

Three directions for further work follow. First, the noise specification could be revisited. Additional noise modes concentrated near the center could be explored, but their operator form and empirical benefit remain to be established. Second, log coordinates keep raw liquidity positive at the cost of a back-transform bias, $\E[e^u]=e^{m+\sigma^2/2}>e^{\E[u]}$, which the lognormal correction in $R_V$ removes only under the Gaussian approximation; the BS/raw-liquidity branch of Section~\ref{sec:BS-model} keeps value linear in the profile, under the nonnegativity assumption stated there, and is the natural next calibration. Lastly, structural drift breaks the stationary scheme, and time-varying forcing $h(t,x)$ and price-aware coordinates are the direct remedies. With $\ell_t(z):=\ln L_t(e^z)$ and $u_t(x)=\ell_t(r_t+x)$, a smooth fixed-frame model whose noise is independent of $Z$ yields the moving-frame terms $\p_xu_t\,dr_t+\tfrac12\p_x^2u_t\,d\langle r\rangle_t$. These terms motivate a price-aware extension of \eqref{eqn:mkt-model-u-OU}. The regressors $\Delta r_j\,\p_xu_j$ and $\tfrac12(\Delta r_j)^2\p_x^2u_j$ offer exploratory small-step residual diagnostics in finite rank, since differentiation preserves $\mathcal P_{K-1}$. They require the profile in log-price coordinates: a price move shifts the profile in log-price, not in rank, so these diagnostics do not apply on the rank grid of Section~\ref{subsec:case-data} without a map from ranks to tick distances.

\section{Conclusion and Discussion} \label{sec:conclusion}

This paper developed a SPIDE framework for the dynamics of liquidity profiles in concentrated liquidity AMMs. The degenerate Sturm--Liouville operator $\cL$ encodes mean-reverting concentration around the market price, while the Hilbert--Schmidt integral operator $V$ captures nonlocal cross-tick redistribution; together they generate a \emph{family} of operator-based models rather than a single specification.

We established the existence, uniqueness, and temporal regularity of the mild solution, showed that the Gegenbauer eigenfunction expansion reduces the SPIDE to a coupled system of linear SDEs with financially interpretable coefficients, and, pairing the SPIDE with a log-price process, obtained tractable expressions for pool reserves, pool value, and loss-versus-rebalancing under both OU-type and BS-type noise specifications. The contribution is this whole family: a single calibration returns jointly interpretable estimates of the concentration rate $a$, the concentration order $\alpha$, and the relocations block $\bV_K$, any of which a practitioner may estimate freely or restrict a priori.

The empirical case study calibrated one branch of this family, the stationary additive-OU specification, on Uniswap ARB5, ETH5, and ETH30 pools, and clarified how the framework behaves in finite rank. Fixing a reference frame makes comparison across $\alpha$ coherent by removing the scale-collapse that an $\alpha$-dependent residual norm would otherwise induce. Within that fixed frame the calibration produces interpretable estimates whose scales match independently documented pool behavior, and diagnostics on ARB5 and ETH5 consistent with mean-reverting structure in the active band; but $\alpha$ and a flexible $\bV_K$ compete for the same low-order Gegenbauer-mode dynamics, and a rich relocations block can absorb damping that belongs on the quadratic eigenvalue ladder. The results support restricting the relocation block, without establishing that $\alpha=1/2$. On ETH30 the diagnostics suggest strain on this stationary branch rather than a failure of the framework: persistence dominates there, consistent with a training-mean mismatch under structural drift, and bounding $V$ correctly leaves the forecasts essentially unchanged (Section~\ref{subsubsec:case-eth30}).

Read this way, the case study is one disciplined calibration of one branch, and the broader framework retains the modularity to go beyond it. A practitioner may choose log or non-log coordinates, additive OU-type or multiplicative BS-type noise (the latter under a nonnegativity assumption on the profile), sparse or richer relocation supports, and stationary or time-varying drift, fitting the rest conditionally. The empirical limitations point directly at the unused branches. First, price-aware coordinates and time-varying parameters are the most direct candidates to address the held-out deterioration and possible structural drift. Second, state-dependent noise operators $\tilde{\sigma}_m(x, u_t)$ could reproduce local mid-price covariance spikes, though new identification and regularity conditions would be required. Third, replacing Brownian innovations with L\'evy or jump-driven noise would accommodate the heavy-tailed increments documented by~\citeA{risk2025dynamics}, at the cost of revisiting the mild-solution representation and moment-based calibration.

\section*{Acknowledgments}
C.-Y. Lee was supported in part by the Shenzhen Peacock grant 2025TC0013. S.-N. Tung is supported by NSTC grant 114-2115-M-007-012-MY3. T.-H. Wang expresses his gratitude for the hospitality of Professors Ching-Lung Lin and Yufen Huang as well as the Department of Mathematics at National Cheng Kung University during his stay in Tainan, Taiwan in August 2026.

\section*{Use of AI Tools}
OpenAI GPT-6.1 Sol and Claude Opus 5.5 were used in preparing this manuscript for coding assistance, manuscript comparison, editorial revision, and checks. The authors reviewed the resulting material and remain responsible for the manuscript's content.

\appendix

\section{Mathematical Technicalities} \label{sec:math-tech}
We collect the technical proofs in the appendix.

\subsection{Proof of Lemma \ref{lem:self-adj}} \label{pf:self-adj} $\mbox{ }$

\emph{(i).} It is clear that $E \subset D(\mathcal{L})$.
That $E$ is dense in $L^2_w$ follows from the Weierstrass approximation theorem together with the properties that each $e_n$ is a degree $n$ polynomial and $w$ is integrable under Assumption \ref{a:ab}.

\emph{(ii).} Consider two cases.

\noindent
{\bf Case 1:} $b/a\in (0,2)$.
In this case, both $w$ and $1/p$ are in $L^2((-L,L),dx)$, hence $\mathcal{L}$ is regular on $[-L,L]$; see, e.g., Definition 2.2 of \citeA{gesztesy2020on}.
Then, it follows from Theorem 2.6 of \citeA{gesztesy2020on} that $\mathcal{L}$ with the Neumann-type boundary condition for $D(\mathcal{L})$ as in \eqref{D(L)} is self-adjoint.

\noindent
{\bf Case 2:} $b/a \in [2,\infty)$.
In this case, $\mathcal{L}$ is in the limit point case at both endpoints $\pm L$; see, e.g., Definition 3.7 of \citeA{gesztesy2020on}.
Indeed, we observe that $\mathcal{L}u=0$, or $(pu')'=0$, has two independent solutions---one is $u(x) = 1$; another can be obtained by variation of parameters:
\[
	\tilde{u}(x)  = u(x) \int_0^x \frac{dy}{p(y) u(x)^2} = \int_0^x \frac{dy}{a(L^2-y^2)^{\frac{b}{2a}}}.
\]
A few lines of calculations and the condition $b/a\ge 2$ show that $u \in L^2((-L,L),w(x)dx)$, whereas $\tilde{u} \not\in L^2((-L,c), w(x)dx)$ and $\tilde{u} \not\in L^2((c, L), w(x)dx)$ for all $c \in (-L,L)$.
Hence, $\mathcal{L}$ is in the limit point case at $\pm L$.
By Theorem 3.8 of \citeA{gesztesy2020on}, the maximal operator is self-adjoint, and $\mathcal{L}$ with $D(\mathcal{L})$ defined in \eqref{D(L)} when $b/a \ge 2$ is precisely the maximal operator; see \citeA[Definition 3.2]{gesztesy2020on}.

The last part of \emph{(ii)} follows from self-adjointness and the fact that $\{e_n : n \in \N_0\}$ is an orthonormal basis of eigenfunctions of $\mathcal{L}$ with eigenvalues $-\lambda_n \le 0$ (see Lemma \ref{lem:eigen:L}).

\emph{(iii).} Since self-adjoint operators are closed, it follows from part \emph{(ii)} that $\mathcal{L}$ is a closed operator, which means that the graph $G(\mathcal{L}) = \{ (u, \mathcal{L}u) : u \in D(\mathcal{L}) \}$ is a closed subspace of $L^2_w \times L^2_w$ with respect to the norm $\|(u, \mathcal{L}u)\| = (\|u\|_w^2 + \|\mathcal{L}u\|_w^2)^{1/2}$; see, e.g., \citeA[Chapter 13]{rudin1991functional}. This implies that $D(\mathcal{L})=H^2_{\mathcal{L}}$ is closed with respect to \eqref{ip:H^2_L}.
Clearly, $\mathcal{L} : H^2_{\mathcal{L}} \to L^2_w$ is bounded.
By \emph{(i)}, $\{e_n : n \in \N_0\} \subset H^2_{\mathcal{L}}$. Note that $\{f_n : n \in \N_0 \}$ forms an orthonormal set since \eqref{eigen:L} and \eqref{orth:L} imply that
\begin{align*}
	\langle e_n, e_m \rangle_{H^2_{\mathcal{L}}}
	= \langle e_n, e_m \rangle_w + \langle \mathcal{L} e_n , \mathcal{L} e_m \rangle_w
	= (1+\lambda_n \lambda_m) \langle e_n, e_m \rangle_w = (1+\lambda_n^2) \delta_{n,m}.
\end{align*}
It remains to prove that $E$ is dense in $H^2_{\mathcal{L}}$ under $\|\cdot\|_{H^2_{\mathcal{L}}}$.
Let $u \in H^2_{\mathcal{L}}$. Then the partial sum $u_N = \sum_{n=0}^N \langle u, e_n\rangle_w e_n$ belongs to $E$ and converges to $u$ in $L^2_w$ by \emph{(i)}.
Applying \eqref{eigen:L} yields $\mathcal{L} u_N = -\sum_{n=0}^N  \lambda_n \langle u, e_n \rangle_w e_n$.
By \eqref{eigen:L} and \eqref{L:symm},
\[
	\sum_{n=0}^\infty \lambda_n^2 |\langle u,  e_n \rangle_w|^2 = \sum_{n=0}^\infty |\langle u, \mathcal{L} e_n \rangle_w|^2 = \sum_{n=0}^\infty |\langle \mathcal{L} u, e_n \rangle_w|^2 \le \|\mathcal{L} u\|_w^2 < \infty,
\]
which implies that $\mathcal{L}u_N$ converges (to $-\sum_{n=0}^\infty \lambda_n \langle u, e_n \rangle_w e_n$) in $L^2_w$.
But since $\mathcal{L}$ is closed, $\mathcal{L}u_N$ must converge to $\mathcal{L}u$ in $L^2_w$, and hence
\[
	\|u_N-u\|_{H^2_{\mathcal{L}}}^2 = \|u_N-u\|_w^2 + \|\mathcal{L} u_N-\mathcal{L}u\|_w^2 \to 0.
\]
This proves that $E$ is dense in $H^2_{\mathcal{L}}$ and completes the proof of \emph{(iii)}.

\emph{(iv).} Every $u \in L^2_w$ can be written as $u = \sum_{n=0}^\infty \langle u, e_n \rangle_w e_n$.
The proof of part \emph{(iii)} shows that $u\in H^2_{\mathcal{L}}$ if and only if $u_N$ and $\mathcal{L} u_N$ both converge in $L^2_w$, which is equivalent to $\sum_{n=0}^\infty |\langle u, e_n \rangle_w|^2 < \infty$ and $\sum_{n=0}^\infty \lambda_n^2 |\langle u, e_n \rangle_w|^2 < \infty$, respectively.
This proves \eqref{u:H^2_L}. Since $\mathcal{L}$ is closed, this also proves that $\mathcal{L} u = \lim_{N\to\infty} \mathcal{L} u_N = - \sum_{n=0}^\infty \lambda_n \langle u, e_n \rangle_w e_n = \sum_{n=0}^\infty \langle \mathcal{L}u, e_n \rangle_w e_n$ in $L^2_w$, thanks to \eqref{eigen:L} and \eqref{L:symm}.
This completes the proof of Lemma \ref{lem:self-adj}.

\subsection{Proof of Lemma \ref{lem:C_0}} $\mbox{ }$ \label{pf:C_0} \\
Note that \eqref{G:op} together with the orthonormality relation \eqref{orth:L} implies that
\[
	\|\mathcal{G}(t)z\|_w^2 = \sum_{n=0}^\infty e^{-2\lambda_n t} |\langle z, e_n \rangle_w|^2 \le \sum_{n=0}^\infty |\langle z, e_n \rangle_w|^2 = \|z\|_w^2.
\]
This shows that $\mathcal{G}(t): L^2_w \to L^2_w$ is a bounded operator with \eqref{G:norm}.
By definition, $\mathcal{G}(0)$ is the identity map.
Next, it follows from \eqref{G:op} and the orthonormality relation \eqref{orth:L} that
\begin{align*}
	\mathcal{G}(t)\mathcal{G}(s)z = \sum_{m=0}^\infty \sum_{n=0}^\infty e^{-\lambda_m t} e^{-\lambda_n s} \langle z, e_n\rangle_w\, \langle e_n, e_m\rangle_w e_m = \sum_{n=0}^\infty e^{-\lambda_n (t+s)} \langle z, e_n\rangle_w e_n = \mathcal{G}(t+s) z.
\end{align*}
This shows the semigroup property.
If $z \in L^2_w$ is fixed, then for any $s, t \ge 0$,
\begin{align*}
	&\|\mathcal{G}(t)z - \mathcal{G}(s)z\|_w^2
	= \sum_{n=0}^\infty |e^{-\lambda_n t} - e^{-\lambda_n s}|^2\, |\langle z, e_n \rangle_w|^2 \to 0
\end{align*}
as $s \to t$ by dominated convergence theorem,
and therefore $t \mapsto \mathcal{G}(t)z$ is continuous.
This proves that $\mathcal{G}(t)$ is a $C_0$-semigroup.
This completes the proof.

\subsection{Proof of Lemma \ref{lem:A}} $\mbox{ }$ \label{pf:lem:A}

\paragraph{\textit{Existence.}}
Let $z \in D(\cA)=D(\mathcal{L})$.
We construct a solution using Picard iteration.
Define $u_0(t)\equiv z$. For any $n \in \N_+$, define iteratively
\begin{align}\label{cp:A:picard}
	u_n(t) = \mathcal{G}(t)z - \int_0^t \mathcal{G}(t-r) V[u_{n-1}(r)] dr.
\end{align}
Fix $T>0$.
By \eqref{G:norm} and Lemma \ref{lem:V}, for any $s \in [0,T]$,
\begin{align*}
	\|u_{n+1}(s)-u_n(s)\|_w &\le \int_0^s \|\mathcal{G}(s-r) V[u_n(r)-u_{n-1}(r)]\|_w\, dr\\
	&\le  \int_0^s \|\mathcal{G}(s-r)\| \|V\| \|u_n(r)-u_{n-1}(r)\|_w dr\\
	&\le \|V\| \int_0^s \|u_n(r)-u_{n-1}(r)\|_w dr.
\end{align*}
Then
\[
	H_n(t):=\sup_{s \in [0,t]}\|u_{n+1}(s)-u_n(s)\|_w \le \|V\| \int_0^t H_{n-1}(s) ds.
\]
By Gronwall's lemma,
\begin{align*}
	H_n(t) \le H_1(T) \frac{(\|V\| T)^{n-1}}{(n-1)!} \quad \text{for all $n \in \N_+$ and $t \in [0,T]$.}
\end{align*}
Since $H_1(T) \le \|\mathcal{G}(t)z\|_w + \|z\|_w + \int_0^T \|V\|\|z\|_w dr < \infty$, we have $\sum_{n=1}^\infty H_n(t)<\infty$, which implies that for every $t \in [0,T]$, $u_n(t)$ converges to a limit $u(t) \in L^2_w$ with
\begin{align*}
	\sup_{t \in [0,T]}\|u_n(t) - u(t)\|_w \to 0.
\end{align*}
The preceding allows us to let $n\to \infty$ in \eqref{cp:A:picard} and obtain
\begin{align}\label{cp:A:mild}
	u(t) = \mathcal{G}(t)z - \int_0^t \mathcal{G}(t-r) V[u(r)] dr.
\end{align}
Hence, we may use \eqref{cp:A:mild} and Lemma \ref{lem:cp} to show that
\begin{align*}
	\frac{d}{dt}u(t) &= \frac{d}{dt}\mathcal{G}(t)z - \int_0^t \frac{d}{dt} \mathcal{G}(t-r) V[u(r)] dr - V[u(t)]\\
	&= \mathcal{L}\mathcal{G}(t)z - \int_0^t \mathcal{L}\mathcal{G}(t-r) V[u(r)] dr - V[u(t)]\\
	&= \mathcal{L}u(t) - V[u(t)] = \cA u(t).
\end{align*}
Hence, $\mathcal{S}(t)z:=u(t)$ is a strong solution to \eqref{cp:A} for any $z \in D(\cA)$.
By \eqref{G:norm} and Lemma \ref{lem:V}, $\mathcal{S}(t)$ is a bounded operator on $D(\cA)$, hence extends to a bounded operator on $L^2_w$.
Since $\mathcal{G}$ is a $C_0$-semigroup on $L^2_w$ whose generator is $\mathcal{L}$, one can verify that $\mathcal{S}$ is a $C_0$-semigroup on $L^2$ whose generator is $\cA$.
It follows that $u=\mathcal{S}(\cdot)z \in C([0,\infty); D(\cA)) \cap C^1([0,\infty); L^2_w)$; see, e.g., \citeA[Proposition A.2]{daprato2014stochastic}.

\paragraph{\textit{Uniqueness.}}
Suppose $\tilde{u}$ is another solution. Let $v=u-\tilde{u}$.
Then $v$ satisfies
\begin{align}\label{cp:v}
	\begin{cases}
		\frac{d}{dt} v = \cA v, & t \ge 0,\\
		v(0)=0.
	\end{cases}
\end{align}
For every $t \ge 0$, define
\[
	E(t) := \frac12 \|v(t)\|_w^2 = \frac12 \int_{-L}^L (v(t,x))^2 w(x) dx.
\]
Then it follows from \eqref{cp:v} and \eqref{eigen:L} that
\begin{align*}
	E'(t) &= \int_{-L}^L v\, v_t \, w \, dx = \int_{-L}^L v \, \mathcal{L}v \, w \, dx - \int_{-L}^L v\, V[v] \, w \, dx\\
	&= - \sum_{n=0}^\infty \lambda_n |\langle v, e_n \rangle_w|^2 - \langle V[v] , v \rangle_w.
\end{align*}
The preceding and Cauchy-Schwarz inequality together imply that
\[
	E'(t) \le \|V[v(t)]\|_w \|v(t)\|_w \le \|V\| \|v(t)\|_w^2 = 2\|V\| E(t).
\]
By Gronwall's lemma, $E(t) \le E(0) e^{2\|V\|t} = 0$, and hence $v=0$.
Uniqueness follows.

\subsection{Proof of Theorem \ref{th:mild:sol}} $\mbox{ }$ \label{pf:mild:sol} \\
The proof is based on a standard Picard iteration procedure.
Define
\[
	u_0(t) = \mathcal{G}(t) u_0 + \int_0^t \mathcal{G}(t-s) h(s) ds.
\]
For $n \in \N_+$, define iteratively the process
\begin{align}\begin{split}\label{picard}
	u_n(t) &= \mathcal{G}(t)u_0 + \int_0^t \mathcal{G}(t-s) h(s) ds\\
	& \quad - \int_0^t \mathcal{G}(t-s) V[u_{n-1}(s)] ds
	+ \int_0^t \mathcal{G}(t-s) B(u_{n-1}(s)) dW_s.
\end{split}\end{align}
Let $v_n(t) = u_n(t)-u_{n-1}(t)$. Then
\begin{align*}
	v_{n+1}(t)&= - \int_0^t \mathcal{G}(t-s)V[v_n(s)] ds + \int_0^t \mathcal{G}(t-s) \tilde{B}(v_n(s)) dW_s,
\end{align*}
where $\tilde{B}:L_w^2 \to L_2^0 = L_2(\R^M, L_w^2)$ is given by
\[
    \tilde{B}(v)x := \sum_{m=1}^M x_m \tilde\sigma_m[v]
\]
since the terms $\sigma_m$ cancel out; see \eqref{B} above.
Define the norm for the space $L^p(\Omega; L^2_w)$ of $L^p$ random variables $X: \Omega \to L^2_w$ by $\|X\|_{L^p(\Omega; L^2_w)} = [\E(\|X\|_w^p)]^{1/p}$.
By Minkowski inequality, \eqref{G:norm}, Lemma \ref{lem:V}, and Cauchy-Schwarz inequality,
\begin{align}\begin{split}\label{picard:bd:1}
	&\left\|\int_0^t \mathcal{G}(t-s)V[v_{n+1}(s)] ds\right\|_{L^p(\Omega; L^2_w)}
	 \le \int_0^t \|\mathcal{G}(t-s) V[v_n(s)]\|_{L^p(\Omega; L^2_w)} ds\\
	&\le \|V\| \int_0^t \|v_n(s)\|_{L^p(\Omega;L^2_w)} ds
	\le \|V\| \sqrt{T} \left( \int_0^t \|v_n(s)\|_{L^p(\Omega;L^2_w)}^2 ds\right)^{1/2}.
\end{split}\end{align}
For any $n \in \N_+$ and $s \in [0,T]$, $\Phi_n(s):= \mathcal{G}(t-s) \tilde{B}(v_n(s))$ defines an linear operator $\R^M \to L^2_w$ by
\[
	\Phi_n(s) : x =(x_1,\dots, x_M) \in \R^M \mapsto \sum_{m=1}^M x_m \mathcal{G}(t-s) \tilde\sigma_m[v_n(s)] \in L^2_w.
\]
Let $\{{\bf e}_1,\dots,{\bf e}_M\}$ denote the standard basis for $\R^M$.
Then, the Hilbert-Schmidt norm of the operator $\Phi_n(s): \R^M \to L^2_w$ satisfies
\begin{align*}
	\|\Phi_n(s)\|_{L^0_2}^2 &= \sum_{m=1}^M \|\Phi_n(s) {\bf e}_m\|_w^2
	= \sum_{m=1}^M \|\mathcal{G}(t-s)\tilde\sigma_m[v_n(s)]\|_w^2
    \le \sum_{m=1}^M \|\tilde\sigma_m\|^2 \|v_n(s)\|_w^2,
\end{align*}
thanks to \eqref{G:norm} and Assumption \ref{a:sigma}.
By Theorem 4.37 of \citeA{daprato2014stochastic},
\begin{align*}
	\E\left[ \left\| \int_0^t \Phi_n(s) dW_s \right\|_w^p \right]
	\le c_p \left[ \int_0^t \left( \E \|\Phi_n(s)\|_{L^0_2}^p\right)^{2/p} ds \right]^{p/2}.
\end{align*}
It follows that
\begin{align}\begin{split}\label{picard:bd:2}
	\left\|\int_0^t \mathcal{G}(t-s) \bar{B}(v_n(s)) dW_s\right\|_{L^p(\Omega; L^2_w)}^2
	&\lesssim \int_0^t \left( \E \|\Phi_n(s)\|_{L^0_2}^p\right)^{2/p} ds\\
	&\lesssim C_{\tilde\sigma} \int_0^t  (\E\|v_n(s)\|_w^p)^{2/p} ds,
\end{split}\end{align}
where $C_{\tilde\sigma} := \sum_{m=1}^M \|\tilde\sigma_m\|^2$.
Define
\[
	H_n(s) := \sup_{r \in [0,s]} \|v_n(s)\|_{L^p(\Omega; L^2_w)}^2.
\]
Combining \eqref{picard:bd:1} and \eqref{picard:bd:2}, we see that there exists $0<C_p<\infty$ such that for all $n \in \N_+$ and $0 \le s \le t \le T$,
\begin{align*}
	\|v_{n+1}(s)\|_{L^p(\Omega;L^2_w)}^2 &\le C_p (T\|V\|^2 + C_{\tilde\sigma}) \int_0^s \|v_n(r)\|_{L^p(\Omega;L^2_w)}^2 dr\\
	&\le C_p (T\|V\|^2 + C_{\tilde\sigma}) \int_0^t H_n(s) ds.
\end{align*}
Hence, there exists $0<C<\infty$ such that
\begin{align}\label{gronwall}
	H_{n+1}(t) \le C \int_0^t H_n(s) ds \quad \text{for all $n \in \N_+$ and $t \in [0,T]$.}
\end{align}
By Gronwall's lemma,
\begin{align*}
	H_n(t) \le H_1(T)\frac{(CT)^{n-1}}{(n-1)!} \quad \text{for all $n \in \N_+$ and $t \in [0,T]$.}
\end{align*}
Moreover, the above estimates also yield
\begin{align*}
	H_1(T)=\sup_{t \in [0,T]}\|v_1(t)\|_{L^p(\Omega; L^2_w)}^2 \le C_p (T\|V\|^2+C_{\tilde\sigma}) T H_0(T),
\end{align*}
where
\begin{align*}
	H_0(T):=\sup_{t \in [0,T]}\|u_0(t)\|_{L^p(\Omega; L^2_w)}^2 \le \left[\|u_0\|_w + \int_0^T \|h(s)\|_{L^p(\Omega; L^2_w)} ds \right]^2
	<\infty,
\end{align*}
thanks to Assumption \ref{a:u0:h}.
The last three displays together imply that
\[
	\sup_{t \in [0,T]}\|u_0(t)\|_{L^p(\Omega; L^2_w)} + \sum_{n=0}^\infty\sup_{t \in [0,T]} \|u_{n+1}(t)-u_n(t)\|_{L^p(\Omega; L^2_w)} < \infty.
\]
Therefore, for every $t \in [0,T]$, $u_n(t)$ converges to a limit $u(t) \in L^p(\Omega; L^2_w)$ such that
\[
	\sup_{t \in [0,T]}\|u_n(t)-u(t)\|_{L^p(\Omega; L^2_w)} \to 0 \quad \text{as $n \to \infty$.}
\]
In particular, $u$ satisfies \eqref{mild:Lp}.
Letting $n \to \infty$ in \eqref{picard} yields
\begin{align}\label{u:mild:form:G}
	u(t) = \mathcal{G}(t)u_0 + \int_0^t \mathcal{G}(t-s)h(s) ds
	- \int_0^t \mathcal{G}(t-s)V[u(s)] ds + \int_0^t \mathcal{G}(t-s)B(u(s)) dW_s.
\end{align}
Note that
\[
	\left\|\int_0^T \|u(s)\|_w\, ds \right\|_{L^p(\Omega)}\le
	\int_0^T \|u(s)\|_{L^p(\Omega; L^2_w)} ds < \infty,
\]
so $\int_0^T \|u(s)\|_w \, ds < \infty$ a.s.
In order to verify that $u$ is a mild solution, it remains to show that $u(t)$ satisfies \eqref{mild}.
Thanks to Lemma \ref{lem:S}, the last equation can be written as
\begin{align}\begin{split}\label{u:rewritten}
	u(t)
	& = \mathcal{S}(t)u_0 + \int_0^t \mathcal{S}(t-s) h(s) ds + \int_0^t \mathcal{S}(t-s) B(u(s)) dW_s\\
	& \quad - \int_0^t \mathcal{S}(t-s) V[u(s)] ds + R(t),
\end{split}\end{align}
where
\begin{align*}
	R(t)&:=\int_0^t \mathcal{S}(t-s) V[\mathcal{G}(s)u_0] ds
	- \int_0^t \left[ \int_0^{t-s} \mathcal{S}(t-s-r) V\big[\mathcal{G}(r) V[u(s)]\big] dr \right] ds\\
	&\quad + \int_0^t \left[ \int_0^{t-s} \mathcal{S}(t-s-r) V[\mathcal{G}(r) h(s)] dr \right]ds\\
	& \quad + \int_0^t \left[ \int_0^{t-s}\mathcal{S}(t-s-r) V\big[\mathcal{G}(r) B(u(s))\big] dr \right] dW_s.
\end{align*}
By Fubini's theorem, a stochastic Fubini theorem \cite[Theorem 4.33]{daprato2014stochastic}, and a change of variable, we deduce that
\begin{align*}
	R(t)
	& = \int_0^t \mathcal{S}(t-r)\, V\left[ \mathcal{G}(r)u_0 - \int_0^r \mathcal{G}(r-s) V[u(s)] ds \right.\\
	&\hskip1.2in \left. + \int_0^r \mathcal{G}(r-s) h(s) ds + \int_0^r \mathcal{G}(r-s) B(u(s)) dW_s \right] dr\\
	& = \int_0^t \mathcal{S}(t-r) V[u(r)] dr,
\end{align*}
where we have used \eqref{u:mild:form:G} to obtain the last equality.
Putting this back into \eqref{u:rewritten} yields
\[
	u(t) = \mathcal{S}(t)u_0 + \int_0^t \mathcal{S}(t-s) h(s) ds + \int_0^t \mathcal{S}(t-s) B(u(s)) dW_s.
\]
This proves that $u$ is a mild solution to \eqref{SPIDE:DZ-form}.

To show uniqueness, suppose $\tilde{u}$ is another mild solution satisfying \eqref{mild:Lp} for $p=2$. Letting $v = u - \tilde{u}$, we have
\[
	v(t) = -  \int_0^t \mathcal{G}(t-s) V[v(s)]ds + \int_0^t \mathcal{G}(t-s) \tilde{B}(v(s)) dW_s.
\]
Then we may use the same calculations that led to \eqref{gronwall} above to deduce that
\[
	H(t) \le C \int_0^t H(s) ds \quad \text{for all $t \in [0,T]$,}
\]
where
\[
	H(s) = \sup_{r \in [0,s]} \|v(r)\|_{L^2(\Omega; L^2_w)}^2.
\]
By Gronwall's lemma, $H(t) = 0$ for all $t \in [0,T]$ and hence $u(t) = \tilde{u}(t)$ in $L^2_w$ a.s.

\subsection{Proof of Theorem \ref{th:sys:sde}} \label{pf:sys:sde} $\mbox{ }$ \\
Recall from Theorem \ref{th:mild:sol} and \eqref{B} that the mild solution to \eqref{SPIDE} satisfies
\begin{align}\begin{split}\label{recall:mild}
	u(t) &= \mathcal{G}(t)u_0 + \int_0^t \mathcal{G}(t-s)h(s) ds\\
	& - \int_0^t \mathcal{G}(t-s)V[u(s)]ds  + \sum_{m=1}^M \int_0^t \mathcal{G}(t-s) B_m(u(s)) dW^m_s,
\end{split}\end{align}
where the heat semigroup $\mathcal{G}$ is given by \eqref{G:op} and $B_m: L_w^2 \to L_w^2$ is the mapping defined by
\[
    B_m(u) = \sigma_m + \tilde\sigma_m[u].
\]
By \eqref{sigma}, \eqref{tsigma:exp} and \eqref{sigma:exp},
\begin{align}\begin{split}\label{Bu:exp}
	B_m(u(t))(x) &= \sigma_m(x) + \tilde\sigma_m[u(t)](x)\\
    &= \sigma_m(x) + \int \tilde\sigma_m(x,y) u(t,y) w(y) dy
    = \sum_{i=0}^\infty \sigma_m^i e_n(x) + \sum_{k=0}^\infty \sigma_m^{ik} u^k_t e_i(x).
\end{split}\end{align}
Moreover, by \eqref{V} and \eqref{V:exp},
\begin{align}\label{Vu:exp}
	V[u(t)](x) = \int V(x,y) u(t,y) w(y) dy = \sum_{i=0}^\infty \left[ \sum_{k=0}^\infty V^{ik} u^k_t \right] e_i(x).
\end{align}
Therefore, for each $n \in \N_0$, we may project \eqref{recall:mild} onto the $n$-th eigenmode, and use \eqref{G:op}, \eqref{u_0:exp}--\eqref{h:exp}, and \eqref{Vu:exp}--\eqref{Bu:exp} to obtain
\begin{align}\begin{split}\label{sde:n-mode}
	u^n_t &= e^{-\lambda_n t} u^n_0 + \int_0^t e^{-\lambda_n(t-s)} h_s^n ds- \sum_{k=0}^\infty \int_0^t e^{-\lambda_n (t-s)} V^{nk} u^k_s ds\\
	& \quad  + \sum_{m=1}^M \int_0^t e^{-\lambda_n (t-s)} \sigma_m^n dW^m_s + \sum_{m=1}^M \sum_{k=0}^\infty \int_0^t e^{-\lambda_n (t-s)} \tilde\sigma_m^{nk} u^k_s dW^m_s.
\end{split}\end{align}
It follows that
\begin{align*}
	e^{\lambda_n t} u^n_t &= u^n_0 + \int_0^t e^{\lambda_n s} h_s^n ds - \sum_{k=0}^\infty \int_0^t e^{\lambda_n s} V^{nk} u^k_s ds\\
    & \quad + \sum_{m=1}^M \int_0^t e^{\lambda_n s} \sigma_m^{n} dW^m_s + \sum_{m=1}^M \sum_{k=0}^\infty \int_0^t e^{\lambda_n s} \tilde\sigma_m^{nk} u^k_s dW^m_s.
\end{align*}
It\^{o}'s lemma and the preceding imply that
\begin{align*}
	&e^{\lambda_n t} d u^n_t + \lambda_n e^{\lambda_n t} u^n_t dt
	= d(e^{\lambda_n t} u^n_t)\\
	& = e^{\lambda_n t} h^n_t dt - \sum_{k=0}^\infty e^{\lambda_n t} V^{nk} u^k_t dt + \sum_{m=1}^M e^{\lambda_n t} \sigma_m^n dW_t^m + \sum_{m=1}^M \sum_{k=0}^\infty e^{\lambda_n t} \tilde\sigma^{nk}_m u^k_t dW^m_t.
\end{align*}
Finally, subtract both sides by $\lambda_n e^{\lambda_n t} u^n_t$, then multiply by $e^{-\lambda_n t}$ to deduce equation \eqref{eq:u^n}.

\subsection{Proof of Theorem \ref{th:temporal-regularity}} \label{pf:temporal-regularity} $\mbox{ }$ \\
Recall the eigensystem $(\lambda_n, e_n)$ given in Lemma \ref{lem:eigen:L}.
\begin{lemma}\label{lem:eigen:bd}
Under Assumption \ref{a:ab}, the following properties hold:
\begin{enumerate}
\item[(i)] There exist $0<c<C<\infty$ such that $cn^2 \le \lambda_n \le Cn^2$ for all $n \in \N_0$;
\item[(ii)] $\sup_{n \in \N_0, x \in [-L+\delta, L-\delta]}|e_n(x)| < \infty$ for any $\delta \in (0,L)$.
\end{enumerate}
\end{lemma}

\begin{proof}
First, part \emph{(i)} follows directly from \eqref{eigen:L}.
For part \emph{(ii)}, fix $\delta \in (0,L)$ and set $\delta' := \delta/L \in (0,1)$.
By \eqref{e_n}, $|e_n(x)| = L^{-\alpha} K_{\alpha,n}^{-1} |C^\alpha_n(\xi)|$ with $\xi = x/L$, and $x \in [-L+\delta, L-\delta]$ if and only if $\xi \in [-1+\delta', 1-\delta']$.
Since $e_0$ is constant, it suffices to bound $e_n$ for $n \ge 1$.

We first bound $K_{\alpha,n}$ from below.
If $\alpha \ne 0$, then \eqref{orth:K} and the standard asymptotics $\Gamma(n+s)/\Gamma(n+r) \sim n^{s-r}$ as $n \to \infty$, which follow from Stirling's formula, give $K_{\alpha,n}^2 \sim \pi 2^{1-2\alpha} \Gamma(\alpha)^{-2} n^{2\alpha-2}$; all factors are positive for $n \ge 1$ since $\alpha > -1/2$.
If $\alpha = 0$, then $K_{0,n} = \sqrt{2\pi}/n$ for $n \ge 1$ by \eqref{orth:K}.
In both cases,
\begin{align}\label{K:lb}
	K_{\alpha, n} \gtrsim n^{\alpha-1} \quad \text{for all $n \ge 1$.}
\end{align}

Next, we bound $C^\alpha_n$ on $[-1+\delta', 1-\delta']$.
If $\alpha = 0$, then $C^0_n(\cos\theta) = \frac2n \cos(n\theta)$, so $|C^0_n(\xi)| \le 2n^{-1}$ for all $\xi \in [-1,1]$.
If $\alpha \ne 0$, recall the relation between Gegenbauer and Jacobi polynomials \cite[\S22]{abramowitz2012handbook}:
\begin{align*}
	C^\alpha_n(\xi) = \frac{\Gamma(n+2\alpha)\Gamma(\alpha+1/2)}{\Gamma(2\alpha)\Gamma(n+\alpha+1/2)} P^{(p,p)}_n(\xi), \quad p := \alpha - \frac12 > -1,
\end{align*}
where $\Gamma(2\alpha)$ is finite and nonzero because $2\alpha \in (-1,0) \cup (0,\infty)$.
Darboux's formula for Jacobi polynomials \cite[Theorem 8.21.8]{szego1975orthogonal} states that, for real $p, q$ and fixed $\varepsilon \in (0,\pi/2)$,
\begin{align*}
	P^{(p, q)}_n(\cos\theta) = n^{-1/2}k(\theta) \cos(N\theta+\gamma) + O_\varepsilon(n^{-3/2}),
\end{align*}
uniformly for all $\theta \in [\varepsilon, \pi-\varepsilon]$, where $k(\theta) = \pi^{-1/2} (\sin(\theta/2))^{-p-1/2}(\cos(\theta/2))^{-q-1/2}$, $N=n+(p+q+1)/2$, and $\gamma = -(p+1/2)\pi/2$.
Choose $\varepsilon$ with $\cos\varepsilon = 1-\delta'$, so that $\xi = \cos\theta \in [-1+\delta', 1-\delta']$ if and only if $\theta \in [\varepsilon, \pi-\varepsilon]$.
Since $k$ is bounded on $[\varepsilon, \pi-\varepsilon]$ and the gamma ratio is of order $n^{\alpha-1/2}$, we obtain
\[
	|C^\alpha_n(\xi)| \lesssim_{\delta} n^{\alpha-1/2} n^{-1/2} = n^{\alpha-1}
\]
uniformly for all $n \ge 1$ and $\xi \in [-1+\delta', 1-\delta']$.

Combining the bounds on $C^\alpha_n$ with \eqref{K:lb} gives $|e_n(x)| \lesssim_\delta L^{-\alpha}$ uniformly for all $n \ge 1$ and $x \in [-L+\delta, L-\delta]$, which proves part \emph{(ii)} for every $\alpha > -1/2$.
\end{proof}

\begin{proof}[Proof of Theorem \ref{th:temporal-regularity}]
Recall from the proof of Theorem \ref{th:sys:sde} that the eigenmodes \eqref{u:exp} of the mild solution satisfy the infinite dimensional system of SDEs \eqref{sde:n-mode}.
Fix $\eta \in (0,L)$ and $x \in [-L+\eta,L-\eta]$.
Fix $t, t'$ such that $0<\eta \le t \le t' \le T$. Set $\varepsilon = t'-t$.
Then, for each $n \in \N_0$, we can write
\begin{align*}
	u(t+\varepsilon,x)-u(t,x) = \sum_{n=0}^\infty (u^n_{t+\varepsilon} - u^n_t) e_n(x)
\end{align*}
and
\begin{align*}
	u^n_{t+\varepsilon} - u^n_t = J_0 + J_1 + \tilde{J}_1 + J_2 + \tilde{J}_2 + J_3 + \tilde{J}_3 + J_4,
\end{align*}
where
\begin{align*}
	&J_0 = (e^{-\lambda_n(t+\varepsilon)} - e^{-\lambda_n t}) u_0^n,\\
	&J_1 = \int_0^t (e^{-\lambda_n(t+\varepsilon-s)}-e^{-\lambda_n(t-s)}) h^n_s ds,\\
	&\tilde{J}_1 = \int_t^{t+\varepsilon} e^{-\lambda_n(t+\varepsilon-s)} h^n_s ds,\\
	&J_2 = \sum_{k=0}^\infty \int_0^t (e^{-\lambda_n(t+\varepsilon-s)} - e^{-\lambda_n(t-s)}) V^{nk} u^k_s ds,\\
	&\tilde{J}_2 = \sum_{k=0}^\infty \int_t^{t+\varepsilon} e^{-\lambda_n(t+\varepsilon-s)} V^{nk} u^k_s ds,\\
	&J_3 = \sum_{m=1}^M \sum_{k=0}^\infty \int_0^t (e^{-\lambda_n(t+\varepsilon-s)} - e^{-\lambda_n(t-s)}) \tilde\sigma^{nk}_m u^k_s dW^m_s,\\
	&\tilde{J}_3 = \sum_{m=1}^M \sum_{k=0}^\infty \int_t^{t+\varepsilon} e^{-\lambda_n(t+\varepsilon-s)} \tilde\sigma^{nk}_m u^k_s dW^m_s,\\
    &J_4 = \sum_{m=1}^M \sigma_m^n (W^m_{t+\varepsilon} - W^m_t).
\end{align*}
We estimate each term one after another.

\noindent
{\bf (1)} Estimating $J_0$:
By Lemma \ref{lem:eigen:bd}\emph{(i)} and \emph{(ii)}, $|J_0| \lesssim e^{-cn^2 t} (1 \wedge n^2 \varepsilon) |u^n_0|$ and
\begin{align*}
	\sum_{n=0}^\infty |J_0 e_n(x)| &\lesssim \sum_{n=0}^\infty e^{-cn^2 t} (1 \wedge n^2 \varepsilon) |u^n_0| \le \left( \sum_{n=0}^\infty e^{-2cn^2 t} (1 \wedge n^4 \varepsilon^2) \right)^{1/2} \left( \sum_{n=0}^\infty |u^n_0|^2 \right)^{1/2}\\
	& \lesssim \left( \int_0^\infty e^{-2cz^2 t} z^4 \varepsilon^2 dz \right)^{1/2} \|u_0\|_w \lesssim t^{-5/4} \|u_0\|_w \varepsilon \le \eta^{-5/4} \|u_0\|_w \varepsilon.
\end{align*}
\noindent
{\bf (2)} Estimating $J_1$:
By Cauchy-Schwarz inequality and Lemma \ref{lem:eigen:bd},
\begin{align*}
	\left|\sum_{n=0}^\infty J_1 e_n(x)\right| & \le \int_0^t \left| \sum_{n=0}^\infty e^{-\lambda_n(t-s)} (1-e^{-\lambda_n \varepsilon}) h^n_s e_n(x) \right| ds\\
	&\lesssim \int_0^t \left( \sum_{n=0}^\infty e^{-2cn^2(t-s)} (1\wedge n^4 \varepsilon^2) \right)^{1/2} \left( \sum_{n=0}^\infty |h^n_s|^2 \right)^{1/2} ds\\
	& \lesssim \varepsilon^{1/2} \int_0^t \left( \int_0^\infty e^{-2cz^2(t-s)} z^2 dz \right)^{1/2} \|h(s)\|_w \, ds\\
	& \lesssim \varepsilon^{1/2} \int_0^t (t-s)^{-3/4} \|h(s)\|_w \, ds.
\end{align*}
Then, Minkowski's inequality and Assumption \ref{a:u0:h:2} yield
\begin{align*}
	\left\|\sum_{n=0}^\infty J_1 e_n(x)\right\|_{L^p(\Omega)}
	&\lesssim \varepsilon^{1/2} \int_0^t (t-s)^{-3/4} [\E(\|h(s)\|_w^p)]^{1/p} ds\\
	&\le \varepsilon^{1/2} \int_0^t (t-s)^{-3/4} ds \lesssim T^{1/4} \varepsilon^{1/2}.
\end{align*}
{\bf (3)} Estimating $\tilde{J}_1$:
Similarly, by Cauchy-Schwarz inequality and Lemma \ref{lem:eigen:bd},
\begin{align*}
	\left| \sum_{n=0}^\infty \tilde{J}_1 e_n(x) \right|
	\lesssim \int_t^{t+\varepsilon} \left( \sum_{n=0}^\infty e^{-2cn^2(t+\varepsilon-s)} \right)^{1/2} \|h(s)\|_w\, ds
	\lesssim \int_t^{t+\varepsilon} (t+\varepsilon-s)^{-1/4} \|h(s)\|_w\, ds
\end{align*}
and hence
\begin{align*}
	\left\| \sum_{n=0}^\infty \tilde{J}_1 e_n(x) \right\|_{L^p(\Omega)}
	& \lesssim \int_t^{t+\varepsilon} (t+\varepsilon-s)^{-1/4}ds \lesssim \varepsilon^{3/4}.
\end{align*}
{\bf (4)} Estimating $J_2$:
Again, by Cauchy-Schwarz inequality and Lemma \ref{lem:eigen:bd},
\begin{align*}
	\left| \sum_{n=0}^\infty J_2 e_n(x) \right|
	& \lesssim \int_0^t \sum_{k=0}^\infty |u^k_s| \left| \sum_{n=0}^\infty e^{-\lambda_n(t-s)} (1-e^{-\lambda_n \varepsilon}) V^{nk} \right| ds\\
	&\lesssim  \int_0^t \sum_{k=0}^\infty |u^k_s| \left( \sum_{n=0}^\infty e^{-2cn^2(t-s)} (1 \wedge n^4 \varepsilon^2)  \right)^{1/2} \left( \sum_{n=0}^\infty |V^{nk}|^2 \right)^{1/2} ds\\
	& \lesssim \varepsilon^{1/2} \int_0^t (t-s)^{-3/4} \sum_{k=0}^\infty |u^k_s| \left( \sum_{n=0}^\infty |V^{nk}|^2 \right)^{1/2} ds\\
	&\le \varepsilon^{1/2} \int_0^t (t-s)^{-3/4} \left( \sum_{k=0}^\infty |u^k_s|^2 \right)^{1/2} \left( \sum_{k=0}^\infty \sum_{n=0}^\infty |V^{nk}|^2 \right)^{1/2} ds\\
	& = \varepsilon^{1/2} \int_0^t (t-s)^{-3/4} \|u(s)\|_w \,  \|V\| \, ds;
\end{align*}
See Remark \ref{rmk:V:sigma} and \eqref{u:l2} for the last identity.
Then, by Minkowski's inequality, Assumption \ref{a:V}, and \eqref{mild:Lp},
\begin{align*}
	\left\| \sum_{n=0}^\infty J_2 e_n(x) \right\|_{L^p(\Omega)}
	\lesssim \varepsilon^{1/2} \int_0^t(t-s)^{-3/4} ds \lesssim T^{1/4} \varepsilon^{1/2}.
\end{align*}
\noindent
{\bf (5)} Estimating $\tilde{J}_2$:
Similarly,
\begin{align*}
	\left| \sum_{n=0}^\infty \tilde{J}_2 e_n(x) \right|
	&\lesssim \int_t^{t+\varepsilon} \sum_{k=0}^\infty |u^k_s| \left| \sum_{n=0}^\infty e^{-\lambda_n(t+\varepsilon-s)} V^{nk} \right| ds\\
	& \lesssim \int_t^{t+\varepsilon} (t+\varepsilon-s)^{-1/4} \sum_{k=0}^\infty |u^k_s|  \left( \sum_{n=0}^\infty |v^{nk}|^2 \right)^{1/2} ds\\
	& \le \int_t^{t+\varepsilon} (t+\varepsilon-s)^{-1/4} \|u(s)\|_w \, \|V\| \, ds
\end{align*}
and hence
\begin{align*}
	\left\| \sum_{n=0}^\infty \tilde{J}_2 e_n(x) \right\|_{L^p(\Omega)} \lesssim \int_t^{t+\varepsilon} (t+\varepsilon-s)^{-1/4} ds \lesssim \varepsilon^{3/4}.
\end{align*}
\noindent
{\bf (6)} Estimating $J_3$:
Fix $\delta \in (0, 1/2)$.
Observe that Lemma \ref{lem:eigen:L}\emph{(i)} implies
\begin{align}\begin{split}\label{exp:eps:bd}
	e^{-\lambda_n(t-s)}(1-e^{-\lambda_n \varepsilon}) &\lesssim e^{-cn^2(t-s)} (1\wedge n^2 \varepsilon)\\
	&\le \varepsilon^{1/2-\delta} e^{-cn^2(t-s)} n^{1-2\delta}
	\lesssim_\delta \varepsilon^{1/2-\delta} (t-s)^{-1/2+\delta} e^{-\frac12 cn^2(t-s)}.
\end{split}\end{align}
By the Burkholder-Davis-Gundy inequality \cite[p.166, Theorem 3.28]{karatzas1991brownian}, the preceding estimate \eqref{exp:eps:bd}, and Cauchy-Schwarz inequality,
\begin{align*}
	\|J_3\|_{L^p(\Omega)}
	& \le \sum_{m=1}^M \E\left[ \left| \int_0^t e^{-\lambda_n(t-s)}(1-e^{-\lambda_n \varepsilon}) \sum_{k=0}^\infty (\tilde\sigma^{nk}_m u^k_s) dW^m_s \right|^p \right]^{1/p}\\
	& \lesssim_p \sum_{m=1}^M \E\left[ \left( \int_0^t | e^{-\lambda_n(t-s)}(1-e^{-\lambda_n \varepsilon})|^2 \left| \sum_{k=0}^\infty \tilde\sigma^{nk}_m u^k_s\right|^2 ds \right)^{p/2} \right]^{1/p}\\
	& \lesssim_\delta \varepsilon^{1/2-\delta} \sum_{m=1}^M \left( \sum_{k=0}^\infty |\tilde\sigma_m^{nk}|^2 \right)^{1/2} \E\left[ \left( \int_0^t (t-s)^{-1+2\delta} e^{-cn^2(t-s)} \|u(s)\|_w^2 \right)^{p/2} ds \right]^{1/p}.
\end{align*}
Then, by Minkowski's inequality, Lemma \ref{lem:eigen:L}\emph{(iii)}, the moment property \eqref{mild:Lp}, a change of variable, together with Assumption \ref{a:sigma:2},
\begin{align*}
	&\left\|\sum_{n=0}^\infty J_3 e_n(x)\right\|_{L^p(\Omega)}\\
	&\lesssim \varepsilon^{1/2-\delta} \sum_{m=1}^M \sum_{n=0}^\infty \left( \sum_{k=0}^\infty |\tilde\sigma_m^{nk}|^2 \right)^{1/2} \left( \int_0^t (t-s)^{-1+2\delta} e^{-cn^2(t-s)} [\E(\|u(s)\|_w^p)]^{2/p} ds \right)^{1/2}\\
	&\lesssim \varepsilon^{1/2-\delta} \sum_{m=1}^M \sum_{n=0}^\infty \left( \sum_{k=0}^\infty |\tilde\sigma_m^{nk}|^2 \right)^{1/2} \left( (1\wedge n^{-1-2\delta}) \int_0^\infty r^{-1+2\delta} e^{-cr} dr \right)^{1/2}
	\lesssim \varepsilon^{1/2-\delta}.
\end{align*}
\noindent
{\bf (7)} Estimating $\tilde{J}_3$:
Similarly,
\begin{align*}
	\|\tilde{J}_3\|_{L^p(\Omega)}
	& \le \sum_{m=1}^M \E\left[ \left| \int_t^{t+\varepsilon} e^{-\lambda_n(t+\varepsilon-s)}\sum_{k=0}^\infty (\tilde\sigma^{nk}_m u^k_s) dW^m_s \right|^p \right]^{1/p}\\
	& \lesssim_p \sum_{m=1}^M \E\left[ \left( \int_t^{t+\varepsilon} e^{-2cn^2(t+\varepsilon-s)} \left| \sum_{k=0}^\infty \tilde\sigma^{nk}_m u^k_s\right|^2 ds \right)^{p/2} \right]^{1/p}\\
	& \lesssim \sum_{m=1}^M \left( \sum_{k=0}^\infty |\tilde\sigma_m^{nk}|^2 \right)^{1/2} \E\left[ \left( \int_t^{t+\varepsilon}e^{-2cn^2(t+\varepsilon-s)} \|u(s)\|_w^2 \right)^{p/2} ds \right]^{1/p}.
\end{align*}
Hence, thanks to Assumptions \ref{a:sigma:2} and \eqref{mild:Lp}, we have
\begin{align*}
	\left\|\sum_{n=0}^\infty \tilde{J}_3 e_n(x)\right\|_{L^p(\Omega)}
	&\lesssim \sum_{m=1}^M \sum_{n=0}^\infty \left( \sum_{k=0}^\infty |\tilde\sigma_m^{nk}|^2 \right)^{1/2} \left( \int_t^{t+\varepsilon} e^{-2cn^2(t+\varepsilon-s)} ds \right)^{1/2}
	\lesssim \varepsilon^{1/2}.
\end{align*}
{\bf (8)} Estimating $J_4$: By Assumption \ref{a:sigma:2},
\[
    \left\| \sum_{n=0}^\infty J_4 e_n(x) \right\|_{L^p(\Omega)} \lesssim
    \sum_{m=1}^M \sum_{n=0}^\infty |\sigma_m^n| \|W^m_{t+\varepsilon}-W^m_t\|_{L^p(\Omega)} \lesssim \varepsilon^{1/2}.
\]
Putting together all the estimates above, we get that, for any $p \ge 2$ and $\delta \in (0,1/2)$,
\[
	\|u(t+\varepsilon, x) - u(t,x)\|_{L^p(\Omega)} \lesssim_{p,\delta} \,\varepsilon^{1/2-\delta}
\]
uniformly for all $x \in [-L+\eta, L-\eta]$ and for all $t$, $t'=t+\varepsilon$ such that $\eta \le t \le t' \le T$.
Therefore, the Kolmogorov continuity theorem \cite[p.53, Theorem 2.8]{karatzas1991brownian} implies that $t \mapsto u(t,x)$ is a.s. locally $\gamma$-H\"older continuous for any $\gamma \in (0,1/2-\delta)$.
Since $\delta \in (0,1/2)$ is arbitrary, this completes the proof.
\end{proof}

\subsection{Proof of Lemma \ref{lem:pointwise}} \label{pf:pointwise}
\begin{proof}
Write $\|\cdot\|$ for the norm of $L^2(\Omega)$, and let $B:=\sup_{t\le T}\E(\|u_t\|_w^2)^{1/2}$, which is finite by Theorem~\ref{th:mild:sol} with $p=2$. Write~\eqref{sde:n-mode} as $u=F+\mathcal K u$, where
\[
(\mathcal K c)^n_t:=-\sum_{k=0}^\infty\int_0^t e^{-\lambda_n(t-s)}V^{nk}c^k_s\,ds
\]
and $F^n_t$ collects the remaining terms of~\eqref{sde:n-mode}, with the solution $u$ inserted in the $\tilde\sigma_m$-term.

\emph{Step 1: the forcing.} Let $\rho_{m,n}:=(\sum_k|\tilde\sigma_m^{nk}|^2)^{1/2}=\|\tilde\sigma_m^*[e_n]\|_w$, so that $|\sum_k\tilde\sigma_m^{nk}u^k_s|\le\rho_{m,n}\|u_s\|_w$. Since $\lambda_n\ge0$, Minkowski's inequality and the It\^o isometry give, for $t\le T$,
\[
\|F^n_t\|\le|u^n_0|+\int_0^T|h^n_s|\,ds+\sqrt T\sum_{m=1}^M\left(|\sigma_m^n|+\rho_{m,n}B\right).
\]
Summing over $n$ and using Assumption~\ref{a:pointwise}(i) together with~\eqref{sigma:l1:l2} and~\eqref{tsigma:l1},
\[
K:=\sup_{t\le T}\sum_{n=0}^\infty\|F^n_t\|\le\sum_n|u^n_0|+T\sup_{s\le T}\sum_n|h^n_s|+\sqrt T\sum_{m=1}^M\Big(\sum_n|\sigma^n_m|+B\sum_n\rho_{m,n}\Big)<\infty.
\]

\emph{Step 2: a fixed point in $\ell^1$.} For families $c=(c^n_t)$ of $L^2(\Omega)$-valued functions measurable in $t\in[0,T]$, let $\|c\|_{(1)}:=\sup_{t\le T}\sum_n\|c^n_t\|$ and $\|c\|_{(2)}:=\sup_{t\le T}(\sum_n\|c^n_t\|^2)^{1/2}$, and let $X_1$ and $X_2$ be the corresponding Banach spaces, so that $X_1\subset X_2$ with $\|c\|_{(2)}\le\|c\|_{(1)}$. Since $\lambda_n\ge0$,
\[
\sum_n\|(\mathcal Kc)^n_t\|\le\|V\|_{\ell^1}\int_0^t\sum_k\|c^k_s\|\,ds,
\qquad
\Big(\sum_n\|(\mathcal Kc)^n_t\|^2\Big)^{1/2}\le\|V\|\int_0^t\Big(\sum_k\|c^k_s\|^2\Big)^{1/2}ds,
\]
the second by the Cauchy--Schwarz inequality and Minkowski's integral inequality, with the Hilbert--Schmidt norm $\|V\|$ of~\eqref{iint:V}. Iterating these Volterra bounds gives $\|\mathcal K^jc\|_{(i)}\le(\kappa_iT)^j\|c\|_{(i)}/j!$, with $\kappa_1=\|V\|_{\ell^1}$ and $\kappa_2=\|V\|$, so $c=F+\mathcal Kc$ has a unique solution in each $X_i$, given by the Neumann series. By Step 1, $F\in X_1$, so the solution $c\in X_1$ is also the unique solution in $X_2$. The mild solution satisfies $u=F+\mathcal Ku$ and $\|u\|_{(2)}=B<\infty$, so $u=c\in X_1$. The first bound and Gronwall's inequality then give $A_T=\|u\|_{(1)}\le Ke^{\|V\|_{\ell^1}T}$.

\emph{Step 3: pointwise values.} Fix $t\in[0,T]$ and $\delta\in(0,L)$, and let $C_\delta:=\sup\{|e_n(x)|:n\in\N_0,\ |x|\le L-\delta\}$, which is finite by Lemma~\ref{lem:eigen:bd}\emph{(ii)}. Since $\E\sum_n|u^n_t|\le A_T$, almost surely $\sum_n|u^n_t|<\infty$, and the series $\sum_n u^n_te_n$ converges absolutely and uniformly on $[-L+\delta,L-\delta]$ to a continuous function. The same series converges to $u_t$ in $L^2_w$, hence almost everywhere along a subsequence, so the two agree almost everywhere on $[-L+\delta,L-\delta]$. At $x=0$ the series also converges absolutely in $L^2(\Omega)$, since $\sum_n\|u^n_t e_n(0)\|\le C_0A_T$; this is the series defining $u(t,0)$ in the proof of Theorem~\ref{th:temporal-regularity}, whose version is a modification of it on $(0,T]$. The same estimate gives $\|u_t(0)\|\le C_0A_T$.
\end{proof}

\section{Empirical Calibration Details}\label{app:empirical-calibration}

This appendix records the technical choices behind the estimation scheme of Section~\ref{subsec:case-estimator}: the relocation supports and the truncation closure, the stability convention, the domain width, the Legendre reference algebra used to represent candidate operators (including the proof of Lemma~\ref{lem:ladder}), and the optimizer settings together with the fitted $\alpha$-profiles.

\subsection{Sparse Relocations and Truncation Closure}

The relocation operator $V$ is assumed Hilbert--Schmidt in Assumption~\ref{a:V}. In finite rank we estimate sparse low-order Gegenbauer truncations of $\bV_K$ (the configurations of Section~\ref{subsec:case-results}). The finite-rank reduction of Section~\ref{subsubsec:case-reduction} closes exactly only when the relocation kernel is supported on the retained Gegenbauer modes; relocation mass outside the retained block would appear as a reduced-model error through cross-boundary coupling. Restricting the support to the low Gegenbauer modes is therefore a model specification, motivated by their readings on the empirical rank grid: the constant mode $e_0$ is the profile level, the linear mode $e_1$ is rank-grid tilt, the quadratic mode $e_2$ is symmetric concentration, and the cubic mode $e_3$ is local skewness; limiting $\bV_K$ to these blocks isolates the interpretable interactions from high-frequency tail noise.

The Frobenius penalty in the objective \eqref{eq:objective},
\[
  \lambda_V\,\|\bV_\Delta\|_F^2=\lambda_V\sum_{ij}(\bV_\Delta)_{ij}^2,
\]
is the finite-rank regularizer motivated by the Hilbert--Schmidt restriction
\[
  \|V\|_{\mathrm{HS}}^2=\sum_{ij}|V^{ij}|^2<\infty .
\]
The penalty limits how much low-order damping is absorbed by $\bV_\Delta$. We report $\lambda_V=0$ and $3.0$, the endpoints of the examined path $\{0,\,0.3,\,1.0,\,3.0\}$.

\subsection{Stability and the Constant Mode}

The transition $\boldsymbol{\Phi}_K=\exp(-\bA_\Delta)$ defines a stationary predictor when all eigenvalues of the total drift $\bA_\Delta$ have positive real part, so that $\rho(\boldsymbol{\Phi}_K)<1$. We impose this condition directly on the total drift, with the stability margin $\eta_{\mathrm{stab}}$ in \eqref{eq:estimator}; the case study uses $\eta_{\mathrm{stab}}=10^{-8}$ (Table~\ref{tab:optimizer-settings}). A symmetric positive-definite restriction on $\bV_\Delta$ would be sufficient but unnecessarily strong: it would force all diagonal entries positive and would exclude negative level-feedback terms that are admissible whenever all eigenvalues of $\bA_\Delta$ have positive real part. The joint condition $\bA_\Delta\succeq0$ is also too restrictive. Since $\lambda_0=0$, the constant-direction quadratic form is
\[
  \boldsymbol{\varepsilon}_0^\top\bA_\Delta\boldsymbol{\varepsilon}_0=(\bV_\Delta)_{00},
  \qquad \boldsymbol{\varepsilon}_0=(1,0,\ldots,0)^\top,
\]
which would force $(\bV_\Delta)_{00}\ge0$.

The constant mode $e_0(x)=\mathrm{const}$ has eigenvalue $\lambda_0=0$, so the Sturm--Liouville generator $\cL$ supplies no direct mean-reverting damping for the profile level. Level stationarity is therefore a property of the combined drift $\bA_\Delta$, not of $\cL$ alone. The constant mode can be stabilized even with $(\bV_\Delta)_{00}<0$ if the remaining entries couple the level direction to damped higher modes. The constant direction is retained in the Legendre reference basis and the level row and column of $\bV_\Delta$ are estimated, never projected out.

This is concrete in the smallest coupled case. Take the level mode ($\lambda_0=0$) and one damped partner ($\lambda>0$), with embedded drift block
\[
  \begin{pmatrix} 0 & c \\ d & \lambda+V_{11} \end{pmatrix}.
\]
Its characteristic polynomial is $\mu^2-(\lambda+V_{11})\,\mu-cd$; with $\lambda+V_{11}>0$ the two roots have positive real part if and only if the determinant $-cd>0$, that is $cd<0$. The level mode can therefore be stabilized with no direct self-damping ($V_{00}=0$) precisely when it is cross-coupled to a damped mode through opposite-signed feedback, which is what makes observed level mean reversion evidence for $V$ rather than for $\cL$ (Section~\ref{subsec:case-estimator}).

\subsection{Domain Width}

The empirical domain half-width $L$ trades boundary truncation against tail-noise contamination. Increasing $L$ moves Christoffel--Darboux boundary amplification farther from the active trading region, but the extreme ticks of real Uniswap pools are sparse, stale, and dominated by microstructure noise. Projecting those tails contaminates the low-order modal coefficients. We therefore restrict the empirical analysis to $[-L,L]$ where the signal-to-noise ratio is high, accepting some boundary regularization to protect the estimator from stale-tail artifacts.

\subsection{Legendre Reference Representation: Algebra and Quadrature}\label{app:score-algebra}

This subsection records the construction of the fixed reference frame and the candidate operators of Section~\ref{subsubsec:case-fixedspace} at implementation level, and states and proves Lemma~\ref{lem:ladder}. Holding a reference space fixed while the operator parameters vary is the standard device of parametrized Galerkin and reduced-basis methods \cite{quarteroni2016reduced,hesthaven2016certified} and of ultraspherical spectral methods built on connection matrices \cite{olver2013fast,bella2015spectral}.

\paragraph{\textit{Basis construction.}}
Let $x_1<\cdots<x_N$ be the tick grid and $P_0,\ldots,P_{r-1}$ the Legendre polynomials. Form the Vandermonde-type matrix $\boldsymbol{V}^{\mathrm{raw}}\in\R^{N\times r}$ with entries $V^{\mathrm{raw}}_{kn}=P_n(x_k)$ and take its reduced QR factorization $\boldsymbol{V}^{\mathrm{raw}}=\boldsymbol{Q}\widetilde{\boldsymbol{R}}$. With $s_n=\operatorname{sign}(\widetilde{R}_{nn})$ (zeros mapped to $1$), set
\[
\boldsymbol{B}=\boldsymbol{Q}\,\mathrm{diag}(s_0,\ldots,s_{r-1}),
\qquad
\boldsymbol{B}^{(q)}=\big(P^{(q)}_n(x_k)\big)_{k,n}\;\widetilde{\boldsymbol{R}}^{-1}\mathrm{diag}(s_0,\ldots,s_{r-1}),\quad q=1,2 ,
\]
where $P^{(q)}_n$ is the $q$th derivative of $P_n$, computed analytically from the Legendre recurrence. Thus the columns of $\boldsymbol{B}$ are the grid samples of polynomials $b_0,\ldots,b_{r-1}$ forming a basis of the space $\mathcal{P}_{r-1}$ of polynomials of degree at most $r-1$, orthonormal in the Euclidean grid inner product, and $\boldsymbol{B}'$, $\boldsymbol{B}''$ are the grid samples of their exact derivatives. The Gegenbauer operator $\mathcal{D}_\alpha=(1-x^2)\partial_x^2-(2\alpha+1)x\,\partial_x$ is represented in Legendre reference coefficients by the Galerkin matrix
\begin{equation}\label{eq:Maffine}
\boldsymbol{M}(\alpha)=\boldsymbol{M}_0+\alpha\,\boldsymbol{M}_1,\qquad
\boldsymbol{M}_0=-\boldsymbol{B}^\top\big[(1-x^2)\circ\boldsymbol{B}''-x\circ\boldsymbol{B}'\big],\qquad
\boldsymbol{M}_1=2\,\boldsymbol{B}^\top\big[x\circ\boldsymbol{B}'\big],
\end{equation}
with $x\circ$ row-wise multiplication by the nodes $x_k$; its affine dependence on $\alpha$ makes the $\alpha$-sweep cheap. The diagnostic quantities recorded with each basis are the condition number of $\widetilde{\boldsymbol{R}}$ and, when the constant direction is handled separately, the leakage of the constant vector onto the basis.

\begin{lemma}\label{lem:ladder}
For every $\alpha > -1/2$, the space $\mathcal{P}_{r-1}$ of polynomials of degree at most $r-1$ is invariant under $\mathcal{D}_\alpha$, and $\boldsymbol{M}(\alpha)$ in \eqref{eq:Maffine} is the matrix of $-\mathcal{D}_\alpha$ restricted to $\mathcal{P}_{r-1}$, expressed in the Legendre reference basis. Consequently $\boldsymbol{M}(\alpha)=\boldsymbol{T}(\alpha)\,\mathrm{diag}\big(n(n+2\alpha)\big)_{n=0}^{r-1}\,\boldsymbol{T}(\alpha)^{-1}$, where the columns of $\boldsymbol{T}(\alpha)$ are the Legendre reference coefficients of the sampled orthonormal eigenfunctions $e^{(\alpha)}_0,\ldots,e^{(\alpha)}_{r-1}$ of Lemma~\ref{lem:eigen:L}; in particular the represented drift $a\,\boldsymbol{M}(\alpha)$ carries the eigenvalue ladder $a\,n(n+2\alpha)$ exactly.
\end{lemma}

\paragraph{\textit{Proof of Lemma~\ref{lem:ladder}.}}
For $p\in\mathcal{P}_{r-1}$, both $(1-x^2)p''$ and $xp'$ have degree at most $\deg p$, so $\mathcal{D}_\alpha$ maps $\mathcal{P}_{r-1}$ into itself. Sampling at $N\ge r$ distinct nodes is injective on $\mathcal{P}_{r-1}$, so the sampled space $\mathrm{colspan}(\boldsymbol{B})$ is an isomorphic copy of $\mathcal{P}_{r-1}$ and the samples of $-\mathcal{D}_\alpha b_n$ lie in $\mathrm{colspan}(\boldsymbol{B})$. Since the columns of $\boldsymbol{B}$ are Euclidean-orthonormal, the Legendre reference coefficient vector of any element of the sampled space is obtained by applying $\boldsymbol{B}^\top$; column $n$ of $\boldsymbol{M}(\alpha)$ in \eqref{eq:Maffine} is exactly $\boldsymbol{B}^\top$ applied to the samples of $-\mathcal{D}_\alpha b_n$. Hence $\boldsymbol{M}(\alpha)$ is the matrix of $-\mathcal{D}_\alpha\restriction_{\mathcal{P}_{r-1}}$ in the basis $(b_n)$. By Lemma~\ref{lem:ggb}, $-\mathcal{D}_\alpha C^\alpha_n=n(n+2\alpha)\,C^\alpha_n$ with $\deg C^\alpha_n=n$, so $C^\alpha_0,\ldots,C^\alpha_{r-1}\in\mathcal{P}_{r-1}$ are eigenvectors; the eigenvalues $n(n+2\alpha)$ are strictly increasing in $n$ for $\alpha>-1/2$, hence distinct, and the eigendecomposition $\boldsymbol{M}(\alpha)=\boldsymbol{T}(\alpha)\,\mathrm{diag}(n(n+2\alpha))\,\boldsymbol{T}(\alpha)^{-1}$ holds with the columns of $\boldsymbol{T}(\alpha)$ the Legendre reference coefficients of the sampled eigenpolynomials. Any nonzero column scaling leaves the decomposition unchanged; we fix the scaling to the orthonormalized eigenfunctions $e^{(\alpha)}_n$ of Lemma~\ref{lem:eigen:L}, $\boldsymbol{T}(\alpha)=\boldsymbol{B}^\top\big(e^{(\alpha)}_n(x_k)\big)_{k,n}$, so that the same matrix appears in the relocation-block similarity below. At the grid point $\alpha=0$ the basis uses the explicit $\alpha=0$ formulas of Lemma~\ref{lem:ggb}, with $C^0_0=1$ and the constant mode normalized directly by $K^2_{0,0}=\int_{-1}^1(1-\xi^2)^{-1/2}d\xi=\pi$. \qed

\paragraph{\textit{Quadrature masses for the singular weight.}}
The analysis side requires inner products against the weight $w_\alpha(x)=(1-x^2)^{\alpha-1/2}$, which is singular at $\pm1$ for $\alpha<1/2$. Restrict to the interior nodes $|x_k|<1$ and assign node $k$ the exact $w_\alpha$-mass of its Voronoi cell $[m_{k-1/2},m_{k+1/2}]$ (cell midpoints between neighbors; outer cells extended to $\mp1$):
\[
\omega_k=F_\alpha(m_{k+1/2})-F_\alpha(m_{k-1/2}),
\qquad
F_\alpha(t)=\operatorname{sign}(t)\,\tfrac12\,B\!\big(\tfrac12,\alpha+\tfrac12\big)\,
\widetilde{I}_{t^2}\!\big(\tfrac12,\alpha+\tfrac12\big),
\]
where $B(\cdot,\cdot)$ is the beta function and $\widetilde{I}$ the regularized incomplete beta function; $F_\alpha$ is the primitive of $w_\alpha$, obtained from the substitution $z=x^2$. These finite-volume masses integrate the boundary singularity exactly within each cell; a sampled rectangle rule degrades precisely in the lower part of the $\alpha$ grid where the Gram solve is most delicate. Write $\boldsymbol{\Omega}_\alpha=\mathrm{diag}(\omega_k)$.

\paragraph{\textit{Analysis map and relocation blocks.}}
Let $\boldsymbol{E}(\alpha)$ hold the interior-node samples of the orthonormal eigenfunctions $e^{(\alpha)}_0,\ldots,e^{(\alpha)}_{\max\mathcal{I}}$ of Lemma~\ref{lem:eigen:L}, and form the Gram matrix and analysis coefficients
\[
\widehat{\boldsymbol{G}}(\alpha)=\boldsymbol{E}^\top\boldsymbol{\Omega}_\alpha\boldsymbol{E}+\varrho\,\boldsymbol{I},
\qquad
\widehat{\boldsymbol{c}}(f)=\widehat{\boldsymbol{G}}(\alpha)^{-1}\boldsymbol{E}^\top\boldsymbol{\Omega}_\alpha\, f,
\qquad \varrho=10^{-8},
\]
for any grid function $f$ (restricted to interior nodes). For a chosen relocation support $\mathcal I\subseteq\{0,\ldots,r-1\}$, the elementary relocation block is
\begin{equation}\label{eq:Rblocks}
\boldsymbol{R}_{ij}(\alpha)=\boldsymbol{p}_i(\alpha)\,\boldsymbol{q}_j(\alpha)^\top,\qquad
\boldsymbol{p}_i(\alpha)=\boldsymbol{B}^\top\boldsymbol{e}_i(\alpha),
\qquad
\big(\boldsymbol{q}_j(\alpha)\big)_m=\big[\widehat{\boldsymbol{c}}(b_m)\big]_j,\quad i,j\in\mathcal{I},
\end{equation}
i.e., $\boldsymbol{q}_j$ collects the $j$th Gegenbauer coefficient of each Legendre reference basis function. The synthesis side is exact: for $i\le r-1$, $e^{(\alpha)}_i$ is a polynomial of degree $i$, so its samples lie in $\mathrm{colspan}(\boldsymbol{B})$ and $\boldsymbol{p}_i(\alpha)$ are its exact Legendre reference coefficients. Assembling the differential and relocation blocks, the implemented reference drift of Section~\ref{subsubsec:case-fixedspace} is
\begin{equation}\label{eq:Dconstruct}
\boldsymbol{D}(\theta_r;\alpha) = a_\Delta\,\boldsymbol{M}(\alpha) + \sum_{(i,j)\in\mathcal{I}\times\mathcal{I}} (\bV_\Delta^{(r)})_{ij}\,\boldsymbol{R}_{ij}(\alpha).
\end{equation}
In the exact-quadrature limit ($\boldsymbol{E}^\top\boldsymbol{\Omega}_\alpha\boldsymbol{E}\to\boldsymbol{I}$, $\varrho\to0$), $\widehat{\boldsymbol{c}}(b_m)_j\to\langle e^{(\alpha)}_j,b_m\rangle_{w_\alpha}$ and a direct computation shows that $\boldsymbol{R}_{ij}(\alpha)$ is the $\boldsymbol{T}(\alpha)$-similarity transform of the matrix unit at $(i,j)$ (single unit entry, zeros elsewhere), so the constructed drift \eqref{eq:Dconstruct} is exactly similar to the rank-$r$ Gegenbauer modal drift $\bA_\Delta^{(r)}$ of \eqref{eq:estimator}, recovering the exact reference representation \eqref{eq:Dtheta}, and the stability constraint transfers between the two representations. With the finite-volume masses the deviation from this limit is measured by the Gram residual $\|\boldsymbol{E}^\top\boldsymbol{\Omega}_\alpha\boldsymbol{E}-\boldsymbol{I}\|_\infty$ and the basis self-projection error, both recorded with every fit as conditioning diagnostics.

\subsection{Optimizer Settings and the \texorpdfstring{$\alpha$}{alpha}-Profile}\label{app:alpha-profile}

Table~\ref{tab:optimizer-settings} collects every numerical setting of the estimator \eqref{eq:estimator} in one place.

\begin{table}[!ht]
\centering
\small
\caption{Numerical settings of the calibration scheme (Section~\ref{subsubsec:case-estdef}).}
\label{tab:optimizer-settings}
\begin{tabularx}{\textwidth}{@{}l>{\raggedright\arraybackslash}X@{}}
\toprule
$\alpha$ grid $\mathcal{G}_\alpha$ & $-0.45$ to $3.00$ in steps of $0.05$ (70 points), profiled \\
Legendre reference dimension $r$ & $6$ \\
Bounds & $a_\Delta \in [10^{-8},\, 0.5]$; $|(\bV_\Delta)_{ij}| \le 0.5$ \\
Stability constraint & $\min \operatorname{Re}\,\mathrm{eig}(\bA_\Delta) \ge 10^{-8}$; reject $\rho(\boldsymbol{\Phi}) \ge 1.05$ \\
Objective ridge & $\epsilon = 10^{-7}$ in \eqref{eq:objective} \\
Gram ridge & $\varrho = 10^{-8}$ in the analysis solve (Appendix~\ref{app:score-algebra}) \\
Multistart & 64 unscrambled Sobol draws: $a_\Delta$ log-uniform over its bounds, relocation magnitudes power-shrunk toward zero ($u^2$ scaling), each draw spawning rank-one, diagonal, and dense start shapes, plus three fixed anchors \\
Local polish & SLSQP on the 4 best starts; tolerance $10^{-9}$, at most 160 iterations; finite-difference gradients; the stability margin enters as a smooth inequality \\
Centering & full temporal mean of the fitting segment \\
Penalty weight & $\lambda_V \in \{0,\, 3.0\}$ (Dense-3 and Dense-3 + penalty, Section~\ref{subsec:case-results}) \\
\bottomrule
\end{tabularx}
\end{table}

Table~\ref{tab:alpha-profile} quantifies the near-flatness of the profiled objective in $\alpha$ discussed in Section~\ref{subsubsec:case-alpha}: for each pool and block it reports the $\alpha$-band spanned by the five best grid points of the dense calibration ($\lambda_V=0$) and the spread of the objective across them. On the branch-compatible pools the five best candidates span a band of width $0.2$ while the objective moves by less than $3\times10^{-3}$; on ETH30 the flat band sits in the lower region of the grid (at the edge on W1), consistent with the diagnosis of Section~\ref{subsubsec:case-eth30}.

\begin{table}[!ht]
\centering
\small
\caption{Flatness of the profiled objective in $\alpha$ (dense calibration, $\lambda_V=0$, $r=6$, forecasting-protocol fits): the band spanned by the five best grid points and the max--min objective spread across them.}
\label{tab:alpha-profile}
\begin{tabular}{llccc}
\toprule
\textbf{Pool} & \textbf{Block} & $\hat\alpha$ & five-best $\alpha$-band & objective spread \\
\midrule
ARB5 & W1 & $0.55$ & $[0.45,\,0.65]$ & $5\times10^{-4}$ \\
ARB5 & Full & $0.35$ & $[0.25,\,0.45]$ & $8\times10^{-4}$ \\
ETH5 & W1 & $1.15$ & $[1.05,\,1.25]$ & $3\times10^{-4}$ \\
ETH5 & Full & $0.40$ & $[0.30,\,0.50]$ & $2.4\times10^{-3}$ \\
ETH30 & W1 & $-0.45$ & $[-0.45,\,-0.25]$ (at grid edge) & $8.4\times10^{-3}$ \\
ETH30 & Full & $-0.30$ & $[-0.40,\,-0.20]$ & $7\times10^{-4}$ \\
\bottomrule
\end{tabular}
\end{table}

\FloatBarrier
\section{Empirical Evaluation Metrics}\label{app:empirical-metrics}

This appendix defines the metrics of Section~\ref{subsec:case-results}. Let $u_j(x_k)$ be the observed log-liquidity profile on the grid, $\hat u_{j|j-1}(x_k)$ its one-step forecast, $U_j=e^{u_j}$, $m_k$ the cell widths of the midpoint partition of the grid (end cells clipped at the boundary nodes), and $\mathcal O$ the held-out index set. Every configuration, the martingale null included, is evaluated by the identical pipeline.

\paragraph{\textit{Active-band ratio.}} With $\mathcal X_A=\{x_k:|x_k|\le0.5\}$, the central half of the rank grid of Section~\ref{subsec:case-data} rather than a fixed price interval,
\[
  \mathrm{RMSE}_A(j)=
  \left(
  \frac{\sum_{x_k\in\mathcal X_A}m_k\{u_j(x_k)-\hat u_{j|j-1}(x_k)\}^2}
       {\sum_{x_k\in\mathcal X_A}m_k}
  \right)^{1/2},
  \qquad
  R_A=\frac{\sum_{j\in\mathcal O}\mathrm{RMSE}_A(j)}
           {\sum_{j\in\mathcal O}\mathrm{RMSE}^{\mathrm{rw}}_A(j)},
\]
where the superscript $\mathrm{rw}$ denotes the martingale null $\hat u_{j|j-1}=u_{j-1}$.

\paragraph{\textit{Value-weighted profile ratio.}} Forecast levels are lognormal-corrected,
\[
  \hat U^{\mathrm{ln}}_{j|j-1}(x_k)=\exp\{\hat u_{j|j-1}(x_k)+\tfrac12\hat\sigma_k^2\},
\]
with $\hat\sigma_k^2$ the per-tick training-residual variance of the configuration being evaluated (the null uses its own). The score applies the bin-integrated weights of the pool-value formula \eqref{eq:value-r-U-finite} at unit spot price to the profile on the grid,
\[
  \mathcal V_j=\sum_k v_k\,U_j(x_k),
  \qquad
  v_k=\int_{\mathrm{cell}\,k} e^{\xi-\xi^-}\,d\xi ,
\]
and $R_V$ is the ratio of the held-out RMSE of $\hat{\mathcal V}^{\mathrm{ln}}_{j|j-1}$ to that of the null; the spot-price factor $e^{2r}$ cancels, so $R_V$ isolates the profile forecast. Because the grid coordinate is the normalized tick rank of Section~\ref{subsec:case-data} rather than log-moneyness, $\mathcal V_j$ is a standardized value-weighted profile score, not the pool value in token units; a pool-value forecast error would require a map from ranks to tick distances and a recomputed score.

\paragraph{\textit{Residual whiteness.}} Whiteness is measured in a shared PCA evaluation space of dimension $d=9$, fitted once per window so that no configuration is evaluated in its own fitted coordinates: center the full-window profiles at their temporal mean, weight grid point $k$ by $\sqrt{m_k}$, and take the top $d$ right singular vectors; $\boldsymbol P\in\R^{d\times N}$ applies the same weighting and projection. With $\boldsymbol y_j=\boldsymbol P(\bu_j-\bar{\bu})$, residuals $\boldsymbol r_j=\boldsymbol y_j-\hat{\boldsymbol y}_{j|j-1}$, $T$ evaluated residuals, and lag cutoff $H=10$,
\[
  \widehat{\boldsymbol C}_\tau=\frac1{T-\tau}\sum_j\boldsymbol r_{j+\tau}\boldsymbol r_j^\top,\qquad
  Q_H=T(T+2)\sum_{\tau=1}^H\frac1{T-\tau}
  \operatorname{tr}\!\left(
  \widehat{\boldsymbol C}_\tau^\top\widehat{\boldsymbol C}_0^{-1}
  \widehat{\boldsymbol C}_\tau\widehat{\boldsymbol C}_0^{-1}
  \right),
  \qquad S_H=\frac{Q_H}{d^2H}.
\]
Under whiteness $Q_H$ is approximately $\chi^2$ with $d^2H$ degrees of freedom, so $S_H\approx1$; the objective \eqref{eq:objective} does not enforce vanishing residual autocorrelation, so a severity near unity is an over-identifying check. Comparing in-sample and out-of-sample severity separates in-sample misspecification from parameter drift; both are tabulated in Table~\ref{tab:forecasting-results}.

\FloatBarrier
\section{Fitted Relocation Matrices}\label{app:V-matrices}

Table~\ref{tab:V-matrices-dense3} reports the Dense-3 relocation matrices. The penalty reduces the dominant off-diagonal entries in magnitude while preserving their signs.

\begin{table}[!ht]
\centering
\small
\setlength{\tabcolsep}{3pt}
\caption{Fitted Dense-3 relocation matrices $\widehat{\bV}_\Delta$ (Gegenbauer modes $\{0,1,2\}$, per sampling step, nominally 8 hours, forecasting-protocol fits) for the ARB5 and ETH5 W1 and Full windows. Left block: no penalty ($\lambda_V = 0$). Right block: with penalty ($\lambda_V = 3.0$). Row $i$, column $j$ gives $\hat{V}_{ij}$.}
\label{tab:V-matrices-dense3}
\resizebox{\textwidth}{!}{\begin{tabular}{l cccc cccc}
\toprule
 & \multicolumn{4}{c}{\textbf{No penalty ($\lambda_V = 0$)}} & \multicolumn{4}{c}{\textbf{Penalty ($\lambda_V = 3.0$)}} \\
\cmidrule(lr){2-5}\cmidrule(lr){6-9}
 & ARB5 W1 & ARB5 Full & ETH5 W1 & ETH5 Full & ARB5 W1 & ARB5 Full & ETH5 W1 & ETH5 Full \\
\midrule
$\hat{V}_{00}$ & $0.118$ & $0.112$ & $0.158$ & $0.031$ & $0.063$ & $0.068$ & $0.081$ & $0.021$ \\
$\hat{V}_{01}$ & $-0.007$ & $0.028$ & $0.055$ & $0.037$ & $-0.004$ & $0.014$ & $0.034$ & $0.023$ \\
$\hat{V}_{02}$ & $-0.181$ & $0.020$ & $-0.091$ & $-0.052$ & $-0.041$ & $0.012$ & $-0.041$ & $-0.022$ \\
$\hat{V}_{10}$ & $0.040$ & $0.096$ & $0.122$ & $-0.000$ & $0.022$ & $0.041$ & $0.033$ & $0.000$ \\
$\hat{V}_{11}$ & $0.067$ & $0.045$ & $0.042$ & $0.016$ & $0.036$ & $0.025$ & $0.030$ & $0.015$ \\
$\hat{V}_{12}$ & $0.021$ & $-0.025$ & $-0.035$ & $0.001$ & $-0.010$ & $-0.002$ & $-0.020$ & $-0.004$ \\
$\hat{V}_{20}$ & $-0.001$ & $-0.008$ & $0.003$ & $-0.019$ & $-0.000$ & $-0.009$ & $0.003$ & $-0.015$ \\
$\hat{V}_{21}$ & $0.041$ & $-0.001$ & $-0.040$ & $-0.021$ & $0.030$ & $-0.002$ & $-0.025$ & $-0.015$ \\
$\hat{V}_{22}$ & $0.103$ & $0.022$ & $0.077$ & $0.003$ & $0.057$ & $0.015$ & $0.033$ & $-0.004$ \\
\bottomrule
\end{tabular}}
\end{table}

\bibliographystyle{apacite}
\bibliography{Reference}
\end{document}